\UseRawInputEncoding
\documentclass[%
reprint,
amsmath,amssymb,
aps,
]{revtex4-2}

\usepackage{graphicx} %
\usepackage{dcolumn}%
\usepackage{xcolor}
\usepackage{booktabs}
\usepackage{array}
\usepackage{booktabs}
\usepackage{siunitx}
\usepackage[caption=false]{subfig}
\usepackage{hyperref}
\usepackage{bm}

\newcommand{\mshiva}{\mbox{M-SHIVA}}

\begin{document}

    \title{\mshiva~1.0: A Relativistic Resistive MHD Code for Heavy-Ion Collisions, with Applications to Momentum Anisotropy
and Joule Heating in QGP}
	\author{Krishanu Sengupta}
	\email{krishanu.sengupta@niser.ac.in}
	\author{Victor Roy}
	\email{victor@niser.ac.in}
	\affiliation{School of Physical Sciences, National Institute of Science Education and Research, An OCC of Homi Bhabha National Institute, Jatni - 752050, India}
	
\date{\today}

\begin{abstract}
We present \mshiva~1.0 (Magneto-resistive Solver for Heavy Ions with
finite-Volume Algorithms), a $(2+1)$-dimensional relativistic resistive
magnetohydrodynamics code for heavy-ion collisions in Cartesian and
boost-invariant Milne coordinates. The present implementation treats inviscid
dynamics and employs an implicit-explicit (IMEX) Runge--Kutta scheme for the
stiff conductivity-dependent terms, together with complementary primitive-variable
recovery procedures and generalized Lagrange multiplier (GLM) divergence
cleaning. The code reproduces standard shock-tube, cylindrical-blast-wave and
relativistic-rotor benchmarks. We use the framework to investigate the momentum
anisotropy and electromagnetic energy transfer in an expanding quark--gluon
plasma (QGP). In a controlled elliptic-fireball setup, the electromagnetic field
modifies the total momentum anisotropy mainly through its direct contribution to
the transverse stress tensor, while the matter anisotropy is only weakly
affected. We further separate the electromagnetic work into mechanical
Lorentz-force work and irreversible Joule heating. Applications to Au+Au
collisions at $\sqrt{s_{\rm NN}} = 62.4$ and $200$~GeV and Pb+Pb collisions
at $\sqrt{s_{\rm NN}} = 2.76$~TeV employ optical-Glauber matter profiles,
spectator-generated electromagnetic fields, and a temperature-dependent
electrical conductivity motivated by lattice-QCD results. Increasing the
conductivity enhances the early Ohmic response while accelerating the relaxation
of the electric field. The ratio of Joule heating to expansion cooling is
approximately $10^{-14}$--$10^{-11}$ for Pb+Pb collisions at $2.76$~TeV,
$10^{-8}$--$10^{-5}$ for Au+Au collisions at $200$~GeV and $10^{-7}$--$10^{-4}$
at $62.4$~GeV, with no appreciable conductivity dependence of the matter or total
momentum anisotropy. These results indicate that, although the relative
importance of resistive heating grows by several orders of magnitude with
decreasing collision energy, it remains energetically negligible compared with
longitudinal expansion cooling for all systems considered.
\end{abstract}

\keywords{resistive relativistic magnetohydrodynamics, IMEX Runge--Kutta,
finite-volume method, GLM divergence cleaning, Milne coordinates, heavy-ion
collisions, Joule heating}

\maketitle
\section{\label{sec:Intro}Introduction}

In relativistic heavy-ion collisions, the charged protons inside the colliding nuclei generate extremely intense electromagnetic fields during the earliest stages of the collision, with magnetic-field strengths typically estimated to be of order $10^{18}\,\mathrm{G}$ at RHIC energies and $10^{19}\,\mathrm{G}$, or larger, at LHC energies~~\cite{doi:10.1142/S0217751X09047570, BZDAK2012171, PhysRevC.83.054911,
doi:10.34133/research.0726, PhysRevC.107.034901, Inghirami2020,
PhysRevC.92.064902, MCLERRAN2014184, PhysRevC.85.044907, PhysRevC.110.024902,
Huang_2016, universe3010007}. In non-central collisions the event-averaged field acquires a preferred direction perpendicular to the reaction plane, while event-by-event fluctuations of the proton positions produce fields of comparable magnitude even in central collisions~\cite{BZDAK2012171, PhysRevC.85.044907}.
The subsequent evolution of these initial strong
fields depends not only on the sources (charged protons in the colliding nuclei)
but also on the electrical response of the produced quark-gluon plasma
(QGP)~\cite{PhysRevC.88.024911, PhysRevC.89.054905, Inghirami2016,
STEWART2021122308, particles9010029}. A quantitatively controlled description of this evolution therefore requires a dynamical framework in which the fluid variables and the electromagnetic fields are evolved consistently at finite electrical conductivity. Relativistic resistive magnetohydrodynamics (RRMHD) provides such
a framework, in which one solves for the energy-momentum tensor of the combined
field and matter (fluid) system self-consistently with the help of appropriate
constitutive equations~\cite{Nakamura2023, Inghirami2016, Mayer2025b, Mayer2025a,
10.1093/mnras/stz1015, PhysRevD.88.044020}.

The numerical solution of RRMHD is challenging because the
conductivity-dependent current in Ohm's law introduces stiff source terms. In
the large-conductivity regime, a purely explicit method would require a time
step much smaller than the usual Courant time step, scaling with the resistive
relaxation time. Implicit-explicit (IMEX) time integration avoids this
restriction by treating the hyperbolic transport explicitly and the stiff
electric-field relaxation implicitly~\cite{ASCHER1997151,
10.1111/j.1365-2966.2009.14454.x, PhysRevD.107.056003,
10.1093/acprof:oso/9780198528906.001.0001}. This strategy is particularly useful
for heavy-ion applications, where one may need to deal with regimes ranging from
weakly conducting to nearly ideal MHD, owing to the temperature-dependent
electrical conductivity of the hot QGP and the comparatively cooler hadronic
phase~\cite{PhysRevLett.110.182301, PhysRevD.95.036018, FRANCIS2012212,
PhysRevD.98.054005, PhysRevD.94.034025, PhysRevD.96.094003}.

A second numerical issue with RRMHD is the preservation of the Maxwell
constraints. In a finite-volume discretization, truncation errors can generate
non-zero values of the divergence of the magnetic field
($\nabla\cdot\mathbf{B} \neq 0$) and of the Gauss law for the electric field
($\nabla\cdot\mathbf{E} - q \neq 0$). We therefore use the standard generalized
Lagrange multiplier (GLM) formulation, in which auxiliary fields $\phi$ and
$\psi$ propagate and damp constraint violations~\cite{DEDNER2002645, TOTH2000605,
MIGNONE20102117, SUSANTO2013141}. This allows the scheme to remain robust near
shocks and strong electromagnetic gradients while preserving a conservative
update for the dynamical variables.

The present work develops \mshiva~1.0 (Magneto-resistive Solver for Heavy Ions
with finite-Volume Algorithms), a finite-volume RRMHD code in $2+1$ dimensions
in Cartesian and boost-invariant Milne coordinates, which has been tested and
validated against several important
benchmarks~\cite{10.1093/mnras/sty419, 10.1111/j.1365-2966.2007.12448.x,Nakamura2023, Inghirami2016}.
The conductivity-dependent electric-field relaxation in Amp\`ere's law is treated
implicitly through a local cell-wise update, while the hyperbolic transport and
non-stiff source terms are evolved explicitly. Particular attention is given to
primitive-variable recovery: an inversion based on the Lorentz-boosted enthalpy
$Z = w\gamma^{2}$ is employed during the explicit evolution, while a
complementary pressure-based inversion is applied after the implicit
electric-field update. The numerical implementation is examined using standard
Cartesian benchmarks and a relativistic rotor in Milne coordinates, which test
shock capturing, the resistive-to-ideal transition, multidimensional magnetic
stresses, and transverse fluid--electromagnetic dynamics in an expanding
background.

We subsequently use our code in Milne coordinates to study the momentum
anisotropy and the electromagnetic energy transfer inside the QGP produced in
relativistic heavy-ion collisions. In a controlled elliptic-fireball setup, we
separate the matter and electromagnetic contributions to the total momentum
anisotropy. We further evaluate the total electromagnetic work done on the
fluid, which has contributions from the mechanical work and from resistive
heating. Here we would like to mention a recent work~\cite{Lier2025} that uses a
higher-form-symmetry formulation of RRMHD, in which the evolution can be studied
without explicitly solving Amp\`ere's law; this formulation therefore provides an
alternative route to avoiding stiff equations in the high-conductivity regime.
There, the electric field is not evolved as an independent field variable, and
finite-resistivity effects are instead encoded constitutively through a gradient
expansion. Another recent
study~\cite{gyk4-trpr} adopts an
ultra-relativistic equation of state without a conserved matter-density current
and solves the BDNK MHD equations numerically for astrophysical systems. The present framework, by contrast, retains a conserved baryon current and can
therefore accommodate a baryon-density-dependent EoS; in the applications
presented here, we use both an ideal conformal EoS and a lattice-QCD EoS at
vanishing baryon chemical potential~\cite{Borsanyi2010}. Moreover, we evolve
the electric field explicitly, solving Amp\`ere's law with an IMEX scheme,
which is well suited to stiff source terms~\cite{Pareschi2005, doi:10.1137/080713562, doi:10.1137/0732037, doi:10.1137/S0036142900375906}, and we evaluate the Joule heating directly.

Finally, we apply \mshiva~1.0 to realistic heavy-ion initial conditions and study the evolution of the momentum anisotropy and the Joule heating in Au+Au collisions at $\sqrt{s_{\rm NN}} = 62.4$ and $200$~GeV and Pb+Pb collisions at
$\sqrt{s_{\rm NN}} = 2.76$~TeV.
The initial energy density is obtained from the
optical Glauber model, while the initial electromagnetic fields are computed
from the spectator protons alone, whose positions are sampled from a Monte
Carlo Glauber model with the same nuclear parameters and impact parameter as
the optical calculation. Using these simulations, we examine how the electrical
conductivity, the equation of state, the collision system, and the impact
parameter affect the relaxation of the electromagnetic fields, the development
of momentum anisotropy, and the transfer of energy from the fields to the
fluid. All heavy-ion results presented here are for a longitudinally
boost-invariant QGP with a constant, isotropic electrical conductivity, and are
obtained with a lattice-QCD-based
equation of state~\cite{Borsanyi2010}.

The manuscript is organized as follows. Section~\ref{RRMHD_eq} summarizes the
RRMHD equations. Section~\ref{sec:numerical_method} describes the numerical
method, including the finite-volume discretization, reconstruction, HLL fluxes,
IMEX splitting, primitive-variable recovery, and numerical admissibility.
Section~\ref{sec:benchmark_strategy} summarizes the benchmark strategy.
Section~\ref{sec:benchmarks_results} presents the Cartesian and Milne-coordinate
benchmark problems together with the magnetically modified anisotropic-expansion
study. Section~\ref{sec:hic_initial_conditions} introduces the heavy-ion initial
conditions and presents the results for Au+Au and Pb+Pb collisions. Finally,
Section~\ref{sum_and_con} summarizes the study and discusses the main
conclusions.

In this work we use natural units, $\hbar = c = k_{B} = \mu_{0} = \epsilon_{0}
= 1$, where $\epsilon_{0}$ and $\mu_{0}$ are the permittivity and permeability of
free space, respectively. The signature of the metric used in this work is mostly
negative, i.e., for flat space-time $\eta_{\mu \nu} = \text{diag}[1,-1,-1,-1]$.
Contravariant tensors are denoted by upper indices, i.e. $A^{\mu}$ and
$A^{\mu \nu}$, where the Greek indices run from $0$ to $3$ and the Latin indices
from $1$ to $3$. The covariant tensor is denoted by lower indices and is defined
as $A_{\mu} := g_{\mu \nu} A^{\nu}$, where $g_{\mu \nu}$ is the metric tensor,
whose determinant is defined as $g := \text{det}[g_{\mu \nu}]$.

Given a contravariant four-vector $A^{\mu}$, the covariant derivative is defined
as
\begin{equation}
	A^{\mu}_{\; ;\nu} := \partial_{\nu}A^{\mu} + \Gamma^{\mu}_{\alpha \nu} A^{\alpha},
	\label{covarient_derivative_con}
\end{equation}
where $\Gamma^{\mu}_{\alpha \nu}$ is the Christoffel symbol, defined as
$\Gamma^{\mu}_{\alpha \nu} = \frac{1}{2}g^{\mu \beta}
(-\partial_{\beta}g_{\alpha \nu} + \partial_{\alpha}g_{\beta \nu}
+ \partial_{\nu}g_{\alpha \beta})$, and $\partial_{\nu} = \partial/\partial
x^{\nu}$ denotes the four-derivative. Similarly, the covariant derivative of a
covariant four-vector is given by
\begin{equation}
	A_{\mu ; \nu} := \partial_{\nu}A_{\mu} - \Gamma^{\alpha}_{\mu \nu}A_{\alpha}.
	\label{covarient_derivative_cov}
\end{equation}
Using Eqs.~\eqref{covarient_derivative_con} and
\eqref{covarient_derivative_cov}, we can calculate the covariant derivative of
any higher-rank tensor. Given a second-rank contravariant tensor
$A^{\mu \nu}$, the covariant derivative is
\begin{equation}
	A^{\mu \nu}_{\quad ;\alpha} = \partial_{\alpha} A^{\mu \nu}
	+ \Gamma^{\mu}_{\alpha \beta} A^{\beta \nu}
	+ \Gamma^{\nu}_{\alpha \beta}A^{\mu \beta},
\end{equation}
and for a second-rank covariant tensor
\begin{equation}
	A_{\mu \nu \; ;\alpha} = \partial_{\alpha}A_{\mu \nu}
	- \Gamma^{\beta}_{\mu \alpha} A_{\beta \nu}
	- \Gamma^{\beta}_{\nu \alpha} A_{\mu \beta}.
\end{equation}
We can decompose vectors and tensors into parts parallel and orthogonal to the
fluid four-velocity $u^{\mu}$, where $u^{\mu}u_{\mu} = 1$. To do so, we define
the transverse projection tensor
\begin{equation}
	\Delta^{\mu \nu} := g^{\mu \nu} - u^{\mu}u^{\nu}.
	\label{transverse_projector_tensor}
\end{equation}
One can easily check that $\Delta^{\mu \nu}u_{\nu} = 0$, and any arbitrary
four-vector $A^{\mu}$ can be written as
\begin{equation}
	A^{\mu} = u^{\mu} u_{\alpha}A^{\alpha} + \Delta^{\mu \alpha}A_{\alpha}.
	\label{Decomposition of Four vector}
\end{equation}
Similarly, the covariant derivative of an arbitrary tensor can be decomposed as
\begin{equation}
	A^{\mu_{1} \mu_{2}\cdots \mu_{n}}_{\qquad\quad ; \nu}
	= u_{\nu} D A^{\mu_{1} \mu_{2}\cdots \mu_{n}}
	+ \nabla_{\nu} A^{\mu_{1} \mu_{2}\cdots \mu_{n}},
	\label{covariant_derivative_arbitraty_tensor}
\end{equation}
where $D$ and $\nabla_{\alpha}$ are known as the convective time derivative and
the spatial gradient operator, respectively, and are defined as
\begin{equation}
	DA^{\mu_{1} \mu_{2}\cdots \mu_{n}} := u^{\alpha}A^{\mu_{1} \mu_{2}\cdots \mu_{n}}_{\qquad\quad ;\alpha},
\end{equation}
\begin{equation}
	\nabla_{\alpha}A^{\mu_{1} \mu_{2}\cdots \mu_{n}} := \Delta^{\beta}_{\alpha}A^{\mu_{1} \mu_{2}\cdots \mu_{n}}_{\qquad\quad ;\beta}.
\end{equation}
Lastly, any three-vector $\mathbf{A}$ is represented using bold Latin letters,
and its components by $\mathrm{A}^{i}$ or $\mathrm{A}_{i}$~\cite{Molnar2010}.

\section{\label{RRMHD_eq}Relativistic Resistive Magnetohydrodynamics}
The primary equations of resistive relativistic
magnetohydrodynamics (RRMHD) consist of the conservation of
 a current $N^\mu$, conservation of the total
energy--momentum tensor $T^{\mu\nu}$, and Maxwell's equations
for the electromagnetic fields.
	Mathematically these equations are given as
	\begin{equation}
		N^{\mu}_{;\mu} = 0,
		\label{conservation_current}
	\end{equation}
	\begin{equation}
		T^{\mu \nu}_{; \mu} = 0,
		\label{conservation_energy_momentum_tensor}
	\end{equation}
	\begin{equation}
		F^{\mu \nu}_{; \mu} = J^{\nu},
		\label{Maxwell_with_source}
	\end{equation}
	\begin{equation}
		\tilde{F}^{\mu \nu}_{;\mu} = 0,
		\label{Maxwell_without_source}
	\end{equation}

Here $N^\mu = n u^\mu$ is the current associated with a conserved quantity,
$u^\mu$ is the fluid four-velocity, and $n$ is the corresponding density in the
local rest frame of the fluid. In heavy-ion collisions $N^\mu$ is naturally
identified with the net-baryon current. Several standard RRMHD benchmark
problems, taken from the astrophysical literature, are instead formulated in
terms of the rest-mass density together with an ideal-gas equation of state;
for those tests we interpret $n$ as the rest-mass density, $n = m\,n_p$, where
$n_p$ is the number density of a conserved species of mass $m$. For the
conformal and lattice-QCD equations of state used in the heavy-ion
simulations, which do not depend on $n$ (vanishing baryon chemical potential),
Eq.~\eqref{conservation_current} decouples from the remaining equations and
does not affect the dynamics. $T^{\mu \nu}$ is the energy-momentum tensor; $F^{\mu \nu}$ is the
	Faraday tensor, and $\tilde{F}^{\mu \nu}$ its dual, which is defined as
	$\tilde{F}^{\mu \nu} = \frac{1}{2} \epsilon^{\mu \nu \alpha \beta}F_{\alpha \beta}$
	with $\epsilon^{\mu \nu \alpha \beta} = (-g)^{-1/2} [\mu \nu \alpha \beta]$.
	Finally, $J^{\mu}$ is the electric four-current. The energy-momentum
	tensor contains two terms: one corresponding to the matter part
	$T^{\mu \nu}_{M}$ and the other due to the electromagnetic field
	$T^{\mu \nu}_{f}$; hence we can write
	$T^{\mu \nu} = T^{\mu \nu}_{M} + T^{\mu \nu}_{f}$. For an ideal fluid,
	$T^{\mu \nu}_{M}$ is defined as
	\begin{equation}
		T^{\mu \nu}_{M} := eu^{\mu}u^{\nu} - P\Delta^{\mu \nu},
		\label{energy_momentum_matter}
	\end{equation}
	where $e = T^{\mu \nu}_{M}u_{\mu}u_{\nu}$ is the proper energy density (local
	rest frame) and $P = -\frac{1}{3}\Delta_{\mu \nu}T^{\mu \nu}_{M}$ is the
	isotropic pressure of the fluid. Further, if we ignore magnetization and
	polarization effects, then
	\begin{equation}
		T^{\mu \nu}_{f} := - F^{\mu \alpha}F^{\nu}_{\; \alpha} + \frac{1}{4}g^{\mu \nu} F^{\gamma \beta} F_{\gamma \beta}.
		\label{energy_momentum_EM}
	\end{equation}
	Using \eqref{conservation_energy_momentum_tensor} together with
	\eqref{energy_momentum_EM} and Maxwell's equations \eqref{Maxwell_with_source},
	one can show that
	\begin{equation}
		T^{\mu \nu}_{M; \mu} = F^{\nu \alpha}J_{\alpha},
		\label{eq: matter_enery_momentum_with_source}
	\end{equation}
	i.e. the matter energy-momentum tensor is not separately conserved; the
	Lorentz four-force $F^{\nu\alpha}J_{\alpha}$ exchanges energy and momentum
	between the fluid and the electromagnetic field.
	The Faraday tensor and its dual tensor are given as
	\begin{equation}
		F^{\mu \nu} = E^{\mu}u^{\nu} - E^{\nu}u^{\mu} + \epsilon^{\mu \nu \alpha \beta}u_{\alpha}B_{\beta},
		\label{Faraday_tensor}
	\end{equation}
	\begin{equation}
		\tilde{F}^{\mu \nu} = B^{\mu}u^{\nu} - B^{\nu}u^{\mu} - \epsilon^{\mu \nu \alpha \beta}u_{\alpha}E_{\beta},
		\label{Dual_Faraday_tensor}
	\end{equation}
	where $E^{\mu} = F^{\mu \nu}u_{\nu}$ and $B^{\mu} = \tilde{F}^{\mu \nu}u_{\nu}$
	are the electric and magnetic fields measured in the co-moving frame of the
	fluid, respectively, in the sense that $E^{\mu}$ and $B^{\mu}$ reduce to
	$(0,\mathbf{E}')$ and $(0,\mathbf{B}')$ in the local rest frame of the fluid.
	One can easily check that $E^{\mu}$ and $B^{\mu}$ are orthogonal to the fluid
	four-velocity $u^{\mu}$,
	\begin{equation}
		E^{\mu}u_{\mu} = 0,  \quad B^{\mu}u_{\mu} =0.
		\label{Orthogonality_conditions}
	\end{equation}
	If we decompose the fluid four-velocity as $u^{\mu} = \gamma(1,\mathbf{v})$,
	where $\gamma$ is the Lorentz factor, then the laboratory-frame electric field
	$\mathbf{E}$ and magnetic field $\mathbf{B}$ are related to $E^{\mu}$ and
	$B^{\mu}$ in the following way,
	\begin{equation}
		E^{\mu} = [\gamma \mathbf{v} \cdot\mathbf{E}, \; \gamma \mathbf{E} + \gamma (\mathbf{v}\times \mathbf{B})],
		\label{Components_of_FourE}
	\end{equation}
	\begin{equation}
		B^{\mu} = [\gamma \mathbf{v} \cdot \mathbf{B}, \; \gamma \mathbf{B} - \gamma (\mathbf{v} \times \mathbf{E})].
		\label{Components_of_fourB}
	\end{equation}

The laboratory-frame fields $\mathbf{E}$ and $\mathbf{B}$, which are the
evolved variables in Sec.~\ref{sec:numerical_method}, are the fields measured
by the Eulerian observer $\mathcal{I}^{\mu}$ at rest with respect to the computational
grid, $\mathcal{I}_{\mu}=(1,0,0,0)$. Defining
$\mathcal{E}^{\mu}\equiv F^{\mu\nu}\mathcal{I}_{\nu}=(0,\mathbf{E})$ and
$\mathcal{B}^{\mu}\equiv\tilde{F}^{\mu\nu}\mathcal{I}_{\nu}=(0,\mathbf{B})$, the Faraday
tensor can equivalently be written as
$F^{\mu\nu}=\mathcal{E}^{\mu}\mathcal{I}^{\nu}-\mathcal{E}^{\nu}\mathcal{I}^{\mu}
+\epsilon^{\mu\nu\alpha\beta}\mathcal{I}_{\alpha}\mathcal{B}_{\beta}$,
so that $F^{i0}=\mathrm{E}^{i}$ and $F^{ij}=-[ijk]\,\mathrm{B}^{k}$, where
$\epsilon^{\mu\nu\alpha\beta}=[\mu\nu\alpha\beta]/\sqrt{-g}$ with $[0123]=+1$.

	The system of equations is closed by specifying an EoS together with Ohm's law. The equation of state
relates the thermodynamic pressure, the proper energy density,
and the conserved charge density,
\begin{equation}
    P := P(e,n).
\end{equation}
For an ideal gas, the EoS is given as 
\begin{equation}
    P=(\Gamma-1)(e-n),
    \label{ideal_EOS}
\end{equation}
where $\Gamma$ is the adiabatic index and, following the standard convention of the astrophysical RRMHD benchmark problems, $n$ is interpreted as the rest-mass density. Here $e$ includes both the rest-mass and internal-energy contributions in the local rest frame, so that $e - n$ is the internal energy density.
In the ultra-relativistic(suitable for heavy ion collisions) limit $n \ll e$, Eq.~\eqref{ideal_EOS}
reduces to $P\simeq(\Gamma-1)e$, and therefore to the
conformal equation of state $P=e/3$ for $\Gamma=4/3$. For the heavy-ion simulations, we use either this conformal EoS or a
lattice-QCD EoS at vanishing baryon chemical potential~\cite{Borsanyi2010}, for which $P = P(e)$ is independent of $n$.
Ohm's law is given as
	\begin{equation}
		J^{\mu} = \sigma E^{\mu} + q_{0}u^{\mu},
		\label{Ohm's law}
	\end{equation}
	where $\sigma$ is the electrical conductivity (assumed isotropic),
	$q_{0} = J^{\mu}u_{\mu}$ is the net local charge density of the fluid in the
	co-moving frame, and $q = J^{0}$ is the time component of the four-current in
	the laboratory frame. Equation~\eqref{Ohm's law} is the simplest
	relativistic generalization of Ohm's law, in which the conduction current is
	taken to be linear in the co-moving electric field; Hall, thermoelectric, and
	anisotropic-conductivity contributions are neglected throughout this work.
	Maxwell's equations lead to the charge conservation law
	\begin{equation}
		J^{\mu}_{; \mu} = 0.
		\label{Charge_conservation}
	\end{equation}
	The ideal-MHD limit corresponds to a very large magnetic Reynolds number,
	$\mathrm{R_{m}} \gg 1$, where $\mathrm{R_{m}}$ is defined as
	$
	\mathrm{R_{m}} = \sigma \mu L U ,
	$
	where $L$ denotes the characteristic length (or time) scale of the fluid (QGP),
	$U$ is the characteristic flow velocity, and $\mu$ is the magnetic permeability
	of the fluid (QGP)~\cite{PhysRevD.104.054004}. If we take
	$\sigma \rightarrow \infty$ while keeping the other three quantities finite, we
	recover the ideal conditions: since the conduction current
	$\sigma E^{\mu}$ must remain finite in this limit, one requires
	\begin{equation}
		E^{\mu} = 0,
		\label{ideal_MHD_condition}
	\end{equation}
	which implies $\mathbf{v} \cdot \mathbf{E} = 0$ and
	$\mathbf{E} = - (\mathbf{v} \times \mathbf{B})$, matching the ideal-MHD
	conditions (Appendix~\ref{app:Ideal_MHD_Limit}).

	\section{\label{sec:numerical_method}Numerical Implementation}
	
	\subsection{\label{subsec:glm_cleaning}GLM divergence cleaning}
    The set of conservation equations for given initial/boundary conditions is solved numerically using total variation diminishing (TVD) reconstruction and  Harten--Lax--van Leer (HLL) approximate Riemann solver. In the discretised evolution, however, the two Maxwell constraints,
	\begin{equation}
		\partial_{i} \left(\sqrt{-g}\mathrm{E}^{i}\right) = \sqrt{-g} q, 
		\label{With_source_constraint}
	\end{equation}
	\begin{equation}
		\partial_{i}\left(\sqrt{-g}\,\mathrm{B}^{i}\right) = 0,
		\label{Without_source_constraint}
	\end{equation}
are not preserved exactly: truncation errors inevitably generate small violations that can accumulate over long integrations. To control these violations we adopt the generalised Lagrange multiplier (GLM) formulation \cite{DEDNER2002645, MUNZ2000484, 10.1111/j.1365-2966.2007.12448.x} and introduce two auxiliary scalar fields, $\phi$ and $\psi$. 
	Below we show how these equations are modified due to the introduction of these auxiliary scalar fields in Minkowski space-time with Cartesian coordinates. In Eqs.~\eqref{Modified_Divergence_B}--\eqref{damping_of_B} all spatial indices are Euclidean, i.e., they are raised and lowered with $\delta_{ij}$ and repeated indices are summed; $\epsilon^{ijk}$ is the three-dimensional Levi-Civita symbol with $\epsilon^{123}=1$, and $\nabla^{2}=\partial_{i}\partial_{i}$. For the divergence-free constraint, we modify the source-free Maxwell equations as     
	\begin{equation}
		\partial_{t} \phi + \partial_{i} \mathrm{B}^{i} = - \kappa \phi,
		\label{Modified_Divergence_B}
	\end{equation}
	\begin{equation}
		\partial_{t} \mathrm{B}^{i} + \epsilon^{ijk} \partial_{j} \mathrm{E}_{k} + \partial^{i} \phi = 0,
		\label{Modified_Faradays_law}
	\end{equation}
    where $\kappa$ is a positive constant, with dimension of inverse time, that sets the damping rate of the constraint violations; the propagation speed of the cleaning waves is set equal to the speed of light. From these two equations \eqref{Modified_Divergence_B}, \eqref{Modified_Faradays_law} we get
    \begin{equation}
		\partial^{2}_{t} \phi - \nabla^{2} \phi + \kappa \partial_{t} \phi = 0,
		\label{telegraph_eq_phi}
	\end{equation}
    
	which is the well known telegrapher's equation. Since, $\kappa > 0$, it suggests $\phi$ is transported by hyperbolic waves with the speed of light and decays. So, with whatever non-zero finite value $\phi$ starts with it decays to $\phi(t,\mathbf{x}) = 0$, and the $\partial_{i}\mathrm{B}^{i}$ constraint is maintained. Similarly, the other two Maxwell's equations with source is written as 
	\begin{equation}
		\partial_{t} \psi + \partial_{i}\mathrm{E}^{i} = q - \kappa \psi,
		\label{Modified_Gauss_law_psi}
	\end{equation}
	
	\begin{equation}
		- \partial_{t} \mathrm{E}^{i} + \epsilon^{ijk} \partial_{j} \mathrm{B}_{k} - \partial^{i}\psi = J^{i}.
		\label{Modified_Maxwell_Amper_eq}
	\end{equation}

From Eqs.~\eqref{Modified_Gauss_law_psi},\eqref{Modified_Maxwell_Amper_eq}, and the charge conservation law \eqref{Charge_conservation} one can show that $\psi$ also satisfies the telegrapher's equation similar to $\phi$ and has the same evolution property, so that numerical errors in Gauss's law are transported out of the computational domain and damped, instead of accumulating. In addition, using Eqs.~\eqref{Modified_Gauss_law_psi},
    \eqref{Modified_Maxwell_Amper_eq}, and \eqref{Charge_conservation} one can show that the Gauss-law
    constraint violation,
$
\partial_i \mathrm{E}^i-q,
$
satisfies the following telegrapher's equation:
\begin{equation}
		\partial^{2}_{t}  (\partial_{j}\mathrm{E}^{j} - q) + \kappa \partial_{t}(\partial_{j}  \mathrm{E}^{j}-q) - \nabla^{2}(\partial_{j} \mathrm{E}^{j}-q) = 0.
		\label{damping_E_field}
	\end{equation}
Similarly,
Eqs.~\eqref{Modified_Divergence_B} and
\eqref{Modified_Maxwell_Amper_eq} imply that
$\partial_i \mathrm{B}^i$ also obeys the telegrapher's equation i.e., 
	\begin{equation}
		\partial^{2}_{t}  (\partial_{j}\mathrm{B}^{j} ) + \kappa \partial_{t}(\partial_{j} \mathrm{B}^{j}) - \nabla^{2}(\partial_{j} \mathrm{B}^{j}) = 0.
		\label{damping_of_B}
	\end{equation}

This construction ensures that the violations of
$\partial_i \mathrm{E}^i=q$ and $\partial_i \mathrm{B}^i=0$ remain bounded by the truncation error of the scheme and do not accumulate in time.
	
Now we write the Maxwell equations with the GLM terms in covariant form, valid in an arbitrary coordinate system, as 
	\begin{equation}
		(F^{\mu \nu} + g^{\mu \nu} \psi)_{; \mu} = - \kappa \psi n^{\nu} + J^{\nu},
		\label{Modified_Maxwell_GLM_with_source}
	\end{equation}
	\begin{equation}
		(\tilde{F}^{\mu \nu} + g^{\mu \nu} \phi)_{; \mu} = - \kappa \phi n^{\nu},
		\label{Modified_Maxwell_GLM_without_source}
	\end{equation}
    where $n^{\mu}$ is the future-directed timelike unit vector normal to the $t=\mathrm{const}$ hypersurfaces, $n^{\mu}=(1,0,0,0)$~\cite{Nakamura2023, 10.1111/j.1365-2966.2007.12448.x}.

	\subsection{\label{subsec:conservative_system}Conservative variables, fluxes, and source terms}
	
	The conservation equations within the GLM framework as expressed in \eqref{conservation_current},\eqref{conservation_energy_momentum_tensor}, \eqref{Charge_conservation} \eqref{Modified_Maxwell_GLM_with_source} and \eqref{Modified_Maxwell_GLM_without_source}
	can be written in the following general form
	\begin{equation}
		\partial_{0}(\sqrt{-g} \mathcal{U}) + \partial_{i}(\sqrt{-g} \mathcal{F}^{i}) = \sqrt{-g} \mathcal{S},
		\label{eq:Evolving_equations}
	\end{equation}
	for a specific choice of metric (discussed in Appendix~\ref{app:metric choice}), where $\mathcal{U}$,$\mathcal{F}^{i}$, and $\mathcal{S}$ corresponds to conserved quantities, fluxes, and source term, respectively \cite{Bona2009, Nakamura2023}. Where 
	\begin{equation}
		\mathcal{U} =  \begin{bmatrix}
			N_{c} \\
			S^{j}\\
			\varepsilon\\
			\mathrm{E}^{j} \\
			\mathrm{B}^{j} \\
			q \\
			\psi \\
			\phi 
		\end{bmatrix},  \qquad \mathcal{F}^{i} = \begin{bmatrix}
			N_{c}v^{i} \\
			T^{ij} \\
			S^{i}\\
			\epsilon^{ijk}\mathrm{B}_{k} + g^{ij} \psi \\
			-\epsilon^{ijk}\mathrm{E}_{k} + g^{ij} \phi \\
			{J}^{i}\\
			\mathrm{E}^{i}\\
			\mathrm{B}^{i}
		\end{bmatrix}
        \label{eq:Conserved_Flux_quantities}
	\end{equation}
	
	\begin{equation}
		\mathcal{S} = \begin{bmatrix}
			0 \\
			- \Gamma^{j}_{\mu \alpha}T^{\mu \alpha} \\
			- \Gamma^{0}_{\mu \alpha}T^{\mu \alpha}\\
			- {J}^{j}\\
			\mathbf{0}\\
			0\\
			q - \kappa \psi - g^{\alpha \beta}\Gamma^{0}_{\alpha \beta}\\
			-\kappa \phi - g^{\alpha \beta}\Gamma^{0}_{\alpha \beta}
		\end{bmatrix}, 
		\label{source_terms}
	\end{equation}
	where
	\begin{equation}
		N_{c} = n \gamma,
		\label{conserved_quatity_baryon}
	\end{equation}
	\begin{equation}
		S^{j} :=T^{0j}, \quad S^{j} = (e + P) \gamma^2 v^{j} + (\mathbf{E \times \mathbf{B}})^{j}, 
		\label{S_j_vector}
	\end{equation}
	
	\begin{align}
		T^{ij} = (e + P)\gamma^2 v^{i}v^{j} - (P + \varepsilon_{\rm em})g^{ij}  \notag\\ -\mathrm{E}^{i}\mathrm{E}^{j} - \mathrm{B}^{i}\mathrm{B}^{j},
	\end{align}
	here $\varepsilon_{\rm em} = 1/2(\mathbf{E}^2 + \mathbf{B}^2)$, and
	\begin{equation}
		\varepsilon := T^{00}, \quad \varepsilon = (e + P)\gamma^2 - P + \varepsilon_{em}.
		\label{total_energy_density}
	\end{equation}

Next, we describe the numerical scheme used to solve equations of the form
\eqref{eq:Evolving_equations}. The scheme proceeds in two steps. In the first
step, the numerical fluxes of the conserved variables are computed at the cell
interfaces from reconstructed primitive variables using the HLL approximate
Riemann solver (Sec.~\ref{subsec:reconstruction_hll}). In the second step,
these fluxes, together with the source terms, are used to update the cell
averages of the conserved variables in a finite-volume scheme
(Sec.~\ref{subsec:fv_discretization}).

\subsection{\label{subsec:reconstruction_hll}Reconstruction and HLL flux}
The primitive variables are reconstructed at cell interfaces using a piecewise-linear total-variation-diminishing reconstruction with the minmod limiter
\cite{10.1093/acprof:oso/9780198528906.001.0001}. For a generic primitive variable $\mathcal{P}_i$, we define
\begin{equation}
    \operatorname{minmod}(a,b)
    =
    \begin{cases}
        \operatorname{sgn}(a)\min(|a|,|b|), & ab>0,\\
        0, & ab\leq 0 ,
    \end{cases}
    \label{eq:minmod_limiter}
\end{equation}
where $a=(\mathcal{P}_i-\mathcal{P}_{i-1})/\Delta x$ and
$b=(\mathcal{P}_{i+1}-\mathcal{P}_i)/\Delta x$ are the neighbouring one-sided
slopes and $\mathrm{sgn}(\cdots)$ denotes the signum function.

The reconstructed primitive variables are converted to the left and right conserved states at an
interface are denoted by $\mathcal{U}_L$ and $\mathcal{U}_R$,
with the corresponding physical fluxes
$\mathcal{F}_L=\mathcal{F}(\mathcal{U}_L)$ and
$\mathcal{F}_R=\mathcal{F}(\mathcal{U}_R)$. The numerical flux
is computed using the HLL approximate
Riemann solver,
\begin{equation}
    \hat{\mathcal{F}}_{\rm HLL}
    =
    \frac{
        \lambda^{+}\mathcal{F}_{L}
        -
        \lambda^{-}\mathcal{F}_{R}
        +
        \lambda^{+}\lambda^{-}
        (\mathcal{U}_{R}-\mathcal{U}_{L})
    }
    {\lambda^{+}-\lambda^{-}},
    \label{eq:hll_flux}
\end{equation}
where $\lambda^{-}\leq0\leq\lambda^{+}$ are estimates of the
minimum and maximum characteristic speeds normal to the cell
interface. In the present implementation, the signal speeds
are constrained to remain subluminal
\cite{10.1063/1.4734434}. The resulting HLL fluxes are then
used to evolve the conserved variables in time.

    \subsection{\label{subsec:fv_discretization}Finite-volume discretization}
	In the second step, the interface fluxes obtained above are used to evolve Eq.~\eqref{eq:Evolving_equations} with a cell-centered finite-volume method. For a uniform Cartesian grid, the cell average of the conserved state is
updated according to
	\begin{align}
		\mathcal{U}_{i,j,k}^{n+1}
		= &\, \mathcal{U}_{i,j, k}^{n}
		- \frac{\Delta t}{\Delta x}
		\left(
		\hat{\mathcal{F}}^{x}_{i+1/2,j,k}
		-
		\hat{\mathcal{F}}^{x}_{i-1/2,j,k}
		\right)
		\notag \\
		&
		- \frac{\Delta t}{\Delta y}
		\left(
		\hat{\mathcal{F}}^{y}_{i,j+1/2,k}
		-
		\hat{\mathcal{F}}^{y}_{i,j-1/2,k}
		\right) \notag \\
		& - \frac{\Delta t}{\Delta z}
		\left(
		\hat{\mathcal{F}}^{z}_{i,j,k+1/2}
		-
		\hat{\mathcal{F}}^{z}_{i,j,k-1/2}
		\right)
		+ \Delta t\,(\mathcal{S}_{i,j,k}),
		\label{eq:finite_volume_update}
	\end{align}
   where the superscript $n$ denotes the discrete time level $t^{n}$ with $t^{n+1}=t^{n}+\Delta t$,
    while $(i,j,k)$ label the spatial cell in the
    $x$, $y$, and $z$ directions, respectively.
    
	For one-dimensional tests, the $y$ and $z$-flux contribution are omitted. In curvilinear coordinates we evolve the volume-weighted conserved variables and fluxes, $\sqrt{-g}\,\mathcal{U}$ and $\sqrt{-g}\,\mathcal{F}^{i}$, so that the finite-volume update remains conservative, while the metric-dependent geometric contributions are treated as source terms. This form is used for the Milne-coordinate implementation, where $\sqrt{-g}=\tau$, and $(x,y,\eta_s)$ replace $(x,y,z)$, with
$\Delta z\to\Delta\eta_s$; for the boost-invariant $(2+1)$-dimensional simulations the $\eta_s$-fluxes vanish identically.

	\subsection{\label{subsec:admissibility}Boundary conditions and numerical admissibility}
	
	For one-dimensional shock-tube calculations, we impose zero-gradient outflow boundary conditions at both ends of the computational domain, implemented by copying the outermost interior cell values into the ghost
cells. For the two-dimensional Cartesian blast and rotor tests, outflow boundary conditions are imposed on all sides. The same prescription is used for the transverse boundaries of the Milne-coordinate rotor test unless otherwise stated.
	At every stage of the time integrator
    the primitive variables are recovered from the conserved variables using a non linear root finding algorithm (see Sec.~\ref{subsec:primitive_recovery}) and during this process one must make sure that the they also fulfill the following physical constraints
	\begin{equation}
		n\geq 0,\qquad P \geq 0,\qquad v^{2}<1 .
	\end{equation}
	If the nonlinear primitive-variable recovery produces an inadmissible state, density and pressure floors are applied as needed. The velocity is subsequently rescaled, when necessary, to ensure that its magnitude remains sub luminal. The conserved fluid variables are then recomputed from the corrected primitive state, while the electric and magnetic fields, which are evolved directly, are left unchanged.
    
    The exact floor values used in the simulations presented here are summarized in Table
	~\ref{tab:numerical_admissibility}.
	
	\begin{table}[h]
		\caption{Numerical admissibility parameters used in the benchmark calculations.}
		\label{tab:numerical_admissibility}
		\begin{ruledtabular}
			\begin{tabular}{ll}
				Quantity & Value used in simulations \\
				\hline
				Density floor & $n_{\rm floor}= 10^{-14}$ \\
				Pressure floor & $P_{\rm floor}=10^{-14}$ \\
				Velocity ceiling & $v^2 < 1-\epsilon_v$, $\epsilon_v = 10^{-14}$ \\
				Primitive-recovery tolerance & $10^{-12}$ \\
				Maximum primitive iterations & $30$ \\
				    Numerical guess &  previous admissible state \\
			\end{tabular}
		\end{ruledtabular}
	\end{table}

\subsection{\label{subsec:primitive_recovery}Primitive-variable recovery}

The finite-volume evolution is performed in terms of the conserved
variables $\mathcal{U}$, whereas the evaluation of the fluxes and
source terms requires the corresponding primitive variables
$\mathcal{P}$. The inversion
$\mathcal{U}\rightarrow\mathcal{P}$ is therefore carried out after
each explicit and implicit stage of the IMEX scheme that we use for the time evolution of the conserved
quantities, the details of which are given in the next section. Since the
electromagnetic fields are treated as conserved quantities Eq.\eqref{eq:Conserved_Flux_quantities} , the nontrivial part of
the inversion consists of recovering only the fluid variables 
$(n,P,v^{i})$ from $N_c$, $S^{i}$, and $\varepsilon$ for a given 
$\mathbf{E}$ and $\mathbf{B}$.
The primitive recovery process starts by subtracting the electromagnetic contribution from the conserved
momentum as 
\begin{equation}
    Q^{i}
    :=
    S^{i}-(\mathbf{E}\times\mathbf{B})^{i}.
    \label{eq:S_tilde_Q}
\end{equation}
Its positive spatial norm is denoted by
\begin{equation}
    Q^{2}:=-g_{ij}Q^{i}Q^{j},
\end{equation}
which reduces to the usual Euclidean norm in Cartesian coordinates.
From Eq.~\eqref{S_j_vector},
\begin{equation}
    Q^{i}=(e+P)\gamma^{2}v^{i}.
\end{equation}
It is therefore convenient to introduce the Lorentz-boosted
enthalpy density
\begin{equation}
    Z:=(e+P)\gamma^{2}.
    \label{eq:Z_def}
\end{equation}
For a trial value of $Z$, the velocity and Lorentz factor follow
directly as
\begin{align}
    v^{i}(Z) &= \frac{Q^{i}}{Z}, \\
    v^{2}(Z) &= \frac{Q^{2}}{Z^{2}}, \\
    \gamma(Z)
    &=\frac{1}{\sqrt{1-v^{2}(Z)}}
     =\frac{1}{\sqrt{1-Q^{2}/Z^{2}}}.
    \label{eq:v_gamma_of_Z}
\end{align}

The conserved energy density,
\begin{equation}
    \varepsilon
    =
    (e+P)\gamma^{2}-P+\varepsilon_{\rm em},
\end{equation}
then gives the pressure associated with the trial value of $Z$,
\begin{equation}
    P(Z)=Z-\varepsilon+\varepsilon_{\rm em}.
    \label{eq:P_of_Z}
\end{equation}
Using $N_c=n\gamma$, the corresponding proper density is
\begin{equation}
    n(Z)=\frac{N_c}{\gamma(Z)},
    \label{Local_rest_mass_interms_Z}
\end{equation}
while Eq.~\eqref{eq:Z_def} gives
\begin{equation}
    e(Z)
    =
    \frac{Z}{\gamma^{2}(Z)}-P(Z).
    \label{eq:e_of_Z}
\end{equation}

For a general equation of state
\begin{equation}
    P=P_{\rm{EoS}}(e,n),
\end{equation}
the primitive-variable recovery is reduced to the scalar consistency
condition
\begin{equation}
    f(Z)
    :=
    P(Z)
    -
    P_{\rm EoS}\!\left(e(Z),n(Z)\right)
    =0.
    \label{eq:fZ_root}
\end{equation}
For the ideal EoS\eqref{ideal_EOS} used in the benchmark
calculations, Eq.~\eqref{eq:fZ_root} becomes
\begin{equation}
    f(Z)
    =
    P(Z)
    -
    (\Gamma-1)
    \left[e(Z)-n(Z)\right].
\end{equation}
The above equation is then used to find root using iterative solver. 
The iterative solution obtained above must satisfy the physical constraint
\begin{equation}
    Z \geq Q,
\end{equation}
which guarantees $v^{2}\leq 1$. We therefore choose the lower bound of $Z (=Z_{\min})$ in the root finding procedure  as
\begin{equation}
    Z_{\min}=Q+\delta,
\end{equation}
where a small cutoff $ \delta \sim 10^{-14}$ ensure the physical constraint.
 The function $f(Z)$ \eqref{eq:fZ_root} has at least one root bracketed in $[Z_{\min},\infty)$ could be checked from the fact that for $Q\ll Z$, i.e., $\gamma\rightarrow1$, implies 
$n\rightarrow N_c$, and $e\rightarrow
\varepsilon-\varepsilon_{\rm em}$, and hence
$f(Z)\sim Z$. Since $Z$ is strictly positive, it follows that $f(Z)>0$ in this scenario. The behavior of $f(z)$ near the lower bound of $Z$ (i.e.,  $Z\sim Q$) is 
\begin{equation}
    f(Z_{\min})
    =
    \Gamma
    \left(
    Q-\varepsilon+\varepsilon_{\rm em}
    \right)
    +
    \mathcal{O}(\sqrt{\delta}).
\end{equation}
Using $Q=Z|\mathbf{v}|$ and
$\varepsilon=Z-P+\varepsilon_{\rm em}$, one finds
\begin{equation}
    Q-\varepsilon+\varepsilon_{\rm em}
    =
    Z(|\mathbf{v}|-1)+P
    =
    \frac{P|\mathbf{v}|-e}{1+|\mathbf{v}|},
\end{equation}
which is negative always if $P<e$. Since $f(Z_{\min})<0$, and 
$f(Z_{\max})>0$, this implies that there exist at least one physical root within the range ($Z_{\min},Z_{\max}$).

Once the root $Z_0$ has been found, the primitive variables are
obtained using Eq.(53),(55),(57) as follows:
\begin{align}
    v^{i} &= \frac{Q^{i}}{Z_0}, \\
    \gamma &= \gamma(Z_0), \\
    P &= P(Z_0), \\
    e &= e(Z_0), \\
    n &= \frac{N_c}{\gamma}.
    \label{eq:prims_final}
\end{align}

	\subsection{Implicit--explicit splitting}
	
	The primary difficulty in any RRMHD solver is the
	presence of stiff relaxation terms when the electrical conductivity is large.
	The stiffness originates from Ohm's law, since the conduction current contains
	terms proportional to \(\sigma\). In the high-conductivity regime, the
	relaxation time scale associated with the electric field becomes much smaller
	than the characteristic hydrodynamic time scale. A fully explicit time
	integration would therefore require a time step constrained by
	\(\Delta t \lesssim \sigma^{-1}\), which becomes prohibitively small for
	\(\sigma \gg 1\).
	
	To avoid this restriction, the usual practice is to split the RRMHD differential equations (or the finite difference version of it) into a non-stiff and a stiff relaxation part,
	\begin{equation}
		\partial_{t}\mathcal{U}
		=
		\mathcal{R}_{\rm exp}(\mathcal{U})
		+
		\mathcal{R}_{\rm imp}(\mathcal{U}).
		\label{eq:imex_split}
	\end{equation}
	The explicit operator \(\mathcal{R}_{\rm exp}\) contains the flux-divergence
	terms, geometrical source terms, and non-stiff electromagnetic contributions.
	The implicit operator \(\mathcal{R}_{\rm imp}\) contains only the stiff
	conductivity-dependent part of Ampere's law.
	
	Following this splitting, it is useful to separate the evolved variables into
	two subsets,
	\begin{equation}
		\mathcal{U}
		=
		\left(
		\mathbf{Y},\mathbf{E}
		\right),
	\end{equation}
	where the electric field \(\mathbf{E}\) contains the stiff relaxation terms,
	while
	\begin{equation}
		\mathbf{Y}
		=
		\left(
		N_{c}, S^{j}, \varepsilon, \mathbf{B}^{j}, q, \psi, \phi
		\right)
	\end{equation}
	denotes the remaining non-stiff variables. The evolution equations can then be
	written schematically as
	\begin{align}
		\partial_{t}\mathbf{Y}
		&=
		[\mathcal{R}_{\rm{exp}}(\mathbf{Y},\mathbf{E})]_{\mathbf{Y}},
		\label{eq:Y_evolution}
		\\
		\partial_{t}\mathbf{E}
		&=
		[\mathcal{R}_{\rm{exp}}(\mathbf{Y},\mathbf{E})]_{\mathbf{E}}
		+
		\mathcal{R}_{\rm{imp}}(\mathbf{Y},\mathbf{E}),
		\label{eq:E_evolution}
	\end{align}
	where $[\mathcal{R}_{\rm{exp}}(\mathbf{Y},\mathbf{E})]_{\mathbf{Y}}$ and $[\mathcal{R}_{\rm{exp}}(\mathbf{Y},\mathbf{E})]_{\mathbf{E}}$
    denote the components of the explicit terms
    associated with $\mathbf{Y}$ and $\mathbf{E}$ respectively, while \(\mathcal{R}_{\rm{imp}}\) denotes the stiff part associated
	with Ohm's law Eq~\eqref{Ohm's law}.
	For an isotropic conductivity, the three current is given as 
	\begin{equation}
		\mathbf{J}
		=
		q\mathbf{v}
		+
		\sigma\gamma
		\left[
		\mathbf{E}
		+
		\mathbf{v}\times\mathbf{B}
		-
		\left(\mathbf{E}\cdot\mathbf{v}\right)\mathbf{v}
		\right].
		\label{eq:ohm_law_imex}
	\end{equation}
	The term proportional to \(\sigma\) is responsible for the stiffness. Therefore, $\mathcal{R}_{\rm{imp}}(\mathbf{Y},\mathbf{E})$ is given as (the source term for the stiff equation containing three current in Eq. \eqref{source_terms})
	\begin{equation}
		\mathcal{R}_{\rm{imp}}(\mathbf{Y},\mathbf{E})
		=
		-\sigma\gamma
		\left[
		\mathbf{E}
		+
		\mathbf{v}\times\mathbf{B}
		-
		\left(\mathbf{E}\cdot\mathbf{v}\right)\mathbf{v}
		\right],
	\end{equation}
	which is treated implicitly, while the remaining terms in Ampere's law are evolved explicitly.
	
	This splitting has two important advantages. First, the time step is determined
	by the usual Courant condition (to be discussed later) associated with the hyperbolic part of the
	system, rather than by the much smaller resistive relaxation time scale.
	Second, in the limit \(\sigma\rightarrow\infty\), the stiff source term drives
	the electric field toward the ideal-MHD constraint
	\begin{equation}
		\mathbf{E}
		=
		-\mathbf{v}\times\mathbf{B},
	\end{equation}
	provided the current remains finite. Thus, the IMEX splitting naturally connects
	the resistive and ideal-RMHD regimes within a single numerical framework. 
	
	\subsection{IMEX Runge--Kutta time integration}
    The time integration is performed using an implicit--explicit
    Runge--Kutta scheme, in which the non-stiff terms are treated
    explicitly and the stiff relaxation terms implicitly \cite{Pareschi2005, doi:10.1137/080713562, doi:10.1137/0732037, doi:10.1137/S0036142900375906}. Within each time step, the solution is advanced through a sequence of intermediate stage states $\mathcal{U}^{(l)}$, which are combined to obtain the solution at the next time level. For a system of
equations of the form Eq.~\eqref{eq:imex_split}, an $s$-stage
IMEX Runge--Kutta method is written as
	\begin{align}
		\mathcal{U}^{(l)}
		=
		\mathcal{U}^{n}
		+ 
		\Delta t
		\sum_{r=1}^{l-1}
		\tilde{a}_{lr}
		\mathcal{R}_{\rm exp}
		\left(\mathcal{U}^{(r)}\right)
		& {\notag} \\ 
		+ \Delta t
		\sum_{r=1}^{l}
		a_{lr}
		\mathcal{R}_{\rm imp}
		\left(\mathcal{U}^{(r)}\right),
		\label{eq:imex_stage}
	\end{align}
	for $l=1,\cdots,s$, and the final update is
	\begin{align}
		\mathcal{U}^{n+1}
		=
		\mathcal{U}^{n}
		+
		\Delta t
		\sum_{r=1}^{s}
		\tilde{\omega}_{r}
		\mathcal{R}_{\rm exp}
		\left(\mathcal{U}^{(r)}\right)
		& {\notag} \\
		+\Delta t
		\sum_{r=1}^{s}
		\omega_{r}
		\mathcal{R}_{\rm imp}
		\left(\mathcal{U}^{(r)}\right).
		\label{eq:imex_final}
	\end{align}
	Here, $\tilde{a}_{lr}$ and $\tilde{\omega}_{r}$ denote the stage coefficients and quadrature weights, respectively, of the explicit Runge--Kutta scheme, while $a_{lr}$ and $\omega_{r}$ denote the corresponding stage coefficients and quadrature weights of the diagonally implicit Runge--Kutta scheme used for the stiff sector. The explicit coefficients satisfy
$\tilde{a}_{lr}=0$ for $r\geq l$, while the implicit
coefficients satisfy $a_{lr}=0$ for $r>l$. In the present implementation, we primarily employ the
IMEX--SSP2(3,3,2) and IMEX--SSP3(3,3,2) schemes, with their detailed
formulations and coefficients provided in Appendix~\ref{app:imex_coefficients} \cite{10.1111/j.1365-2966.2009.14454.x}.
	
	For the present RRMHD system, at each IMEX stage, we first
	compute an explicit intermediate state,
	\begin{equation}
		\mathbf{Y}^{*}
		=
		\mathbf{Y}^{n}
		+
		\Delta t
		\sum_{r=1}^{l-1}
		\tilde{a}_{lr} \bigg[
		\mathcal{R}_{\rm{exp}}
		\left(\mathcal{U}^{(r)}\right) \bigg]_{\mathbf{Y}},
	\end{equation}
	\begin{align}
		\mathbf{E}^{*}
		=
		\mathbf{E}^{n}
		+
		\Delta t
		\sum_{r=1}^{l-1}
		\tilde{a}_{lr}
		\bigg[ \mathcal{R}_{\rm{exp}}
		\left(\mathcal{U}^{(r)}\right)\bigg]_{\mathbf{E}}
		& {\notag}\\
		+
		\Delta t
		\sum_{r=1}^{l-1}
		a_{lr}
		\mathcal{R}_{\rm{imp}}
		\left(\mathcal{U}^{(r)}\right).
	\end{align}
	The non-stiff variables are then set to
	\begin{equation}
		\mathbf{Y}^{(l)} = \mathbf{Y}^{*},
	\end{equation}
	while the electric field is obtained from the implicit relation
	\begin{equation}
		\mathbf{E}^{(l)}
		=
		\mathbf{E}^{*}
		+
		a_{ll}\Delta t\,
		\mathcal{R}_{\rm{imp}}
		\left(\mathbf{Y}^{*},\mathbf{E}^{(l)}\right).
		\label{eq:implicit_E_stage}
	\end{equation}
	
	Since the stiff source term is linear in $\mathbf{E}^{(l)}$ for fixed
	fluid velocity and magnetic field, Eq.~\eqref{eq:implicit_E_stage} can be
	solved locally in each computational cell
    \begin{align}
    \mathbf{E}^{(l)} = \mathbf{E}^{*} - \gamma \sigma a_{ll}\Delta t \bigg[ \mathbf{E}^{(l)} + \mathbf{v}^{*}\times\mathbf{B}^{*} - (\mathbf{E}^{(l)}\cdot \mathbf{v}^{*}) \; \mathbf{v}^{*} \bigg], \notag \\
    \mathbf{E}^{(l)} \bigg[\mathbf{I} + \gamma\sigma a_{ll}\Delta t (\mathbf{I} - \mathbf{v}^{*} \otimes \mathbf{v}^{*})   \bigg] = \mathbf{E}^{*} - \sigma \gamma a_{ll} \Delta t (\mathbf{v}^{*} \times \mathbf{B}^{*}),
    \end{align}
    here, $\mathbf{I}$ is the $3 \times 3$ identity matrix and
	$\mathbf{v}^{*} \otimes \mathbf{v}^{*}$ denotes the dyadic product of the velocity. For example in cartesian coordinates $(\mathbf{I} - \mathbf{v}^{*}\otimes \mathbf{v}^{*})$ given as 
    \begin{equation}
        \renewcommand{\arraystretch}{1.2}
        \mathbf{I} - \mathbf{v}^{*}\otimes \mathbf{v}^{*} =\begin{bmatrix}
         1 - (v^{*}_{x})^{2} & - v^{*}_{x}v^{*}_{y} & - v^{*}_{x}v^{*}_{z}\\
         -v^{*}_{y}v^{*}_{x} & 1-(v^{*}_{y})^{2}  & -v^{*}_{y}v^{*}_{z} \\
         -v^{*}_{z}v^{*}_{x} & -v^{*}_{z}v^{*}_{y} & 1 - (v^{*}_{z})^{2} 
        \end{bmatrix}.
    \end{equation}
    From here we can write 
	\begin{align}
		\mathbf{E}^{(l)}
		= &
		\bigg[ \mathbf{E}^{*}
		-
		a_{ll}\Delta t\,\sigma\gamma
		\left(
		\mathbf{v}^{*}\times\mathbf{B}^{*}
		\right)\bigg] \mathbf{H}^{-1}, \\ 
		\mathbf{H}  = & \left[
		\mathbf{I}
		+
		a_{ll}\Delta t\,\sigma\gamma
		\left(
		\mathbf{I}
		-
		\mathbf{v}^{*} \otimes \mathbf{v}^{*}
		\right)
		\right], \notag
		\label{eq:implicit_E_linear_system} 
	\end{align}
    where we calculated the $\mathbf{H}^{-1}$ numerically. 
	This implicit solver is purely local and therefore does not require any
	global matrix inversion \cite{10.1093/acprof:oso/9780198528906.001.0001}.
	After the electric field has been updated at a given stage, the primitive
	variables are recovered from the conserved variables using the pressure based method given in Appendix~\ref{app:pressure_based_recovery}. This recovery is necessary
	because the fluxes, source terms and the stiff electric-field update depend on
	the primitive variables
	\begin{equation}
		\left(n,P,v^{j},\mathrm{E}^{j},\mathrm{B}^{j},q,\psi,\phi\right).
	\end{equation}
	The primitive recovery is therefore performed at each IMEX stage before
	constructing the next stage fluxes and source terms \cite{10.1111/j.1365-2966.2009.14454.x}.
	
    Further the time step $\Delta t$ is determined from the Courant--Friedrichs--Lewy
    (CFL) condition,
\begin{equation}
    \Delta t
    =
    C_{\rm CFL}
    \min
    \left(
    \frac{\Delta x}{\lambda_{\max}^{x}},
    \frac{\Delta y}{\lambda_{\max}^{y}}
    \right),
\end{equation}
where $C_{\rm CFL}<1$ is the CFL number that is fixed on case by case basis, and
$\lambda_{\max}^{i}$ denote the largest signal speeds along
the corresponding spatial directions. %

\section{\label{sec:benchmark_strategy}Benchmark strategy}
    
The benchmark problems are chosen to test different features of the
RRMHD code. The Brio--Wu shock-tube \cite{BRIO1988400} problem examines the
one-dimensional shock-capturing capability of the HLL scheme and its
approach to the ideal-relativistic MHD limit with increasing electrical
conductivity. On the other hand the cylindrical blast provides a multidimensional
test of numerical robustness in the presence of strong shocks, anisotropic
magnetic stresses, and steep electromagnetic gradients. Furthermore a third test 
carried out here is the Cartesian rotor that examines the coupling between relativistic fluid motion and
magnetic stresses. Finally, the Milne-coordinate rotor tests the
extension to curvilinear coordinates and the treatment of geometric
source terms associated with boost-invariant longitudinal expansion.

	\begin{table*}[t]
		\caption{Main numerical parameters used in the benchmark calculations.}
		\label{tab:benchmark_parameters}
		\begin{ruledtabular}
			\begin{tabular}{lccccc}
				Test & Domain & Grid & Final time & $\Gamma$ & $\sigma$ \\
				\hline
				Brio--Wu & $[-0.5,0.5]$ & $500$ & $0.4$ & $2$ & $0\rightarrow10^6$ \\
				Cylindrical blast & $[-6,6]^2$ & $200^2$ & $4.0$ & $4/3$ & $55.55\rightarrow2000$ \\
				Cartesian rotor & $[-0.5,0.5]^2$ & $300^2$ & $0.3$ & $4/3$ & $10^6$ \\
				Milne rotor & $[-0.5,0.5]^2$ & $400^2$ & $\tau_f=1.4$ & $4/3$ & $10\rightarrow 10^{6}$ \\
			\end{tabular}
		\end{ruledtabular}
	\end{table*}
	
	\section{\label{sec:benchmarks_results}Benchmarks and results}
    In this section we discuss the above mentioned benchmark tests and show our result.
	In the following we write the RRMHD equations in conservative form for  different $(1+1)$D and $(2+1)$D geometries. For all test problems, the two Lagrange multipliers, $\phi$ and $\psi$, are initialized to zero. The GLM damping parameter (see Eq.\eqref{telegraph_eq_phi}) is prescribed as
	$
	\kappa = (\alpha \lambda_{\mathrm{max}})/h,
	$
	where $\lambda_{\mathrm{max}}=1$ and $h=\min(\Delta x,\Delta y)$. The dimensionless coefficient $\alpha$ is required to satisfy $\alpha<1$ and is fixed at $\alpha=1/3$ throughout all test problems \cite{10.1111/j.1365-2966.2007.12448.x, Nakamura2023}.
	
	\subsection{$(1+1)$-dimensional Cartesian coordinates}
	
	We first consider the one-dimensional Cartesian case, where all quantities depend only on $x$ and $t$. In Cartesian coordinates, all Christoffel symbols vanish
	$\Gamma^{\alpha}_{\beta\gamma}=0$, so the system of RRMHD equations \eqref{eq:Evolving_equations} reduces to
	\begin{equation}
		\partial_t \mathcal{U}
		+ \partial_x \mathcal{F}^{x}
		= \mathcal{S},
	\end{equation}
	where $\mathcal{U},\mathcal{F}^{x},$ and $\mathcal{S}$ are functions of $x,t$, for brevity we do not write them explicitly.  
	 The source vectors (Eq.\eqref{source_terms}) in this case reduces to
\begin{align}
    \mathcal{S} = [ 0,0,0,0, 0, &-J^{x}, -J^{y}, -J^{z}, 0,0,0, \\ \notag & 0, q-\kappa\psi,\, -\kappa\phi ]^{T}, 
\end{align}
and $\mathcal{S'}$ with all zero entries, here the superscript $[\cdots]^{T}$ denotes the transpose.
The conserved state vector $\mathcal{U}(x,t)$ is given as
\begin{align}
    \mathcal{U}
    =
    \big[
    N_c,\,
    S^{x},S^{y},S^{z},\,
    \varepsilon,\,
    \mathrm{E}^{x}, \mathrm{E}^{y}, & \mathrm{E}^{z},\,
    \mathrm{B}^{x},\mathrm{B}^{y},\mathrm{B}^{z},\, \notag \\  &
    q,\psi,\phi
    \big]^{T},
    \label{eq:conserved_state_cartesian}
\end{align}
where \begin{align} 
N_c &= n\gamma, \\ 
S^{j} &= (e+P)\gamma^{2}v^{j} +(\mathbf{E}\times\mathbf{B})^{j}, \\
\varepsilon &= (e+P)\gamma^{2} -P +\varepsilon_{\rm em}, 
\end{align} 
with $ \varepsilon_{\rm em} = \frac{1}{2} \left( \mathbf{E}^{2}+\mathbf{B}^{2} \right)$.
The corresponding flux $\mathcal{F}^{x}(x,t)$ in the $x$-direction is
\begin{align}
    \mathcal{F}^{x}
    =
    \big[
    N_c v^{x},\,
     T^{xx},T^{xy}, & T^{xz},\,
    S^{x},\,  
    \psi,\mathrm{B}^{z},-\mathrm{B}^{y},\,
    \notag \\ & \phi,\, -\mathrm{E}^{z},\mathrm{E}^{y},\,
    J^{x},\mathrm{E}^{x},\mathrm{B}^{x}
    \big]^{T},
    \label{eq:flux_x_cartesian}
\end{align}
where
\begin{align}
    T^{xj}
    =
    (e+P)\gamma^{2}v^{x}v^{j} &
    +
    (P+\varepsilon_{\rm em})\delta^{xj}\notag  \\ 
   & -
    \mathrm{E}^{x}\mathrm{E}^{j}
    -
    \mathrm{B}^{x}\mathrm{B}^{j},
\end{align}
and $j=x,y,z$. 
 Next we discuss numerical results for this case for a specific initial condition.
	\subsubsection{Brio--Wu shock-tube test}
	To assess the shock-capturing capability of the numerical scheme, we consider
	the Brio--Wu shock-tube problem, a standard one-dimensional benchmark for MHD
	codes \cite{BRIO1988400,10.1046/j.1365-8711.1999.02244.x,10.1063/1.4734434,GIACOMAZZO_REZZOLLA_2006}. The initial discontinuity is placed at $x=0$. The left and right
	states are given by
	\begin{equation}
		\left(n, P, \mathrm{B}^{y}, \mathrm{B}^{x}\right)
		=
		\begin{cases}
			\left(1.0, 1.0, 1.0, 0.5\right), & x<0, \\
			\left(0.125, 0.1, -1.0, 0.5\right), & x>0 .
		\end{cases}
	\end{equation}
	 All other primitive variables are initially
	set to zero,
	\begin{align}
		v^{j}=0, \qquad &\mathrm{E}^{j}=0,  \qquad \mathrm{B}^{z}=0,{\notag}\\ 
		&q_{0}=0,
	\end{align}
	  The normal magnetic field $\mathrm{B}^{x}$ is therefore continuous
	across the initial discontinuity, while $n$, $P$, and $\mathrm{B}^{y}$ are
	discontinuous. We use an ideal-gas EoS \eqref{ideal_EOS} with adiabatic index $\Gamma=2.0$.
	
	\begin{figure}[h]
		\centering
		\hspace*{ - 5.0 mm}\includegraphics[width=1.08\linewidth]{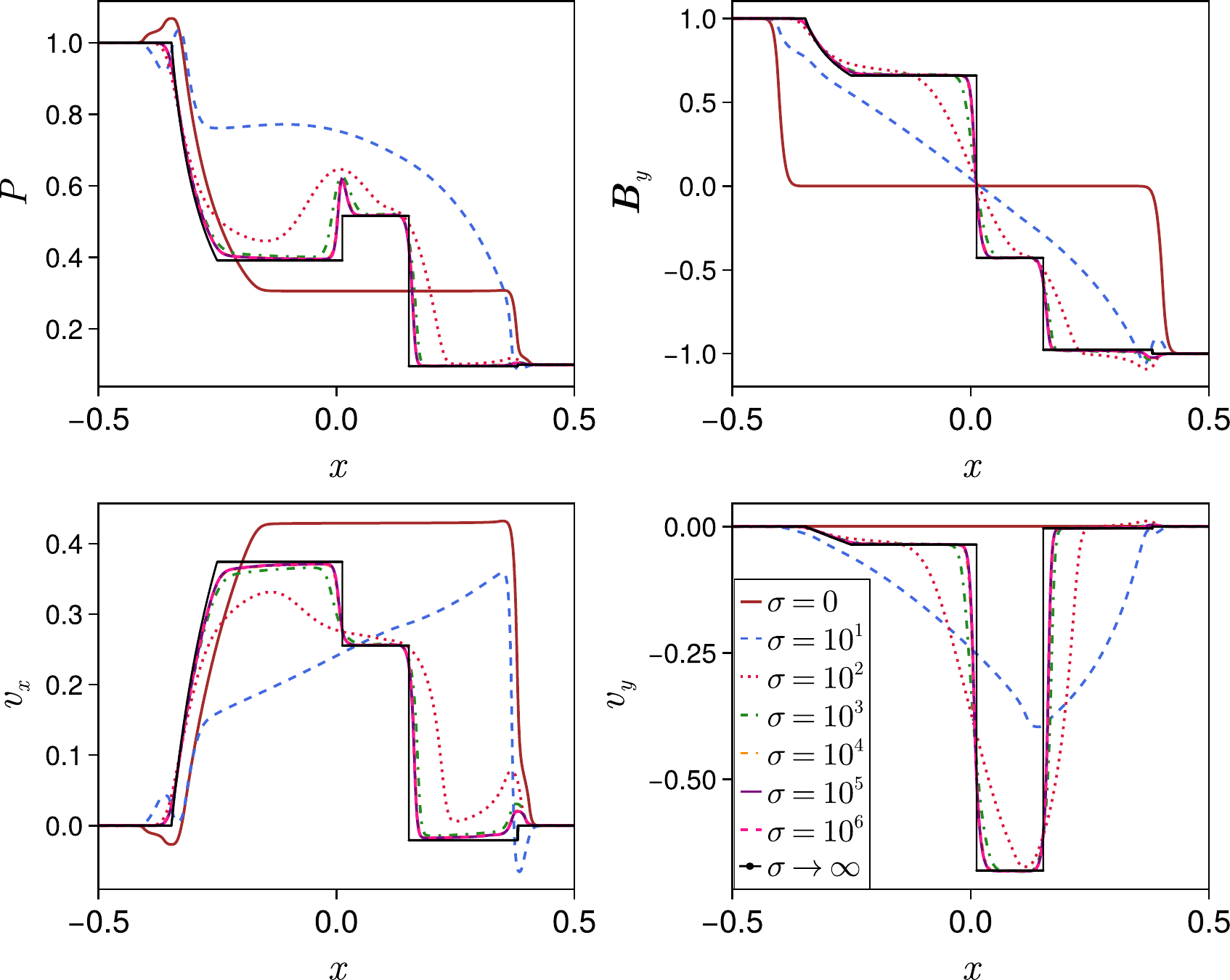}
		\caption{The Brio-Wu test is evolved up to $t=0.4$ on a uniform grid with
			$N=500$ points. The spatial reconstruction is carried out using the minmod
			slope limiter. To study the dependence on resistivity and the recovery of the
			ideal-RMHD limit, we vary the electrical conductivity in the range
			$ 0 \leq \sigma \leq 10^{6}$.}
		\label{fig:brio_wu}
	\end{figure}

 Figure~\ref{fig:brio_wu} compares the RRMHD solutions obtained for
different electrical conductivities with the exact ideal-MHD solution
of Ref.~\cite{GIACOMAZZO_REZZOLLA_2006}, shown by black circular
markers. The conductivity-dependent results span
$\sigma=0$--$10^{6}$ and are distinguished by color and line style:
$\sigma=0$ (brown solid), $10$ (blue dashed), $10^{2}$ (crimson
dotted), $10^{3}$ (green dash--dotted), $10^{4}$ (orange
dash--dot--dotted), $10^{5}$ (purple solid), and $10^{6}$ (pink
solid).
    
	To examine the recovery of the ideal-MHD limit, we vary the electrical
	conductivity $\sigma$ in the range $0 \rightarrow 10^6$. For this test, the CFL coefficient is set to $C_{\rm CFL}=0.4$ for all values of $\sigma$, except for the highly conductive cases; for example, with $\sigma=10^{6}$, we set $C_{\rm CFL}=0.1$.
	As $\sigma$ increases, the effective resistivity decreases and the magnetic
	Reynolds number becomes large, $R_m \gg 1$. For sufficiently large $\sigma$, the dissipative corrections associated with
	finite resistivity become subdominant, allowing a direct comparison with the
	ideal-RMHD reference solution \cite{GIACOMAZZO_REZZOLLA_2006}.
	\subsection{(2+1)-dimensional Cartesian coordinates }
	The implementation for $(2+1)$-dimensional Cartesian geometry is  straightforward: one only needs to include the additional flux contribution along the $y$ direction, while the remaining structure of the equations remains unchanged. The evolution system can therefore be written in conservative form as
	\begin{equation}
		\partial_{t} \mathcal{U}
		+ \partial_{x}\mathcal{F}^{x}
		+ \partial_{y}\mathcal{F}^{y}
		= \mathcal{S},
	\end{equation}
	like the one dimensional case, here we do not explicitly write the $(x,y,t)$ dependency of $\mathcal{U}$, $\mathcal{F}^{x}$, $\mathcal{F}^{y}$,  $\mathcal{S}$ and $\mathcal{S'}$. The additional flux in the y-direction $(\mathcal{F}^{y})$ is given by
	\begin{align}
		\mathcal{F}^{y} = [N_{c}v^{y}, T^{yx}, T^{yy}, T^{yz}, S^{y}, -\mathrm{B}^{z},\;\psi, \mathrm{B}^{x}, \notag \\ \mathrm{E}^{z}, \; \phi, -\mathrm{E}^{x},\; J^{y}, \; \mathrm{E}^{y},\; \mathrm{B}^{y}]^{T}
	\end{align}
	where $T^{yj}$ ($j=x,y,z$)
    \begin{align} T^{yj} = (e + P)\gamma^2 v^{y}v^{j} + (P +\varepsilon_{em})\delta^{yj} & {\notag}\\ - \mathrm{E}^{y}\mathrm{E}^{j} - \mathrm{B}^{y}\mathrm{B}^{j}. 
	\end{align}

	\subsubsection{Strong cylindrical explosion}
	
	We next consider a two-dimensional strong cylindrical explosion test. This
	problem is commonly used as a multidimensional benchmark for relativistic
	magnetohydrodynamic solvers, since the initially over-pressured region
	generates strong shocks propagating both parallel and transverse to the
	background magnetic field \cite{10.1046/j.1365-8711.1999.02244.x, 10.1111/j.1365-2966.2007.12448.x}. It therefore provides a rigorous test of the
	shock-capturing capability, robustness, and multidimensional consistency of
	the numerical scheme.
	
	The computational domain is a Cartesian box,$
	(x,y) \in [-6,6]\times[-6,6],$
	discretized uniformly using $(200\times 200)$ grid points and we set $C_{\text{CFL}} = 0.4$. The explosion is
	centered at the origin, and the initial thermodynamic variables 
 are assumed to have a radially symmetric profile, prescribed in terms of the radial distance from the center $ r = \sqrt{x^{2}+y^{2}}$. But, the radial symmetry is broken by the constant magnetic field taken along the $x$-direction. The initial condition is divided into three regions: (i) the central region $(0 \leq r < 0.8)$, (ii) transition region $(0.8 \leq r \leq 1.0 )$, and (iii) the rest part as background or ambient region. 
	Inside the central region, we set the number density $n_{\rm in}$ and pressure $P_{\rm in}$ as
	\begin{equation}
	    n_{\rm in}=0.01, 
		\qquad 
		  P_{\rm in}=1.0.
	\end{equation}
   For the ambient medium the number density $n_{\rm bg}$ and pressure $P_{\rm bg}$  initialized as
	\begin{equation}
	    n_{\rm bg}=10^{-3}, 
		\qquad 
		P_{\rm bg}=10^{-3}.
	\end{equation}
	In the transition region, the number density and pressure are
	smoothly connected to the ambient region's number density and pressure. We further define the normalized radial
	coordinate
	\begin{equation}
		\zeta = \frac{r-0.8}{1.0-0.8},
	\end{equation}
	and use the interpolation function
	\begin{equation}
		h(\zeta)
		=
		\frac{\exp(-a\zeta)-\exp(-a)}
		{1-\exp(-a)} .
	\end{equation}
    Where $a$ is a free parameter that controls the sharpness of the transition, here we set $a=1$.
	The number density and pressure in the transition region are then given by
	\begin{align}
		n(r)
		&=
		n_{\rm bg}
		+
		h(\zeta)
		\left(n_{\rm in}-n_{\rm bg}\right),\\
		P(r)
		&=
		P_{\rm bg}
		+
		h(\zeta)
		\left(P_{\rm in}-P_{\rm bg}\right).
	\end{align}
	
	The initial velocity is set to zero everywhere,
	\begin{equation}
		\mathbf{v}=(0,0,0),
	\end{equation}
	and the magnetic field is initialized as a uniform background field aligned
	with the $x$-direction,
	\begin{equation}
		\mathbf{B}=(0.1,0,0).
	\end{equation}
	The electric field and the local charge density are initially set to
	\begin{equation}
		\mathbf{E}=(0,0,0),
		\qquad
		q_{0}=0 .
	\end{equation}
	
	For the resistive RMHD evolution, we use a constant electrical conductivity
	$\sigma = 55.55$, (this value is taken from \cite{10.1046/j.1365-8711.1999.02244.x} for a one-to-one comparison). The
	ideal-gas equation of state \eqref{ideal_EOS} is taken with the adiabatic index
	$\Gamma = 4/3$.
	We set the
	$C_{\rm CFL} = 0.4 ,$
	and the simulation is evolved up to
	$t_{\rm final} = 4.0 .$
	This setup produces a strongly anisotropic blast-wave structure. The resulting evolution contains strong shock waves and deforms the initial constant magnetic fields. 
	\begin{figure}[h]
		\centering
		\hspace*{ - 9 mm}
		\includegraphics[width=1.14\linewidth]{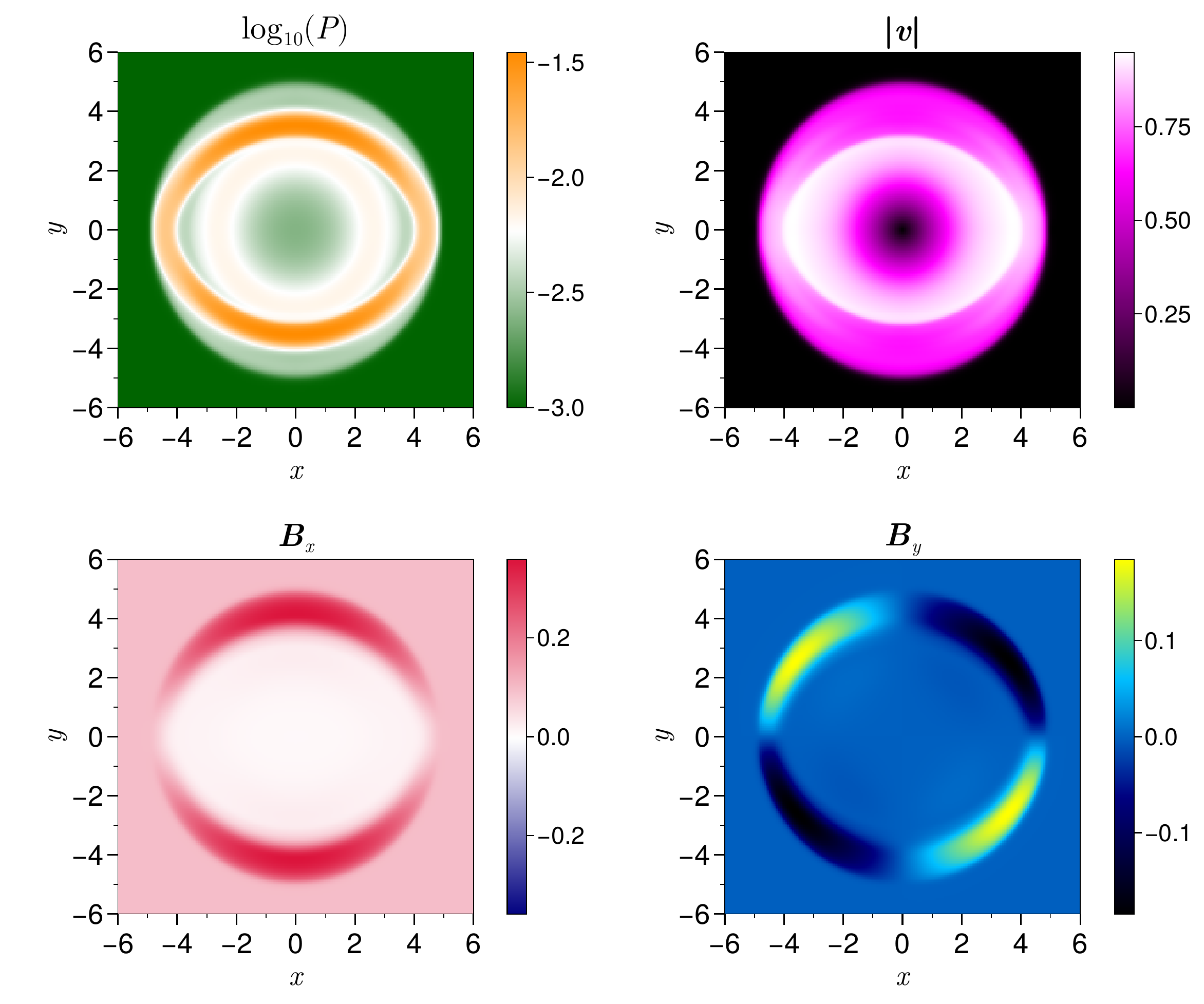}
		\caption{Two-dimensional strong cylindrical blast-wave test at $t_{\rm final}=4.0$.
			The panels show the spatial distributions of $\log_{10}(P)$, the velocity
			magnitude $|\mathbf{v}|$, and the magnetic-field components $\mathrm{B}_{x}$ and
			$\mathrm{B}_{y}$. The initially circular over-pressurized region expands into a
			uniformly magnetized medium, producing an anisotropic blast structure governed
			by the background magnetic field.}
		\label{fig:cylindrical_blast}
	\end{figure}
	
	As an additional robustness check, we repeated the Cartesian
cylindrical blast test at a higher conductivity,
$\sigma=2000$, using $C_{\rm CFL}=0.1$ while keeping all other
initial parameters unchanged. This is the largest conductivity
for which the standard blast configuration could be evolved
without primitive-variable recovery failure, while preserving
the expected anisotropic blast structure. The corresponding
results are shown in Appendix~\ref{app:Extra cylindrical blast and rotor tests}.

We further examined the primitive-variable recovery in a
weaker-field configuration, $\mathbf{B}=(0.05,0,0)$, at
$\sigma=10^{6}$ with $C_{\rm CFL}=0.1$, again keeping the
remaining initial parameters unchanged. This configuration was
successfully evolved without any primitive-variable recovery
failure, demonstrating that the recovery procedure remains
stable at very large conductivity for the weaker magnetic field
considered here. The corresponding results are also presented
in Appendix~\ref{app:Extra cylindrical blast and rotor tests}.
	
	The anisotropy of the blast is a physical consequence of the imposed background magnetic field. Expansion perpendicular to the field is affected by magnetic pressure, while magnetic tension modifies the propagation of disturbances along the field direction. The strong deformation of the magnetic field and the presence of
    sharp discontinuities also make this problem a useful test of the
    divergence-cleaning procedure.
	
	\subsubsection{Cartesian rotor test}
	
	Next, we consider a two-dimensional Cartesian rotor test. This problem is a
	standard multidimensional benchmark for RMHD solvers, since it probes
	the ability of the numerical scheme to follow the interaction between a rapidly
	rotating dense fluid and an initially uniform magnetic field \cite{flash_architecture, Nakamura2023}. The rotation twists
	the magnetic field lines and launches torsional Alfv\'en waves into the ambient
	medium, thereby testing the robustness of the scheme in the presence of strong
	velocity shear, magnetic stresses, and multidimensional wave propagation.
	
	The computational domain is a Cartesian box,
	$(x,y) \in [-0.5,0.5]\times[-0.5,0.5],
	$
	discretized uniformly using $(300\times 300)$ grid points. The rotor is centered
	at the origin, and the initial conditions are prescribed in terms of the radial distance from the center $ r = \sqrt{x^{2}+y^{2}}$.
	The number density and velocity fields are defined in terms of radial distance $r$. We
	introduce the function
	\begin{equation}
	g(r) =
		\begin{cases}
			1, & r \leq r_{0}, \\[4pt]
			\displaystyle \frac{r_{1}-r}{r_{1}-r_{0}}, & r_{0}<r<r_{1}, \\[8pt]
			0, & r \geq r_{1},
		\end{cases}
		\label{eq:rotor_profile}
	\end{equation}
	with
	\begin{equation}
		r_{0}=0.1,
		\qquad
		r_{1}=0.12 .
	\end{equation}
	The number density is then initialized as
	\begin{equation}
		n(r)
		=
		n_{\rm out}
		+
		g(r)\left(n_{\rm in}-n_{\rm out}\right),
	\end{equation}
	where
	\begin{equation}
		n_{\rm in}=10.0,
		\qquad
		n_{\rm out}=1.0 .
	\end{equation}

    \begin{figure}[h]
		\centering
		\hspace*{- 9 mm}
		\includegraphics[width=1.15\linewidth]{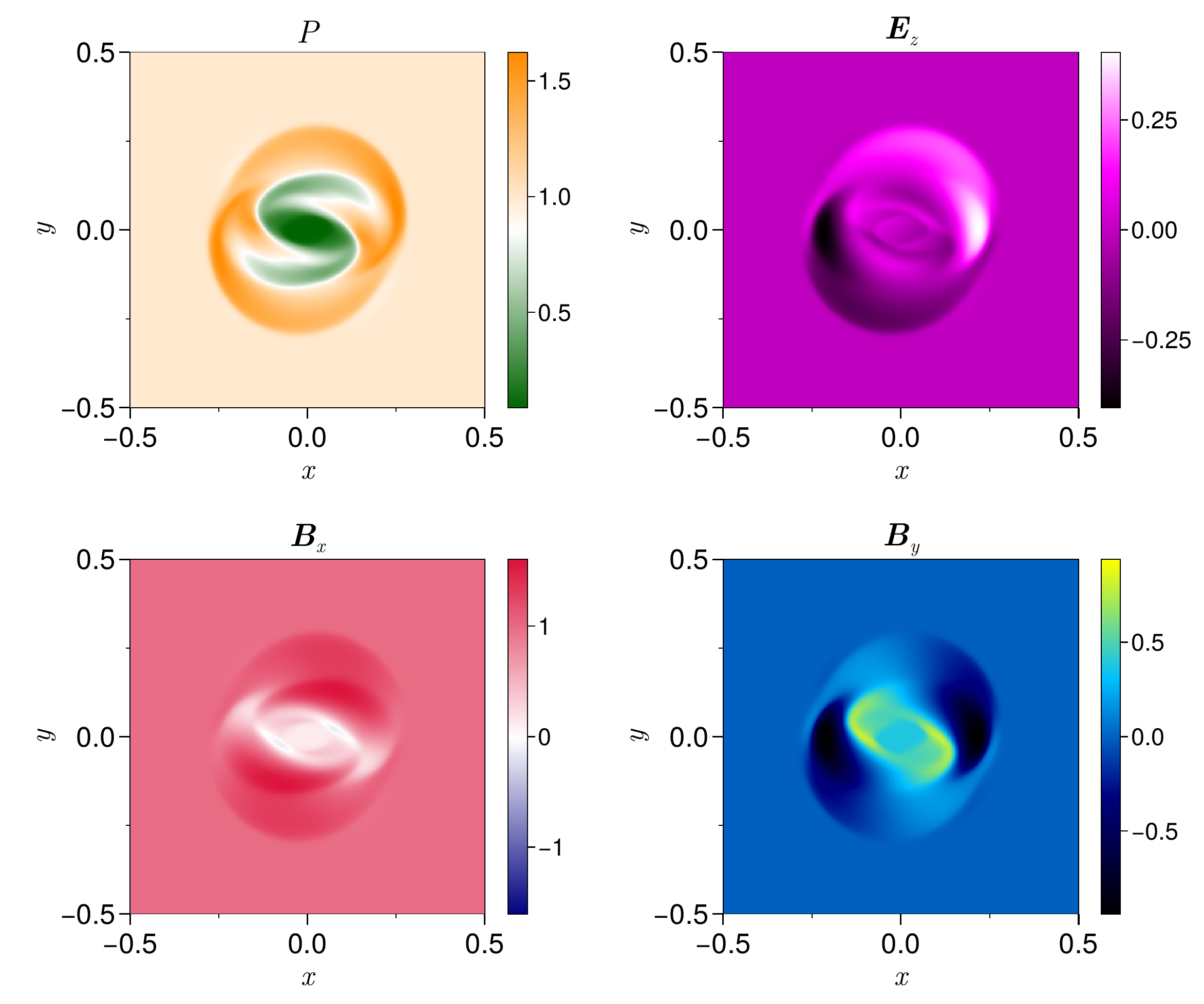}
		\caption{Two-dimensional Cartesian rotor test at $t_f=0.3$. The panels show the
			pressure $P$, electric-field component $\mathrm{E}_z$, and magnetic-field components
			$\mathrm{B}_x$ and $\mathrm{B}_y$. The dense rotating core twists the initially uniform
			magnetic field, generating the characteristic rotor structure in the pressure
			and electromagnetic fields.}
		\label{fig:cartesian_rotor}
	\end{figure}

	Thus, the central rotor has a density \(n=10.0\), while the ambient medium has
	density \(n=1.0\), with a linear transition between \(r_{0}\) and \(r_{1}\).
	
	The velocity field initialized as a rigid-body rotation whose angular profile varies smoothly with radius,
	\begin{equation}
		v^{x} = - g(r)\,\Omega\, y,
		\quad
		v^{y} = g(r)\,\Omega\, x,
		\quad
		v^{z}=0,
	\end{equation}
	with angular velocity $\Omega = 7.0 $.
	The pressure is initialized uniformly as $P = 1.0$ ,
	and the adiabatic index of the ideal-gas equation of state is
	$\Gamma = 4/3$. The magnetic field is initially uniform and aligned with the \(x\)-direction,
	\begin{equation}
		\mathbf{B} = (\mathcal{B}_{0},0,0),
	\end{equation}
    where $\mathcal{B}_{0} =1.0$. We set the electrical conductivity $\sigma = 	10^6$, since the conductivity is very high the initial electric field specified using the ideal MHD condition
	$\mathbf{E}=-\mathbf{v}\times\mathbf{B}$. For the present configuration this
	gives
	\begin{equation}
		\mathbf{E} = (0, 0, v^{y}\mathcal{B}_{0})
	\end{equation}
	The net local charge density is set to zero, $q_{0}=0$. 
	The time step is determined from the condition with
	$C_{\rm CFL}=0.4,$
	and the system is evolved up to $t_{f}=0.3$.
	
	The dense rotating core winds the initially uniform magnetic field and transfers angular momentum to the surrounding medium through magnetic tension. The generated $\mathrm{E}_z$ component provides a direct diagnostic of the coupling between transverse motion and the magnetic field. Since $\sigma=10^6$ is used, the solution should remain close to the ideal-RMHD rotor while still being evolved by the resistive IMEX algorithm. This results matches with the standard ideal MHD results \cite{10.1093/mnras/sty419}.

	\subsection{(2+1)-dimensional Milne coordinates}
	The Milne coordinate system is well suited for describing boost-invariant longitudinal expansion in high-energy heavy-ion collisions. Milne coordinates $(\tau,x,y,\eta)$, is related to the Cartesian coordinate via 
	\begin{equation}
		\tau = \sqrt{t^2-z^2},
		\qquad
		\eta = \frac{1}{2}
		\ln\left(\frac{t+z}{t-z}\right),
	\end{equation}
    where \(\tau\) is the longitudinal proper time  and  \(\eta\) is the spacetime rapidity.
	The metric tensor in this case is
	\begin{align}
		g_{\mu\nu}
		&=
		\mathrm{diag}\left(1,-1,-1,-\tau^2\right).
	\end{align}
	The only non-vanishing Christoffel symbols are
	\begin{equation}
		\Gamma^{\tau}_{\eta\eta}=\tau,
		\qquad
		\Gamma^{\eta}_{\tau\eta}
		=
		\Gamma^{\eta}_{\eta\tau}
		=
		\frac{1}{\tau}.
	\end{equation}
 The determinant of the metric is $g = -\tau^2$, and $\sqrt{-g}=\tau$.
	
	In Milne coordinates, a Lorentz boost along the beam direction corresponds to
	a translation in the spacetime rapidity $\eta$, while
	$\tau$ remains invariant. Therefore, Bjorken symmetry is implemented by
	assuming that all physical quantities are independent of $\eta$, that also implies 
	$\partial_\eta(\cdots)=0$.
	Then the of system
	equations given in ~\eqref{eq:Evolving_equations} for $(2+1)$D Milne coordinates becomes
	\begin{equation}
		\partial_{\tau}\left(\tau\mathcal{U}\right)
		+
		\partial_x\left(\tau\mathcal{F}^{x}\right)
		+
		\partial_y\left(\tau\mathcal{F}^{y}\right)
		=
		\tau\mathcal{S}.
		\label{eq:milne_conservative}
	\end{equation}
	Here, $\mathcal{S}$ contains the geometric source terms arising from the
	non-zero Christoffel symbols \cite{DenicolRischke2021}. The $\mathcal{U}$, $\mathcal{F}^{x}$ and $\mathcal{F}^{y}$ in this case are
	\begin{align}
		\mathcal{U}
		= [ N_c,
		S^x ,
		S^y ,
		S^\eta ,
		\varepsilon ,
		\mathrm{E}^x ,
		\mathrm{E}^y ,
		\mathrm{E}^\eta , & \notag \\
		\mathrm{B}^x ,
		\mathrm{B}^y ,
		\mathrm{B}^\eta,
		q ,
		\psi ,
		\phi ]^{T}, 
	\end{align}
	\begin{align}
		\mathcal{F}^{x} = [ N_{c}v^{x}, T^{xx}, T^{xy},T^{x \eta}, S^{x} , \psi, \mathrm{B}^{\eta}, - \mathrm{B}^{y}, & \notag \\
		\phi, -  \mathrm{E}^{\eta}, \mathrm{E}^{y}, J^{x}, \mathrm{E}^{x},\mathrm{B}^{x}]^{T}
	\end{align}
	\begin{align}
		\mathcal{F}^{y} = [N_{c}v^{y}, T^{yx}, T^{yy}, T^{y \eta}, S^{y}, - \mathrm{B}^{\eta}, \psi, \mathrm{B}^{x},  & \notag \\
		\mathrm{E}^{\eta}, \phi , -\mathrm{E}^{x}, J^{y},\mathrm{E}^{y},\mathrm{B}^{y}]^{T}
	\end{align}
	where the quantities appearing in these expressions are
	\begin{align}
		N_{c} & = n \gamma, \\
		T^{xj}
		& =
		(e+P)\gamma^{2}v^{x}v^{j}
		+
		\left(P+\varepsilon_{\rm em}\right)\delta^{xj}
		{\notag} \\ &
		- 
		\mathrm{E}^{x}\mathrm{E}^{j}
		-
		\mathrm{B}^{x}\mathrm{B}^{j},\\
		T^{yj}
		& =
		(e+P)\gamma^{2}v^{y}v^{j}
		+
		\left(P+\varepsilon_{\rm em}\right)\delta^{yj}
		{\notag}\\ & 
		- 
		\mathrm{E}^{y}\mathrm{E}^{j}
		-
		\mathrm{B}^{y}\mathrm{B}^{j},\\    
		S^{j} & = (e + P) \gamma^2 v^{j} + (\mathbf{E} \times \mathbf{B})^{j}, 
	\end{align}
	here $j = x,y,\eta$. The source terms  $\mathcal{S}$ and $\mathcal{S'}$ is given as 
	\begin{align}
		\mathcal{S}  = [0, 0, 0, &-2\Gamma^{\eta}_{\tau \eta} T^{\tau \eta},  - \Gamma^{\tau}_{\eta \eta}T^{\eta \eta}, 
		-J^{x}, -J^{y}, - J^{\eta} \notag, \\ 
		& 0, 0, 0, 0,  q-\kappa \psi + (\psi/\tau), -\kappa \phi + (\phi/\tau)]^{T},
	\end{align}
	clearly we need the $T^{\tau \eta}$ and $T^{\eta \eta}$ component of the energy-momentum tensor whivh are given by
	\begin{equation}
		T^{\tau \eta} = (e + P) \gamma^2 v^{\eta} + (\mathbf{E} \times \mathbf{B})^{\eta},
	\end{equation}
	\begin{align}
		T^{\eta \eta}  = (e + P) \gamma^2 &(v^{\eta})^{2} + \frac{P + \varepsilon_{\text{em}}}{\tau^2} {\notag} \\   & - (\mathrm{E}^{\eta})^{2} - (\mathrm{B}^{\eta})^{2}.  
	\end{align}

	\subsubsection{Rotor test in Milne coordinates}

	We next consider the rotor test in Milne coordinates. This test is used to examine the multidimensional coupling between transverse rotation, magnetic stresses, and the longitudinal expansion \cite{Inghirami2016}. 
	The rotor is centered at the origin of the transverse plane. We have taken the plane size as $x \in [-0.5,0.5]$ and $y \in [-0.5,0.5]$ with $(400 \times 400)$ uniform grids. The initial data are prescribed in terms of the radius 
	$r = \sqrt{x^2+y^2}$ similar to $(2+1)$D cartesian system.
    Similar to the Cartesian rotor test, we choose a larger value of $\sigma$ and the high inside pressure matched with the much lower outside pressure through a thin layer where both side pressure was matches via a smooth interpolating function Eq~\eqref{eq:rotor_profile}. 
	\begin{equation}
        P(r) = P_{\rm out} + g(r) (P_{\rm in} - P_{\rm out})
	\end{equation}
	with
	\begin{equation}
		P_{\rm in}=5.0,\qquad
		P_{\rm bg}=1.0.
	\end{equation}
	The number density is taken to be numerically negligible
and is initialized uniformly at the density floor, This choice preserves the positivity condition required by
the primitive-variable recovery while ensuring that the
dynamics is dominated by the pressure gradient, rotation,
and electromagnetic field.
	
	Inside the disk, the transverse velocity is initialized as a rigid rotation,
	\begin{equation}
		v^x = - g(r) \frac{\Omega y}{r_0} ,
		\qquad
		v^y = g(r)\frac{\Omega x}{r_0},
		\qquad
		v^\eta = 0,
	\end{equation}
    
    we set $r_{0}=0.1$ and $r_{1} = 0.12$ for $g(r)$ and angular velocity is set to $\Omega = 0.70$.
	Outside the disk, the fluid is initially at rest,
	\begin{equation}
		v^x=v^y=v^\eta=0.
	\end{equation}
	The condition $v^\eta=0$ in Milne coordinates corresponds to a Bjorken-type longitudinal expansion, $v^z=z/t$, while the imposed rotor motion is purely transverse.
	
	The magnetic field is initially uniform and aligned with the $x$-direction,
	\begin{equation}
		\mathbf{B} = (\mathcal{B}_{0},0.0,0.0),
	\end{equation}
	with $\mathcal{B}_{0} = 2.0$. We set the electrical conductivity $\sigma = 10^6$ and 
	the electric field is initialized using the ideal-MHD relation 
	\begin{equation}
		\mathbf{E}
		=
		-\mathbf{v}\times\mathbf{B}.
	\end{equation}
	In the Milne coordinate-basis formulation if we write the components of $\mathbf{v}$ and $ \mathbf{B}$ they are given as,
	\begin{equation}
		\mathbf{v}
		=
		(v^x,v^y,\tau v^\eta), \quad
		\mathbf{B}
		=
		(\mathrm{B}^x,\mathrm{B}^y,\tau \mathrm{B}^\eta).
	\end{equation}
	For the present initial condition,
	\begin{equation}
		\mathrm{E}^x=0,\qquad
		\mathrm{E}^y=0,
	\end{equation}
	and the physical longitudinal electric field is
	\begin{equation}
		\mathrm{E}^{z} = v^y \mathcal{B}_0 .
	\end{equation}
	Corresponding elecetric field component along $\eta$ is initialized as 
	\begin{equation}
		\mathrm{E}^\eta = \frac{\mathrm{E}^{z}}{\tau_0}
		= \frac{v^y \mathcal{B}_0}{\tau_0}.
	\end{equation}
	\begin{figure}[h]
		\centering
		\hspace*{-0.3 cm}\includegraphics[width=1.10\linewidth]{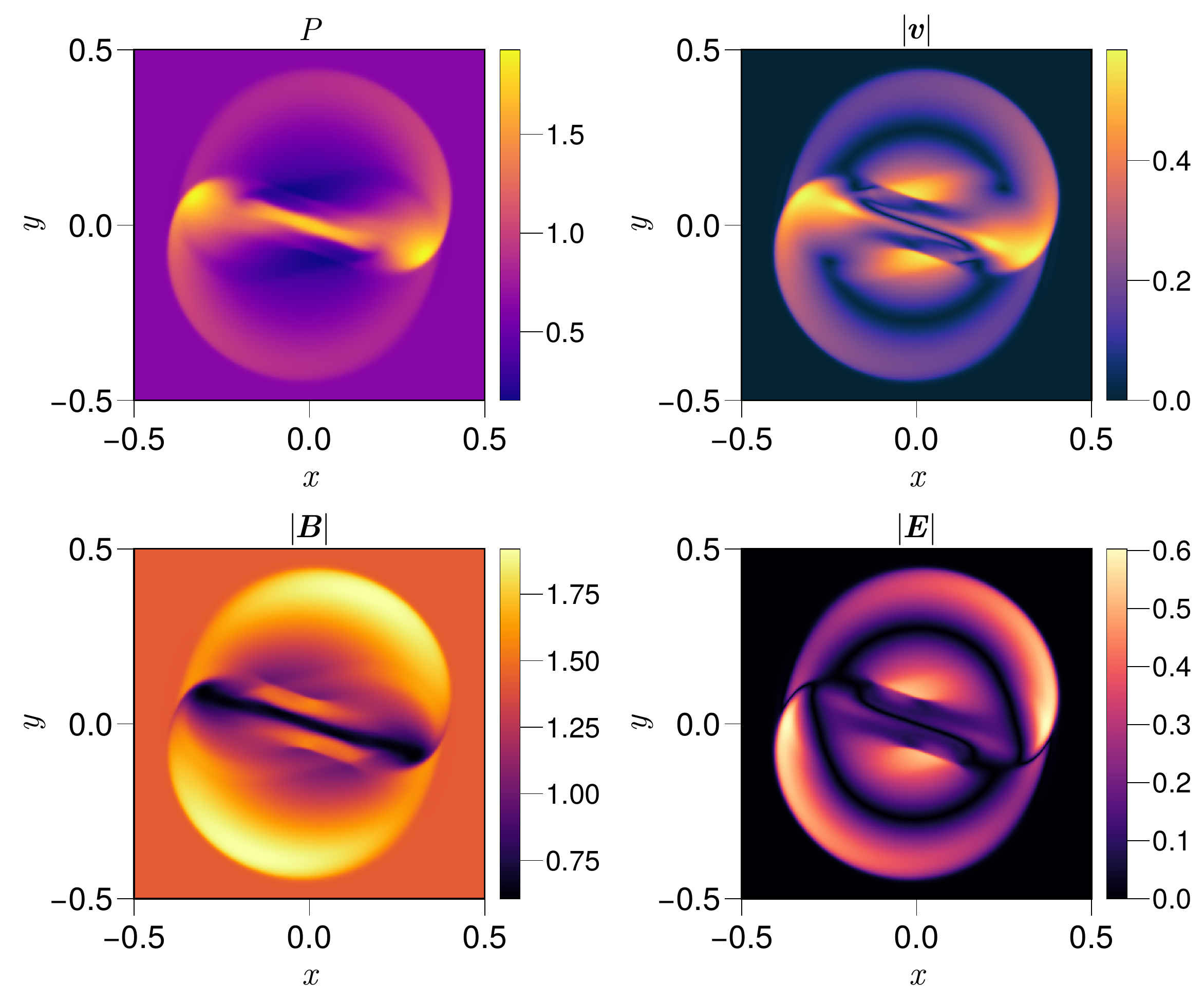}
		\caption{Milne-coordinate rotor test at $\tau=1.4$. The panels show the pressure $P$, physical velocity magnitude $|\mathbf{v}|$,  magnetic-field magnitude $|\mathbf{B}|$, and  electric-field magnitude $|\mathbf{E}|$ for $\sigma = 10^{6}$.}
		\label{fig:milne_rotor_sigma1e6}
	\end{figure}

	For the present test, the simulation is initialized at $\tau_0=1.0$ and evolved until $\tau_f=1.4$. The CFL coefficient is set to $C_{\rm CFL}=0.25$, while the adiabatic index is fixed at $\Gamma=4/3$.
	This setup produces a rapidly rotating, over-pressured disk embedded in a uniform magnetic field. Magnetic stresses transfer angular momentum from the rotating core to the surrounding medium. 
    If we compare the Cartesian and Milne-coordinate rotor tests, they differ in both their initial conditions and the treatment of longitudinal expansion. In the Cartesian test, the fluid is initialized with a uniform pressure and a dense rotating core, without longitudinal expansion. In contrast, the Milne-coordinate test employs a higher initial pressure inside the rotating region, a negligible rest-mass density, and boost-invariant longitudinal expansion. The latter introduces additional dilution through the expanding geometry, characterized by $\sqrt{-g}=\tau$, resulting in a decrease in the fluid and magnetic pressures across the computational domain. Thus, the Milne-coordinate test examines the combined effects of transverse rotation, magnetic stresses, and longitudinal expansion.
    To further compare our results with previously reported works such as \cite{Nakamura2023, Inghirami2016}, we initialized our Milne rotor test that matches with the cited works and corresponding  results are presented in Appendix~\ref{app:Extra cylindrical blast and rotor tests}

	\subsection{ Magnetically modified anisotropic expansion }
    To understand the influence of electromagnetic fields on the transverse expansion of QGP, first we consider a simplified model of boost-invariant, elliptically deformed fireball with initial conditions similar to HIC. This setup allows us to
examine how initial magnetic field affects the evolution of momentum
anisotropy and the energy transfer between the electromagnetic
field and the fluid. A more realistic heavy-ion initialization is considered in the following section \ref{sec:hic_initial_conditions}.
	  The system initialized at $\tau_{0} = 0.6 \; \rm{fm}$ with an elliptic Gaussian matter-energy density profile,
	\begin{equation}
		e(\tau_{0},x,y) = e_{\rm bg} + (e_{0} - e_{\rm bg}) \exp{\bigg[ - \frac{x^2}{2 \sigma^{2}_{x}} - \frac{y^{2}}{2 \sigma^{2}_{y}}   \bigg]}.
	\end{equation}
	For our test problem we choose $e_{0} = 30 \;  \rm{GeV/fm}^{3}$ with $\sigma_{x} = 3.0 \; \rm{fm}$ and $\sigma_{y} = 4.5 \; \rm{fm}$. Here $e_{\rm bg}$ is the numerical background chosen as $e_{\rm bg} = 10^{-6} \; \text{GeV/fm}^{3}$.
	Since $\sigma_{y}>\sigma_{x}$, the matter
	distribution is initially elongated along the $y$ direction. The net
	baryon density is taken to be negligible, and the conformal equation 
	$P=e/3 $ is employed. 
	The fluid is initially at rest, which corresponds to Bjorken longitudinal expansion in Milne coordinates. Along with that the initial
	electric field and local charge density are set to zero.
    In the magnetized setup, the initial transverse magnetic field is generated from a vector potential $(\mathcal{A})$ directed along the beam axis,
	\begin{equation}
		{\mathcal A}_{\parallel}(x,y)
		=
		-\mathcal{B}_{0}x
		\exp\left[
		-\frac{x^{2}}{2L_{x}^{2}}
		-\frac{y^{2}}{2L_{y}^{2}}
		\right],
		\label{eq:magnetic_vector_potential}
	\end{equation}
	with
	$
		L_{x}=4.0~{\rm fm},
    $ and $
		L_{y}=5.5~{\rm fm}
	$.
	The corresponding field components are
	\begin{align}
		\mathrm{B}^{x}(x,y)
		&=
		\mathcal{B}_{0}\frac{xy}{L_{y}^{2}}
		\exp\left[
		-\frac{x^{2}}{2L_{x}^{2}}
		-\frac{y^{2}}{2L_{y}^{2}}
		\right],
		\\
		\mathrm{B}^{y}(x,y)
		&=
		\mathcal{B}_{0}\left(1-\frac{x^{2}}{L_{x}^{2}}\right)
		\exp\left[
		-\frac{x^{2}}{2L_{x}^{2}}
		-\frac{y^{2}}{2L_{y}^{2}}
		\right],
		\\
		\mathrm{B}^{\eta}(x,y)&=0.
	\end{align}
	This construction satisfies
	\begin{equation}
		\partial_{x}\mathrm{B}^{x}+\partial_{y}\mathrm{B}^{y}=0
	\end{equation}

	analytically and produces a field predominantly directed along the $y$ direction in the central region \cite{FRAGILE2019100020}.
	The central field strength ($\mathcal{B}_{0}$) is specified through the magnetic-to-thermal
	pressure ratio
	\begin{equation}
		\beta_{B}
		:=
		\frac{\mathcal{B}_{0}^{2}}{2P_{0}}.
		\label{eq:beta_B_definition}
	\end{equation}
	We compare a zero-field reference, $\beta_{B}=0$, with a magnetized
	case by varying $\beta_{B}$. The electrical conductivity is kept constant at $\sigma = 5.8 \; \rm{MeV}$, inspired by Lattice QCD results \cite{PhysRevC.93.014905, Aarts2021,Inghirami2020}.
	\begin{table}[t]
		\centering
		\caption{Values of the dimensionless magnetic-field parameter $\beta_B$ and the corresponding peak magnetic-field strength expressed in units of $m^{2}_{\pi}$.}
		\label{tab:magnetic_field_strength}
		\begin{tabular}{
				S[table-format=1.5]
				S[table-format=1.5]
			}
			\toprule
			{$\beta_B$} & {$e_{\rm EM}\mathcal{B}_0/m_\pi^2$} \\
			\midrule
			0.10 & 2.05 \\
			0.50 & 4.59 \\
            1.0  & 6.50 \\
            1.5  & 7.96 \\
           2.0  & 9.19 \\
			\bottomrule
		\end{tabular}
	\end{table}
	
	The deformation of the matter distribution is characterized by the real part of  
    \begin{equation}
        \mathrm{Re}\bigg[- \frac{ \displaystyle\int d A(x+iy)^{n} e(x,y,\tau)}{\displaystyle\int dA ( x^2 + y^2)^{n/2} e(x,y,\tau)} \bigg],
    \end{equation}
    where $dA\, := dx\, dy\,$. The second $(n=2)$
	\begin{equation}
		\varepsilon_{2}(\tau)
		=
		\frac{
			\displaystyle
			\int dA\,
			(y^{2}-x^{2}) \; e(\tau,x,y)
		}{
			\displaystyle
			\int dA\,
			(y^{2}+x^{2}) \; e(\tau,x,y)
		},
		\label{eq:spatial_eccentricity_toy}
	\end{equation}
    and fourth $(n=4)$ 
    \begin{equation}
        \varepsilon_{4} = \frac{\displaystyle\int dA (-x^{4} + 6x^2y^2 -y^4)e(\tau,x,y)}{\displaystyle\int dA (x^2 + y^2)^2 e(\tau,x,y)}
    \end{equation}
	order spatial eccentricity, since we started from a elliptic Gaussian profile  $\varepsilon_{2}(\tau_{0})$ and $\varepsilon_{4}(\tau_{0})$ will have non-zero value. 

\begin{figure}[h]
    \centering
    \hspace{-1.0 cm}
    \includegraphics[width=1.10\linewidth]{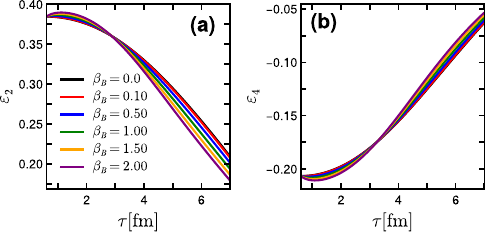}
    \caption{Evolution of the spatial anisotropies (a) $\varepsilon_{2}(\tau)$
    and (b) $\varepsilon_{4}(\tau)$ for different initial values of
    $\beta_{B}$.}
    \label{fig:spatial_aniso}
\end{figure}

The time evolution of $\varepsilon_{2}$ and $\varepsilon_{4}$ for different
initial values of $\beta_{B}$ is shown in Fig.~\ref{fig:spatial_aniso}. In all
cases, $\varepsilon_{2}$ decreases monotonically after a brief initial
plateau, from $\varepsilon_{2}\simeq 0.39$ at early times to
$\varepsilon_{2}\simeq 0.2$ at $\tau \approx 7$~fm, while $\varepsilon_{4}$,
which is initially negative, approaches zero from below. This is the expected
behavior of an expanding anisotropic fireball because the stronger pressure gradients
along the short axis drive faster expansion in that direction and gradually
wash out the initial spatial deformation, converting it into momentum
anisotropy. The dependence on the magnetic field is weak. The curves for
different $\beta_{B}$ cross at $\tau \approx 3$~fm for $\varepsilon_{2}$ and at
$\tau \approx 4$~fm for $\varepsilon_{4}$; at later times, $\varepsilon_{2}$
decreases slightly faster for larger $\beta_{B}$, indicating that the
anisotropic magnetic stress modestly accelerates the decay of the spatial
eccentricity. The spatial geometry of the fluid is thus only weakly modified by
the magnetic field in this setup, which will help in interpreting the
momentum anisotropy discussed below.

    The momentum space anisotropy generated in the fluid is measured using the transverse stress tensor of fluid as 
	
	\begin{equation}
		\varepsilon_{p}^{\rm M}(\tau)
		=
		\frac{
			\displaystyle
			\int dA\,
			\left(T_{\rm M}^{xx}-T_{\rm M}^{yy}\right)
		}{
			\displaystyle
			\int dA\,
			\left(T_{\rm M}^{xx}+T_{\rm M}^{yy}\right)
		}.
		\label{eq:matter_momentum_anisotropy_toy}
	\end{equation}
	For the ideal-fluid matter tensor \eqref{energy_momentum_matter},
	\begin{equation}
		T_{M}^{xx}-T_{M}^{yy}
		=
		(e+P)\gamma^{2}
		\left[(v^{x})^{2}-(v^{y})^{2}\right]
	\end{equation}
	
	The quantity $\varepsilon_{p}^{\rm M}$ therefore measures the
	difference between the matter momentum fluxes along the $x$ and $y$
	directions. 

	For completeness, we also calculate the momentum anisotropy associated
	with the complete matter-plus-electromagnetic energy--momentum tensor,
	\begin{equation}
		T^{\mu\nu}
		=
		T_{M}^{\mu\nu}
		+
		T_{f}^{\mu\nu},
	\end{equation}
	which is defined as
	\begin{equation}
		\varepsilon_{p}^{\rm tot}(\tau)
		=
		\frac{
			\displaystyle
			\int dA\,
			\left(T^{xx}-T^{yy}\right)
		}{
			\displaystyle
			\int dA\,
			\left(T^{xx}+T^{yy}\right)
		}.
		\label{eq:total_momentum_anisotropy_toy}
	\end{equation}
	
	    \begin{figure*}[t]
		\centering
		\hspace{-20 mm}
		\includegraphics[
		width=\textwidth
		]{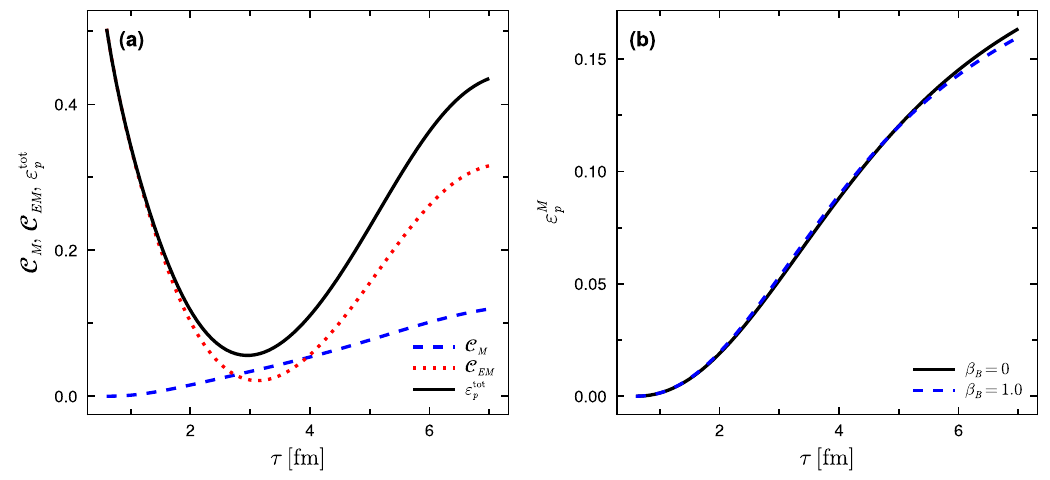}
		\caption{Panel (a) shows the total momentum anisotropy
			$\varepsilon_{p}^{\rm tot}$ together with its matter and
			electromagnetic contributions, ${\cal C}_{\rm M}$ and
			${\cal C}_{\rm EM}$, for $\beta_{B}=1.0$. Panel (b) compares the
			matter momentum anisotropy $\varepsilon_{p}^{\rm M}$ with and without
			the magnetic field.}
		\label{fig:anisotropy_1}
	\end{figure*}

	To distinguish the matter and electromagnetic parts of this quantity,
	we introduce
	\begin{equation}
		N_{\rm M}
		=
		\int dA\,
		\left(T_{\rm M}^{xx}-T_{\rm M}^{yy}\right), \; \;
		D_{\rm M}
		=
		\int dA\,
		\left(T_{\rm M}^{xx}+T_{\rm M}^{yy}\right),
	\end{equation}
    \begin{figure}[h]
		\centering
		\hspace*{-1.3 cm}\includegraphics[width=1.15\linewidth]{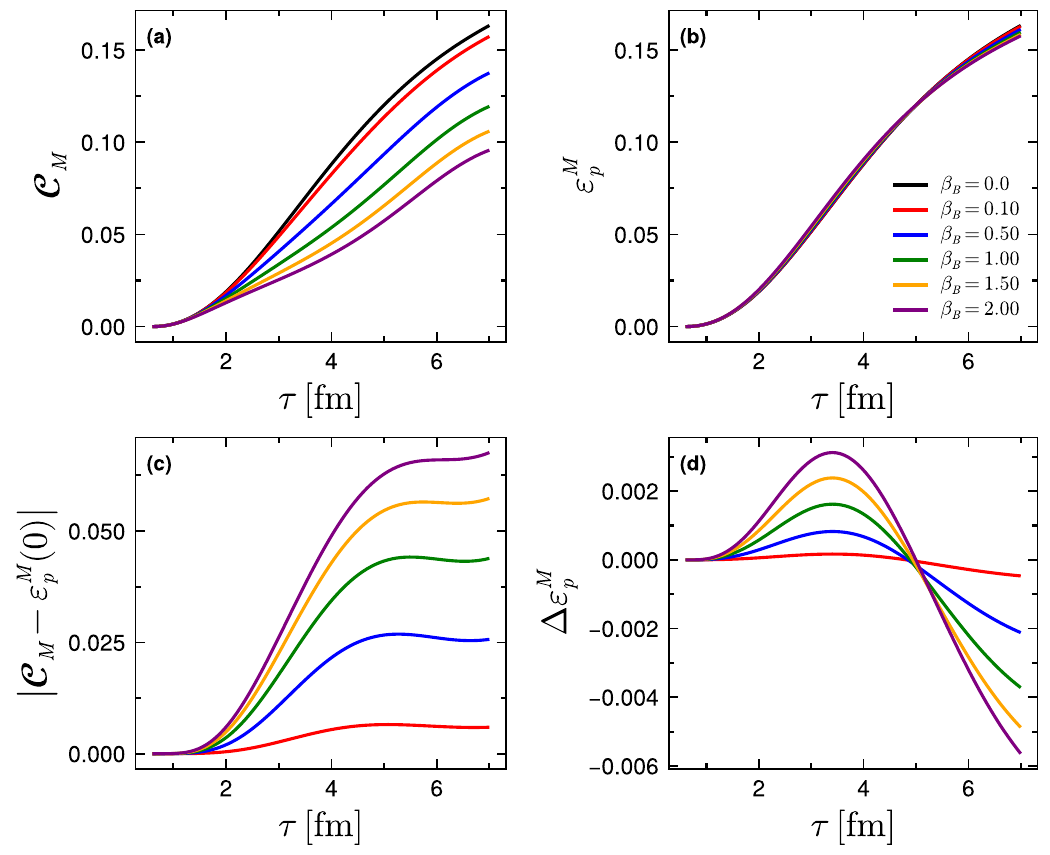}
		\caption{Panels (a) and (b) show ${\cal C}_{\rm M}$ and
			$\varepsilon_p^{\rm M}$ for different values of $\beta_B$.
			Panels (c) and (d) show the corresponding deviations from the
			zero-field reference,
			$\left|{\cal C}_{\rm M}(\beta_B)-\varepsilon_p^{\rm M}(0)\right|$
			and $\Delta\varepsilon_p^{\rm M}$, respectively.}
		\label{fig:anisotropy_2}
	\end{figure}

	\begin{equation}
		N_{\rm EM}
		=
		\int dA\,
		\left(T_{\rm EM}^{xx}-T_{\rm EM}^{yy}\right),
		\; \;
		D_{\rm EM}
		=
		\int dA\,
		\left(T_{\rm EM}^{xx}+T_{\rm EM}^{yy}\right).
	\end{equation}

	The matter and total momentum anisotropies can then be written as
	\begin{equation}
		\varepsilon_{p}^{\rm M}
		=
		\frac{N_{\rm M}}{D_{\rm M}},
		\qquad
		\varepsilon_{p}^{\rm tot}
		=
		\frac{N_{\rm M}+N_{\rm EM}}
		{D_{\rm M}+D_{\rm EM}}.
	\end{equation}

	We further define the matter and electromagnetic contributions to the
	total normalized momentum anisotropy as
	\begin{equation}
		\mathcal{C}_{\rm M}
		=
		\frac{N_{\rm M}}
		{D_{\rm M}+D_{\rm EM}},
		\qquad
		\mathcal{C}_{\rm EM}
		=
		\frac{N_{\rm EM}}
		{D_{\rm M}+D_{\rm EM}}.
		\label{eq:momentum_anisotropy_contributions}
	\end{equation}
	By construction,
	\begin{equation}
		\varepsilon_{p}^{\rm tot}
		=
		\mathcal{C}_{\rm M}
		+
		\mathcal{C}_{\rm EM}.
		\label{eq:total_anisotropy_decomposition}
	\end{equation}

	It is important to emphasize that ${\mathcal C}_{\rm M}$ is not the same
	quantity as $\varepsilon_{p}^{\rm M}$. Their relation is
	\begin{equation}
		\mathcal{C}_{\rm M}
		=
		\varepsilon_{p}^{\rm M}
		\frac{D_{\rm M}}
		{D_{\rm M}+D_{\rm EM}}.
		\label{eq:matter_contribution_relation}
	\end{equation}

Since the spatial anisotropies in Fig.~\ref{fig:spatial_aniso} are only weakly affected by the magnetic field, the collective flow of the fluid, and hence the matter momentum anisotropy, is also expected to change only slightly. The electromagnetic field can nevertheless modify the total momentum anisotropy directly, through its own contribution to the transverse stress tensor. Figures~\ref{fig:anisotropy_1} and \ref{fig:anisotropy_2} confirm that this direct contribution is the dominant effect.

For finite $\beta_B$, the electromagnetic contribution $\mathcal{C}_{\rm EM}$
is nonzero and therefore changes the total momentum anisotropy,
$\varepsilon_p^{\rm tot}=\mathcal{C}_{M}+\mathcal{C}_{\rm EM}$. By contrast,
the matter momentum anisotropy $\varepsilon_p^{M}$ deviates only weakly from
its zero-field evolution. The difference
$\Delta\varepsilon_p^{M}=\varepsilon_p^{M}(\beta_B)-\varepsilon_p^{M}(0)$
changes sign from positive to negative as $\tau$ increases, and its magnitude
grows with the initial magnetic-field strength. This
effect remains small, however, compared with the direct electromagnetic contribution $\mathcal{C}_{\rm EM}$ shown in Fig.~\ref{fig:anisotropy_1}.

	Next, we consider the Joule heating due to the presence of the current density and electric field. Throughout this subsection, $E^{\mu}$ and $B^{\mu}$ denote the comoving fields defined through Eq.~\eqref{Faraday_tensor}, while the bold symbols $\mathbf{E}$, $\mathbf{B}$, $\mathbf{J}$, $\mathbf{v}$, together with $q=J^{0}$, denote the corresponding quantities measured by the laboratory observer $\mathcal{I}_{\mu}$, expressed in the orthonormal basis of that observer. To distinguish irreversible resistive heating from the reversible
	exchange of energy between the fluid and the electromagnetic field,
	we additionally monitor the covariant Joule-heating source. Using Ohm's law~\eqref{Ohm's law} and the orthogonality condition $E^{\mu}u_{\mu}=0$, we define the irreversible Joule-heating density as
	\begin{equation}
		\mathcal{Q}_{J}
		=
		-E_{\mu}J^{\mu}
		=
		- E_{\mu}j^{\mu}
		=
		-\sigma E_{\mu}E^{\mu}
		\geq 0,
		\label{eq:covariant_joule_heating}
	\end{equation}
	where $j^{\mu} = \sigma E^{\mu}$ is the conduction current. The inequality holds because $E^{\mu}$ is spacelike, so that $E_{\mu}E^{\mu}=-|\mathbf{E}'|^{2}\leq 0$ for the mostly negative metric, where $\mathbf{E}'$ is the electric field in the local rest frame of the fluid. Hence $\mathcal{Q}_{J}=\sigma|\mathbf{E}'|^{2}$ is a Lorentz scalar, equal to the heating rate per unit proper volume and unit proper time.
	Using Eq.~\eqref{Components_of_FourE}, $\mathcal{Q}_{J}$ can be expressed in terms of the laboratory-frame quantities as
	\begin{equation}
		\mathcal{Q}_{J}
		=
		\sigma\gamma^{2}
		\left[
		\left|
		\mathbf{E}
		+
		\mathbf{v}\times\mathbf{B}
		\right|^{2}
		-
		\left(
		\mathbf{E}\cdot\mathbf{v}
		\right)^{2}
		\right].
		\label{eq:three_vector_joule_heating}
	\end{equation}
	Projecting Eq~\ref{eq: matter_enery_momentum_with_source} onto $u_{\nu}$ and using $u_{\nu}F^{\nu\lambda}J_{\lambda}=-E_{\lambda}J^{\lambda}=\mathcal{Q}_{J}$, the matter internal-energy equation for an ideal fluid can then be written as
	\begin{equation}
		D e
		+
		(e+P)\theta
		=
		\mathcal{Q}_{J},
		\qquad \text{ where }
		\theta=\nabla_{\mu}u^{\mu}.
		\label{eq:internal_energy_with_joule}
	\end{equation}
	Thus, $\mathcal{Q}_{J}$ is a positive definite source that
	opposes the reduction of the matter internal energy caused by expansion. In the laboratory frame, the four-force density $f^{\nu}\equiv F^{\nu\lambda}J_{\lambda}$ exerted by the field on the matter has the components $f^{\nu}=(\mathbf{J}\cdot\mathbf{E},\,\mathbf{f}_{\rm EM})$, where $\mathbf{J}\cdot\mathbf{E}$ is the electromagnetic power delivered to the matter per unit laboratory volume and the Lorentz force density is
    \begin{equation}
        \mathbf{f}_{\rm{EM}} = q \mathbf{E} + \mathbf{J} \times \mathbf{B}.
        \label{eq:Lorentz_force_density}
    \end{equation}
   
\begin{figure*}[t]
		\centering
		\hspace{-20 mm}
		\includegraphics[
		width=\textwidth
		]{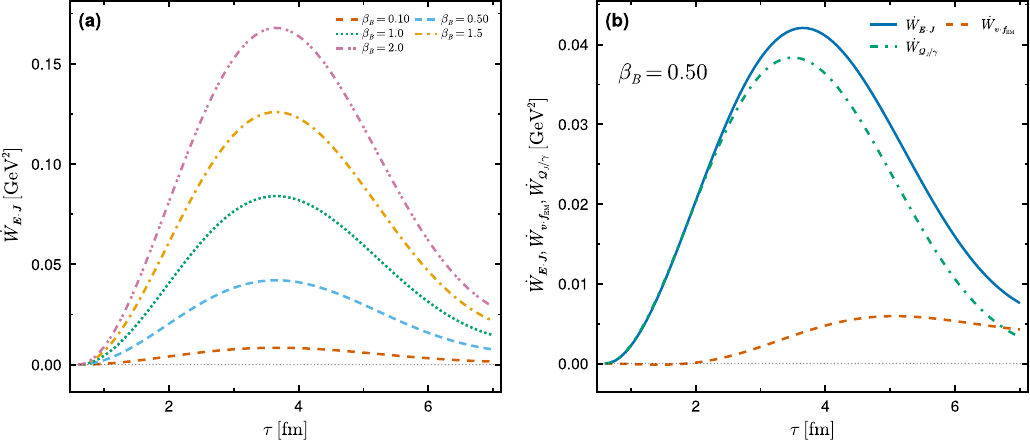}
		\caption{(a) Evolution of the electromagnetic power $\dot{W}_{\mathbf{E}\cdot\mathbf{J}}(\tau)$ for different initial magnetic-field strengths $\beta_{B}=0.10$, $0.50$, $1.0$, and $2.0$ (see
 Table~\ref{tab:magnetic_field_strength}). (b) Decomposition of $\dot{W}_{\mathbf{E}\cdot\mathbf{J}}(\tau)$ (solid blue line) into the mechanical contribution $\dot{W}_{\mathbf{v} \cdot \mathbf{f}_{\rm{EM}}}$ (dotted red line), and the Joule-heating contribution $\dot{W}_{\mathcal{Q}_{J}/\gamma}$ (dashed dot green line).}
		\label{fig:different_PEJ_diff_power}
	\end{figure*}
    
    The power delivered by $\mathbf{f}_{\rm{EM}}$ to the moving fluid is given as 
    \begin{equation}
        \mathbf{v} \cdot \mathbf{f}_{\rm{EM}} = q \mathbf{E} \cdot \mathbf{v} - \mathbf{J} \cdot (\mathbf{v} \times \mathbf{B}).
        \label{eq:Lorentz_force_power}
    \end{equation}
    Since $\mathcal{Q}_{J}=u_{\nu}f^{\nu}=\gamma\left(\mathbf{J}\cdot\mathbf{E}-\mathbf{v}\cdot\mathbf{f}_{\rm EM}\right)$, one obtains
    \begin{equation}
        \mathbf{J} \cdot (\mathbf{v} \times \mathbf{B}) = q \mathbf{v} \cdot \mathbf{E} + \frac{\mathcal{Q}_{J}}{\gamma} - \mathbf{J} \cdot \mathbf{E},
        \label{eq:second_Joule_heating_equation}
    \end{equation}
    and, combining Eqs.~\eqref{eq:second_Joule_heating_equation} and \eqref{eq:Lorentz_force_power},
    \begin{equation}
        \mathbf{E}\cdot \mathbf{J}   = \mathbf{v} \cdot \mathbf{f}_{\rm{EM}} + \frac{\mathcal{Q}_{J}}{\gamma}.
        \label{eq:Diff_power_terms}
    \end{equation}

Equation~\eqref{eq:Diff_power_terms} splits the electromagnetic power delivered
to the matter into the work done by the Lorentz force on the moving fluid and
the irreversible Joule heating.

In Milne coordinates, with $g_{\eta\eta}=-\tau^{2}$, the electromagnetic power
density takes the form
\begin{equation}
    \mathbf{E}\cdot\mathbf{J}
    =
    \mathrm{E}^{x}\mathrm{J}^{x} + \mathrm{E}^{y}\mathrm{J}^{y} + \tau^{2}\mathrm{E}^{\eta}\mathrm{J}^{\eta},
    \label{eq:eulerian_ej_power}
\end{equation}
where $E^{i}$ and $J^{i}$ are the contravariant components measured by the
laboratory observer. On a constant-$\tau$ hypersurface, the volume element per
unit space-time rapidity is $\tau\,dA$, and we monitor the terms of
Eq.~\eqref{eq:Diff_power_terms} integrated over this volume element,
\begin{equation}
    \dot{W}_{(\cdots)}(\tau)
    =
    \int_{\Omega_{\mathrm{M}}(\tau)}
    \tau\,dA\,(\cdots),
    \label{eq:integrated_scalar_joule}
\end{equation}
where $\Omega_{\mathrm{M}}(\tau)$ denotes the physical matter region. In the
numerical calculation, this region is selected using the energy-density
criterion
\begin{equation}
    \Omega_{\mathrm{M}}(\tau)
    =
    \left\{
    (x,y)\,:\,
    e(\tau,x,y)>e_{\rm bg}
    \right\}.
    \label{eq:matter_region_joule}
\end{equation}
Following Eq.~\eqref{eq:Diff_power_terms}, we consider three integrated
quantities: the electromagnetic power delivered to the matter,
$\dot{W}_{\mathbf{E}\cdot\mathbf{J}}$; the mechanical power,
$\dot{W}_{\mathbf{v}\cdot\mathbf{f}_{\rm EM}}$; and the Joule-heating contribution, $\dot{W}_{\mathcal{Q}_{J}/\gamma}$.

  Unlike $\mathcal{Q}_{J}$, the power density $\mathbf{E}\cdot\mathbf{J}$ is not
sign-definite, because the mechanical term $\mathbf{v}\cdot\mathbf{f}_{\rm EM}$
can have either sign. A positive value of $\mathbf{E}\cdot\mathbf{J}$
corresponds to a net transfer of energy from the electromagnetic field to the
matter, and a negative value to a transfer in the opposite direction.
Integrating Eq.~\eqref{eq:Diff_power_terms} according to
Eq.~\eqref{eq:integrated_scalar_joule} gives
\begin{equation}
    \dot{W}_{\mathbf{E}\cdot\mathbf{J}}
    =
    \dot{W}_{\mathbf{v}\cdot\mathbf{f}_{\rm EM}}
    +
    \dot{W}_{\mathcal{Q}_{J}/\gamma}.
    \label{eq:integrated_em_work_decomposition}
\end{equation}
This decomposition is also useful for interpreting the momentum anisotropy.
The non-negative term $\dot{W}_{\mathcal{Q}_{J}/\gamma}$ increases the internal
energy and hence the pressure of the fluid, and can therefore modify
$\varepsilon_{p}^{\mathrm{M}}$.

Panel (a) of Fig.~\ref{fig:different_PEJ_diff_power} shows
$\dot{W}_{\mathbf{E}\cdot\mathbf{J}}$ for the values of $\beta_{B}$ listed in Table~\ref{tab:magnetic_field_strength}. In all cases, $\dot{W}_{\mathbf{E}\cdot\mathbf{J}}$ is close to zero initially, increases with $\tau$, reaches a maximum around $\tau \simeq 3.5$--$4.0$~fm, and
decreases at later times. Its magnitude increases systematically with $\beta_{B}$, so a stronger initial magnetic field enhances the electromagnetic energy-transfer rate at all times. The time dependence of $\dot{W}_{\mathbf{E}\cdot\mathbf{J}}$ is governed by the evolution of the electric field and current within the matter region.  
Panel (b) of Fig.~\ref{fig:different_PEJ_diff_power} shows, for
$\beta_{B}=0.50$, the decomposition of $\dot{W}_{\mathbf{E}\cdot\mathbf{J}}$
into the mechanical and Joule-heating contributions according to
Eq.~\eqref{eq:integrated_em_work_decomposition}. At early times
($\tau \lesssim 2.5$~fm), the electromagnetic power is almost entirely dissipated as Joule heat, $\dot{W}_{\mathbf{E}\cdot\mathbf{J}} \simeq
\dot{W}_{\mathcal{Q}_{J}/\gamma}$, because the transverse flow has not yet developed and the mechanical work $\dot{W}_{\mathbf{v}\cdot\mathbf{f}_{\rm EM}}$
remains close to zero. As the flow builds up, the mechanical contribution grows, reaching a maximum around $\tau \simeq 5$~fm, later than the Joule-heating term, which peaks around $\tau \simeq 3.5$~fm and then decreases more rapidly. Consequently, the relative weight of the mechanical work increases with time, and at the latest times shown it becomes comparable to the Joule-heating contribution. Over most of the evolution, however, the energy transferred from the electromagnetic field to the matter is dominated by irreversible Joule heating rather than by work done on the collective flow, consistent with the weak modification of $\varepsilon_{p}^{\mathrm{M}}$
discussed above.

	\section{\label{sec:hic_initial_conditions}Heavy-ion collision initial conditions}

    Ultra-relativistic heavy-ion collisions undergo rapid longitudinal expansion and are approximately boost invariant near mid-rapidity, which makes Milne
coordinates a natural framework for describing their space-time evolution.
Having validated our RRMHD solver in these coordinates, we now apply it to
initial conditions relevant to heavy-ion collisions. We employ a hybrid
initialization: the matter energy-density profile is constructed from a smooth
optical Glauber model~\cite{Miller_2007,shukla2001glaubermodelheavyion}, while
the initial electromagnetic fields are obtained, following
Ref.~\cite{PhysRevC.85.044907}, as the superposition of the fields generated by
individual protons whose positions are sampled from a Monte Carlo Glauber
model. In the present calculation, only the spectator protons are included in
this superposition, since they dominate the field in the overlap region at
early times~\cite{PhysRevC.85.044907}.
	\subsection{Matter initialization}
	\label{subsec:optical_glauber}
	
	For a nucleus with mass number $A$, the nucleon density is parametrized by a Woods--Saxon distribution,
	\begin{equation}
		\rho_A(r)
		=
		\frac{\rho_{0,A}}
		{1+\exp\!\left[(r-R_A)/a_A\right]},
		\label{eq:woods_saxon_density}
	\end{equation}
	where $R_A$ is the nuclear radius, $a_A$ is the surface diffuseness, and $\rho_{0,A}$ is fixed through
		$\int d^3r\,\rho_A(r)=A.$
	The corresponding nuclear thickness function is
		$T_A(x,y)
		=
		\int_{-\infty}^{\infty} dz\,
		\rho_A\!\left(\sqrt{x^2+y^2+z^2}\right),$
	which satisfies
	$\int dx\,dy\,T_A(x,y)=A$.
	For a collision with impact parameter $b$ directed along the $x$ axis, the two nuclear centres are placed at $x=+b/2$ and $x=-b/2$.  We therefore define
	\begin{equation}
		T_A^{(+)}(x,y;b)=T_A(x-b/2,y),
		\label{eq:shifted_tickness_function_A}
	\end{equation}
	\begin{equation}
		T_B^{(-)}(x,y;b)=T_B(x+b/2,y).
		\label{eq:shifted_thickness_function_B}
	\end{equation}
	
	Using the optical Glauber prescription, the participant densities associated with the two nuclei ($A$ and $B$) are
	\begin{align}
		n_{\mathrm{part}}^{A}(x,y;b)
		& =
		T_A^{(+)}
		\left[
		1-
		\left(
		1-\frac{\sigma_{\rm{NN}}^{\mathrm{inel}}T_B^{(-)}}{B}
		\right)^{B}
		\right],
		\label{eq:npart_A}
		\\
		n_{\mathrm{part}}^{B}(x,y;b)
		& =
		T_B^{(-)}
		\left[
		1-
		\left(
		1-\frac{\sigma_{\rm{NN}}^{\mathrm{inel}}T_A^{(+)}}{A}
		\right)^{A}
		\right],
		\label{eq:npart_B}
	\end{align}
	where $\sigma_{\rm{NN}}^{\mathrm{inel}}$ denotes the inelastic nucleon--nucleon cross section.  The total participant density and the binary-collision density are then
		$n_{\mathrm{part}}
		=
		n_{\mathrm{part}}^{A}+n_{\mathrm{part}}^{B}, $
and 
		$n_{\mathrm{coll}}(x,y;b)
		=
		\sigma_{\rm{NN}}^{\mathrm{inel}}
		T_A^{(+)}(x,y;b)
		T_B^{(-)}(x,y;b)$. 
The transverse deposition profile is taken in the conventional
   two-component form
   \begin{equation}
       \mathcal{W}(x,y;b)
       =
       \frac{1-\alpha_H}{2}\,
       n_{\mathrm{part}}(x,y;b)
       +
       \alpha_H\,n_{\mathrm{coll}}(x,y;b),
       \label{eq:two_component_deposition}
   \end{equation}       
	where $\alpha_H$ controls the relative binary-collision contribution. The initial matter-energy density at impact parameter $b$ is
	defined by
	\begin{equation}
		e(\tau_0,x,y;b)
		=
		e_0\,
		\frac{\mathcal{W}(x,y;b)}{\mathcal{W}_{\rm ref}},
		\label{eq:initial_energy_density_glauber}
	\end{equation}
	where $\mathcal{W}_{\rm ref} = \mathcal{W}(0,0;b=0)$ is the maximum
	deposition weight at the center of a head-on collision, and
	$e_0$, a free parameter, denotes the corresponding peak energy density which is adjusted to reproduce experimental multiplicity. A dilute numerical background is introduced ($e_{\rm{bg}} \sim 0.01 \;\mathrm{MeV/fm^{3}}$ ) to avoid exact-vacuum cells in the finite-volume evolution. The fluid is initially at rest in Milne coordinates,
    \begin{figure}[h]
        \centering
        \hspace{-0.6 cm}
        \includegraphics[width=0.9\linewidth]{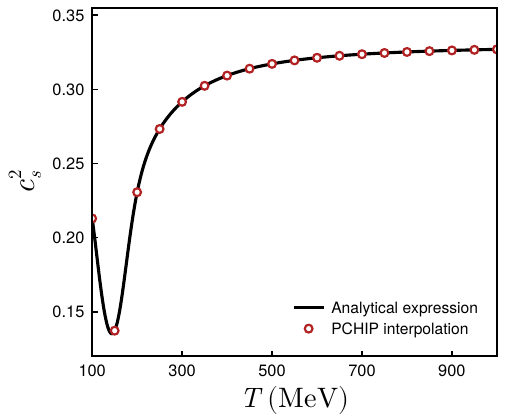}
        \caption{Comparison of $c^{2}_{s}(T)$ with the analytical expression given in \cite{Borsanyi2010} from expression and numerically computed from the interpolating equation $P(e)$.}
        \label{fig:comparison of c_2_s}
    \end{figure}
	\begin{equation}
		u^{\mu}(\tau_0)=(1,0,0,0),
		\qquad
		v^x=v^y=v^{\eta}=0,
		\label{eq:glauber_initial_velocity}
	\end{equation}
	which represents longitudinal Bjorken expansion with no initial transverse flow. 
We employ the lattice-QCD EoS at vanishing baryon chemical potential of
Ref.~\cite{Borsanyi2010}, whose parametrization provides the pressure and
energy density as functions of the temperature $T$. Eliminating $T$ yields
the relation $P(e)$ required by the RRMHD solver. 
For the numerical implementation, $P(e)$ is tabulated and interpolated with a
piecewise cubic Hermite interpolating polynomial (PCHIP), which preserves the monotonicity of $P(e)$ and therefore guarantees a non-negative squared speed of sound,
\begin{equation}
    c_{s}^{2} = \frac{dP}{de},
    \label{eq:cs2_lattice}
\end{equation}
obtained directly from the derivative of the interpolant.
Figure~\ref{fig:comparison of c_2_s} compares $c_{s}^{2}(T)$ obtained in this way
with that computed directly from the parametrization of
Ref.~\cite{Borsanyi2010} over $100 \le T \le 1000$~MeV. The two are in
excellent agreement over the entire range, including the pronounced minimum
of $c_{s}^{2}$ near the crossover temperature $T \simeq 145$~MeV.
    
	\subsection{EM field due to spectator protons}
	\label{subsec:spectator_em_initialization}
	We assumed the initial electromagnetic fields are sourced exclusively by spectator protons. For each incoming nucleus, nucleon positions are sampled from the corresponding Woods--Saxon distribution, and exactly $Z$ nucleons are randomly labelled as protons. Participants are identified using the transverse black-disk criterion: a pair of nucleons undergoes an inelastic collision if
	\begin{equation}
		(x_i-x_j)^2+(y_i-y_j)^2
		<
		\frac{\sigma_{\rm{NN}}^{\mathrm{inel}}}{\pi}.
		\label{eq:black_disk_collision}
	\end{equation}
	A nucleon is classified as a participant if it undergoes at least one such collision; otherwise, it is classified as a spectator. The initial electromagnetic fields are then obtained by summing the contributions from all protons classified as spectators.
	
	The beam Lorentz factor and velocity are given as
	\begin{equation}
		\gamma_{\mathrm{beam}}
		=
		\frac{\sqrt{s_{\rm{NN}}}}{2m_N},
		\qquad
		\beta_{\mathrm{beam}}
		=
		\sqrt{1-\frac{1}{\gamma_{\mathrm{beam}}^{2}}},
		\label{eq:beam_gamma_beta}
	\end{equation}
	where $m_{N}$ is the mass of nucleon.
	For a spectator proton located at $(x_a,y_a)$ and moving with velocity $s_a\beta_{\mathrm{beam}}\hat{\boldsymbol z}$, where $s_a=+1$ for the nucleus moving along $+z$ and $s_a=-1$ for the nucleus moving along $-z$, the transverse electric field at $z=\eta=0$ and $t=\tau_0$ is initialized as
	\begin{align}
		\mathbf{E}_{(a)}^x(\tau_0,x,y)
		& =
		\frac{\alpha_{\mathrm{em}}\gamma_{\mathrm{beam}}(\hbar c)^2}{e_{\rm EM}}
		\bigg[\frac{x-x_a}{\mathcal D_a^{3/2}}\bigg],
		\label{eq:spectator_Ex}
		\\
		\mathbf{E}_{(a)}^y(\tau_0,x,y)
		& =
		\frac{\alpha_{\mathrm{em}}\gamma_{\mathrm{beam}}(\hbar c)^2}{e_{\rm EM}}
		\bigg[ \frac{y-y_a}{\mathcal D_a^{3/2}} \bigg],
		\label{eq:spectator_Ey}
	\end{align}
	where $e_{\rm EM} = \sqrt{4 \pi \alpha_{\rm EM}}$ with
	\begin{equation}
		\mathcal D_a
		=
		(x-x_a)^2+(y-y_a)^2
		+\gamma_{\mathrm{beam}}^2\beta_{\mathrm{beam}}^2\tau_0^2
		+\epsilon_q^2.
		\label{eq:spectator_denominator}
	\end{equation}
	The parameter $\epsilon_q$ regularizes the point-charge field at short transverse distance and it is set to $\epsilon_{q} = 0.30 \; \rm{fm}$.  The corresponding magnetic field follows from the relation $\mathbf{B}_{(a)}= \mathbf{v}_{(a)}\times \mathbf{E}_{(a)}$,
	\begin{equation}
		\mathbf{B}_{(a)}^x=-s_a\beta_{\mathrm{beam}}\mathbf{E}_{(a)}^y,
		\qquad
		\mathbf{B}_{(a)}^y= s_a\beta_{\mathrm{beam}}\mathbf{E}_{(a)}^x.
		\label{eq:spectator_B_components}
	\end{equation}
	The initial fields supplied to the RRMHD solver are obtained by summing over all spectator protons,
	\begin{equation}
		\mathbf{E}^{i}=\sum_{a\in\mathrm{spec.\;p}}\mathbf{E}_{(a)}^i,
		\qquad
		\mathbf{B}^{i}=\sum_{a\in\mathrm{spec.\;p}}\mathbf{B}_{(a)}^i,
		\qquad i=x,y.
		\label{eq:spectator_field_sum}
	\end{equation}

	At the initial time $\tau_{0}$, we retain only the transverse
    components of the spectator-generated electromagnetic fields
    at midrapidity and set
    \begin{equation}
     \mathrm{E}^\eta(\tau_0,x,y)=0,
     \qquad
     \mathrm{B}^\eta(\tau_0,x,y)=0.
    \end{equation}
    The condition $\mathrm{B}^\eta=0$ follows from
    $\mathbf{B}_{(a)} =\mathbf{v}_{(a)}\times\mathbf{E}_{(a)}$ for spectator
protons moving along the beam direction. The choice
$\mathrm{E}^\eta=0$ is an additional transverse-field approximation
adopted in the present $(2+1)$-dimensional calculation. The spectator protons are used only to initialize the
electromagnetic fields at $\tau_0$; no external spectator
source is retained during the subsequent RRMHD evolution. After initialization, the electromagnetic fields evolve self-consistently with the conducting fluid.
	
	To satisfy the reduced $2+1$ dimensional Gauss constraint at the initial time, the local charge variable is initialized as
	\begin{equation}
		q_0(\tau_0,x,y)
		=
		\partial_x \mathrm{E}^x+\partial_y \mathrm{E}^y.
		\label{eq:initial_q0_gauss}
	\end{equation}
	Since the initial fluid velocity vanishes, $\gamma=1$ and $\boldsymbol E\cdot\boldsymbol v=0$, so the laboratory-frame charge variable initially satisfies $q=q_0$ \eqref{eq:q0_from_q}.  
The quantity $q_0$ in Eq.~\eqref{eq:initial_q0_gauss} should not be
interpreted as a physical net-charge distribution of the QGP. Rather, it is the
effective charge density required to satisfy Gauss's law when the
$(3+1)$-dimensional spectator-generated electromagnetic field is represented
within the boost-invariant $(2+1)$-dimensional description used in the present
calculation. In the full spectator-generated field, the longitudinal variation
of the electric field also contributes $\partial_{\eta}E^{\eta}\neq 0$ to
Gauss's law; this contribution is absent once Bjorken symmetry,
$\partial_{\eta}(\cdots)=0$, is imposed. The transverse integral of $q_0$
vanishes identically for each spectator proton, so no net charge is
introduced. 
Each $\mathbf{E}_{(a)}$ is a central field in the transverse plane, and hence
curl-free there, $\partial_x E^y_{(a)}-\partial_y E^x_{(a)}=0$, consistent with
Faraday's law and $B^{\eta}=0$. Since
$\mathbf{B}_{(a)}=s_a\beta_{\rm beam}\hat{\boldsymbol z}\times\mathbf{E}_{(a)}$,
the transverse divergence of $\mathbf{B}_{(a)}$ is proportional to this curl,
so the initial magnetic field is exactly divergence-free. The GLM cleaning
fields are therefore initialized to zero,
\begin{equation}
    \psi(\tau_0,x,y)=0,
    \qquad
    \phi(\tau_0,x,y)=0.
    \label{eq:initial_glm_zero}
\end{equation}

 The fluid is initially at rest, $v^x=v^y=v^\eta=0$, with $P=P(e)$ obtained from
the EoS. The complete primitive state
\begin{align}
    \mathcal{P}
    =
    [
    n,P,&v^x, v^y,v^{\eta}, \notag \\
    &\mathrm{E}^x,\mathrm{E}^y,\mathrm{E}^{\eta},
    \mathrm{B}^x,\mathrm{B}^y,\mathrm{B}^{\eta},
    q_0,\psi,\phi ],
    \label{eq:glauber_primitive_state}
\end{align}
is then passed to the RRMHD solver and evolved up to $\tau_{f}=7$~fm,
comparable to the typical lifetime of the QGP phase in heavy-ion collisions.
The same $\tau_f$ is used for all collision systems to allow a direct
comparison.

\begin{table*}[t]
    \centering
    \caption{Parameters employed for the optical-Glauber initialization and the
    spectator-field calculation.}
    \label{tab:nuclear_Glauber_model_parameters}
    \begin{tabular}{lccccccccccc}
        \toprule
        System &
        $\sqrt{s_{\rm NN}}$ &
        $A$ &
        $Z$ &
        $R_A$ &
        $a_A$ &
        $\sigma_{\rm NN}^{\rm inel}$ &
        $\gamma_{\rm beam}$ &
        $\alpha_H$ &
        $e_0$ & $T_0$ &
        \\
        &
        (GeV) &
        &
        &
        (fm) &
        (fm) &
        (mb) &
        & &
        (GeV/$\rm{fm}^{3}$)
        & (MeV) &  
        \\
        \midrule
        Au+Au & 62.4 & 197 & 79 & 6.38 & 0.535 & 36 & 33.3  & 0.10 & 19.5 & 335 \\
        Au+Au & 200  & 197 & 79 & 6.38 & 0.535 & 42 & 106.6 & 0.14 & 30.0 &  371\\
        Pb+Pb & 2760 & 208 & 82 & 6.62 & 0.546 & 64 & 1471  & 0.08 & 103.0 &  500\\
        \bottomrule
    \end{tabular}
\end{table*}

\subsection{Collision systems and reference parameters}
\label{subsec:glauber_parameter_sets}

The nuclear and collision parameters used for the Au+Au and Pb+Pb systems are
summarized in Table~\ref{tab:nuclear_Glauber_model_parameters}. The parameters
$e_0$ and $\alpha_H$ are adjusted to reproduce the measured charged-particle
multiplicity at midrapidity in central collisions~\cite{PhysRevC.97.054910,ROY2011313, PhysRevC.93.024901,PhysRevC.77.064901}. All simulations
are initialized at $\tau_0= 0.6$~fm. For each collision system, we consider
two impact parameters, $b=7$~fm (mid-central collisions) and $b=12$~fm
(peripheral collisions).

The electrical conductivity does not enter the initialization and is specified
separately as an input to the RRMHD evolution. We take it to be isotropic and
temperature dependent,
\begin{equation}
\sigma(T)=\zeta\, C_{\rm EM}T ,
\end{equation}
where
$
C_{\rm EM}=\sum_{f} e_f^{\,2},
$
$f = u,d,s$ with $e_f$, denoting the electric charge of quark flavor $f$. 
Motivated by the range of estimates obtained by
different lattice-QCD groups~\cite{PhysRevLett.99.022002,PhysRevD.83.034504,
FRANCIS2012212,GUPTA200457}, we consider
$\zeta=1/3$, $0.4$, $1.0$, $3.5$, and $7.0$. The local temperature is obtained from the local rest-frame energy density by inverting the relation $e(T)$ of the lattice-QCD EoS~\cite{Borsanyi2010}, and is used to evaluate $\sigma(T)$ at each time step. For each collision system and impact parameter, all other parameters are kept fixed while $\zeta$ is
varied, thereby isolating the effect of the electrical conductivity.

\subsection{Results}
To compare irreversible Joule heating with the cooling caused by the hydrodynamic expansion, we define the expansion cooling power per unit space-time rapidity,
	\begin{equation}
		\dot{W}_{\rm exp}(\tau)
		=
		\int_{\Omega_{\rm M}(\tau)}
		\tau\,dA\,
		\frac{(e+P)\theta}{\gamma}.
		\label{eq:expansion_cooling_power_hic}
	\end{equation}
    where $\theta=\nabla_\mu u^\mu=\partial_\tau u^\tau+\partial_x u^x+\partial_y u^y+u^\tau/\tau$
is the expansion rate. 
	The relative importance of
	Joule heating is then quantified through
	\begin{equation}
		R_{J/{\rm exp}}(\tau)
		=
		\frac{\dot{W}_{\mathcal{Q}_{J}/\gamma} }
		{\dot{W}_{\rm exp}}.
		\label{eq:joule_to_expansion_ratio_hic}
	\end{equation}
	
	We also follow the transverse average of the electric-field magnitude,
	\begin{equation}
		\left\langle |\mathbf E| \right\rangle_{xy}
		=
		\frac{1}{A_{\perp}}
		\int_{\Omega} dA\,
		\sqrt{
			(\mathrm{E}^x)^2+(\mathrm{E}^y)^2+\tau^2(\mathrm{E}^\eta)^2
		},
		\label{eq:E_average_hic}
	\end{equation}
where $\mathrm{E}^i$ are the components in the Milne coordinate frame and $\Omega$ is
the full computational domain  $\Omega_{\rm M}(\tau)$. We report it through the dimensionless combination $e_{\rm EM}\langle|\mathbf E|\rangle_{xy}/m_\pi^2$.

\begin{figure*}[t]
    \centering

    \begin{minipage}{0.49\textwidth}
        \centering
        \includegraphics[width=\linewidth]
        {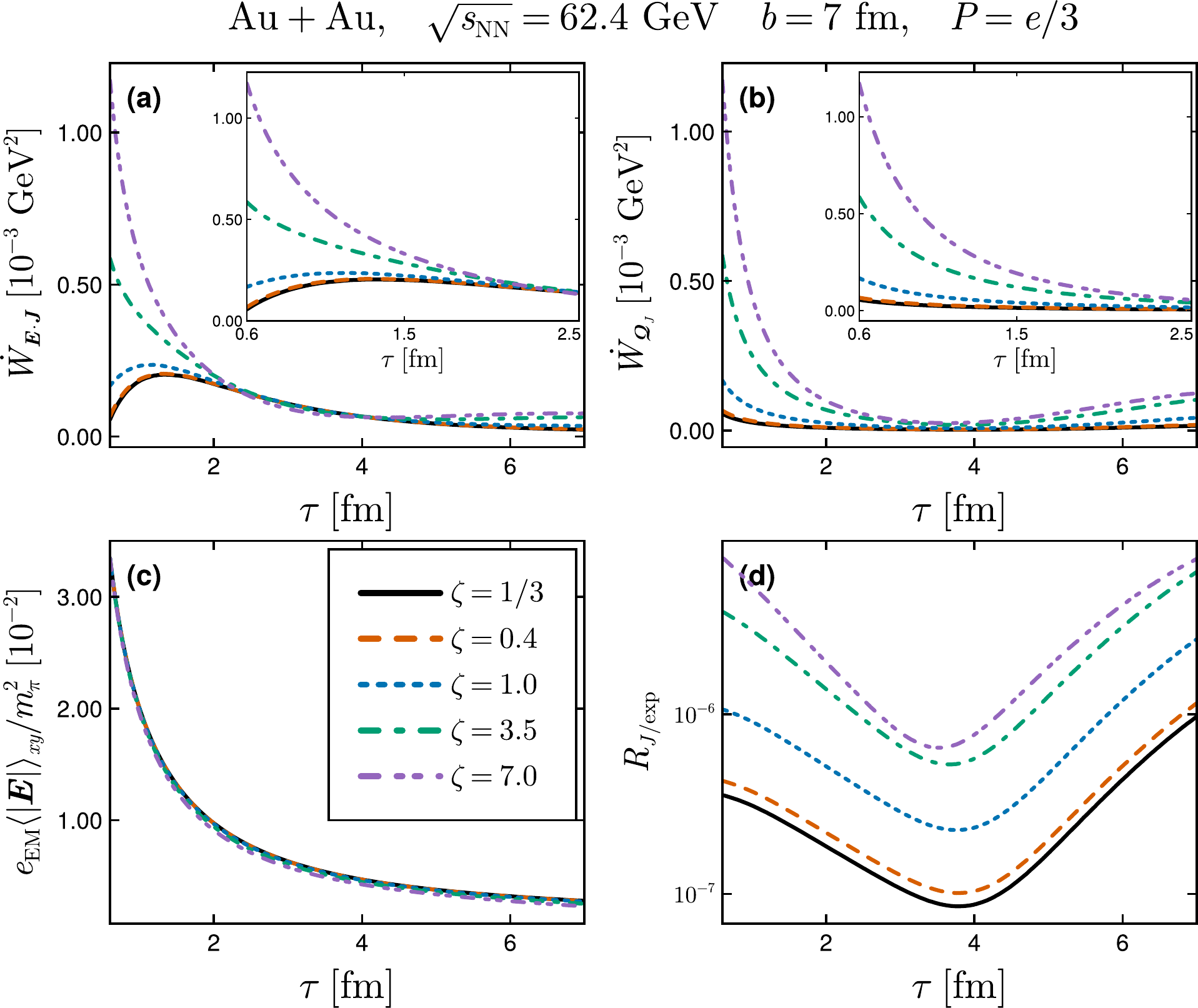}
    \end{minipage}
    \hfill
    \begin{minipage}{0.49\textwidth}
        \centering
        \includegraphics[width=\linewidth]
        {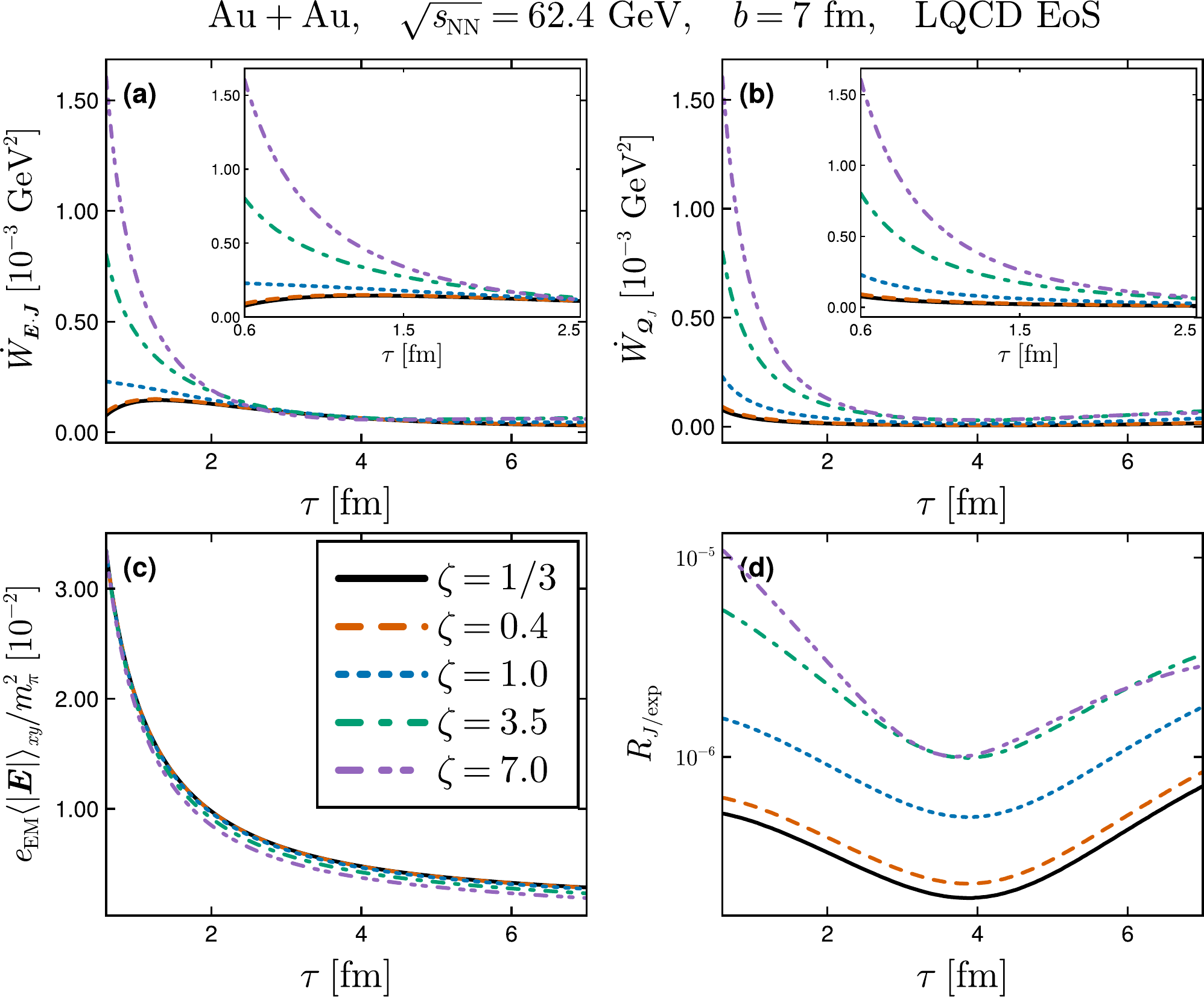}
    \end{minipage}

    \vspace{3mm}

    \begin{minipage}{0.49\textwidth}
        \centering
        \includegraphics[width=\linewidth]
        {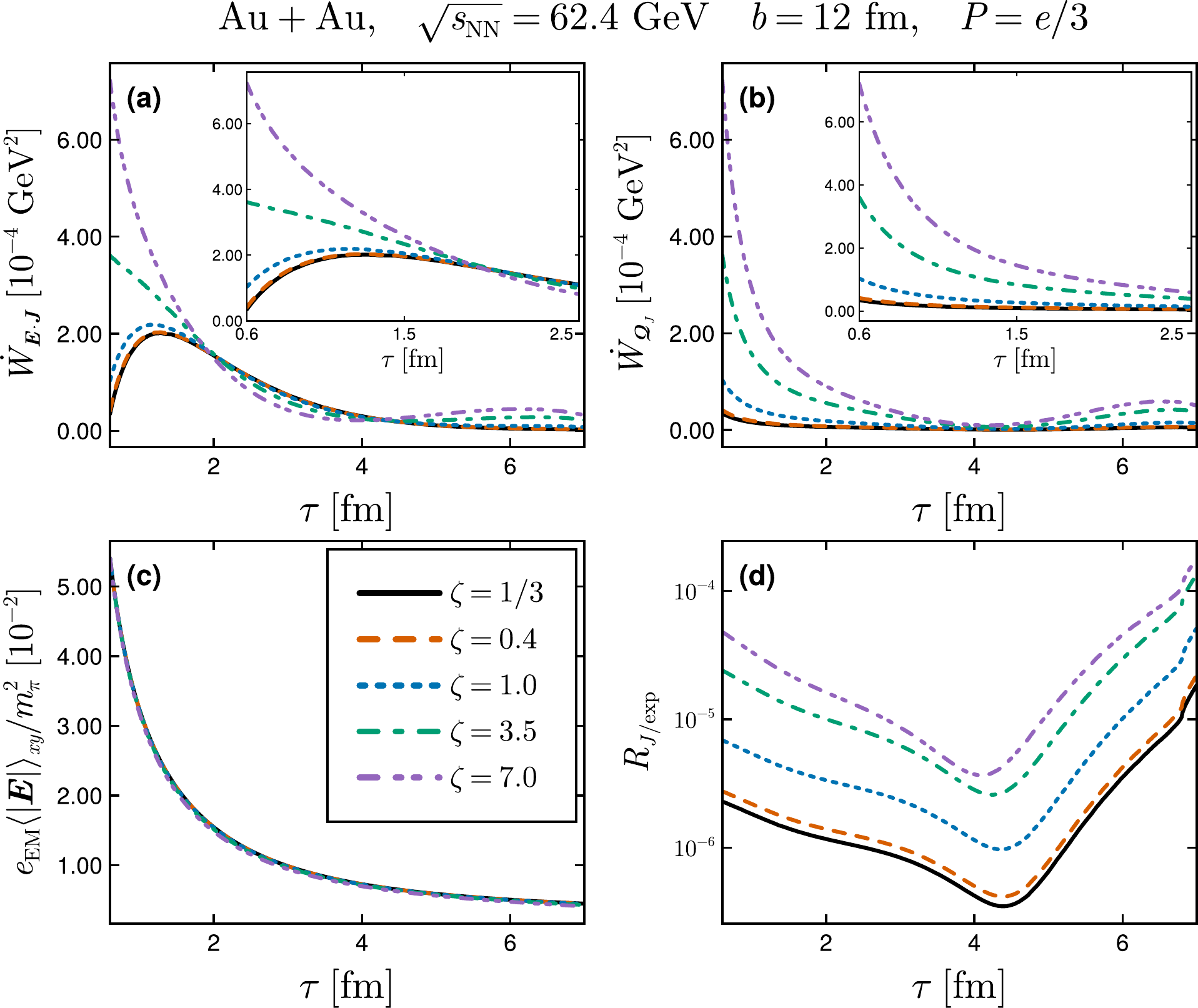}
    \end{minipage}
    \hfill
    \begin{minipage}{0.49\textwidth}
        \centering
        \includegraphics[width=\linewidth]
        {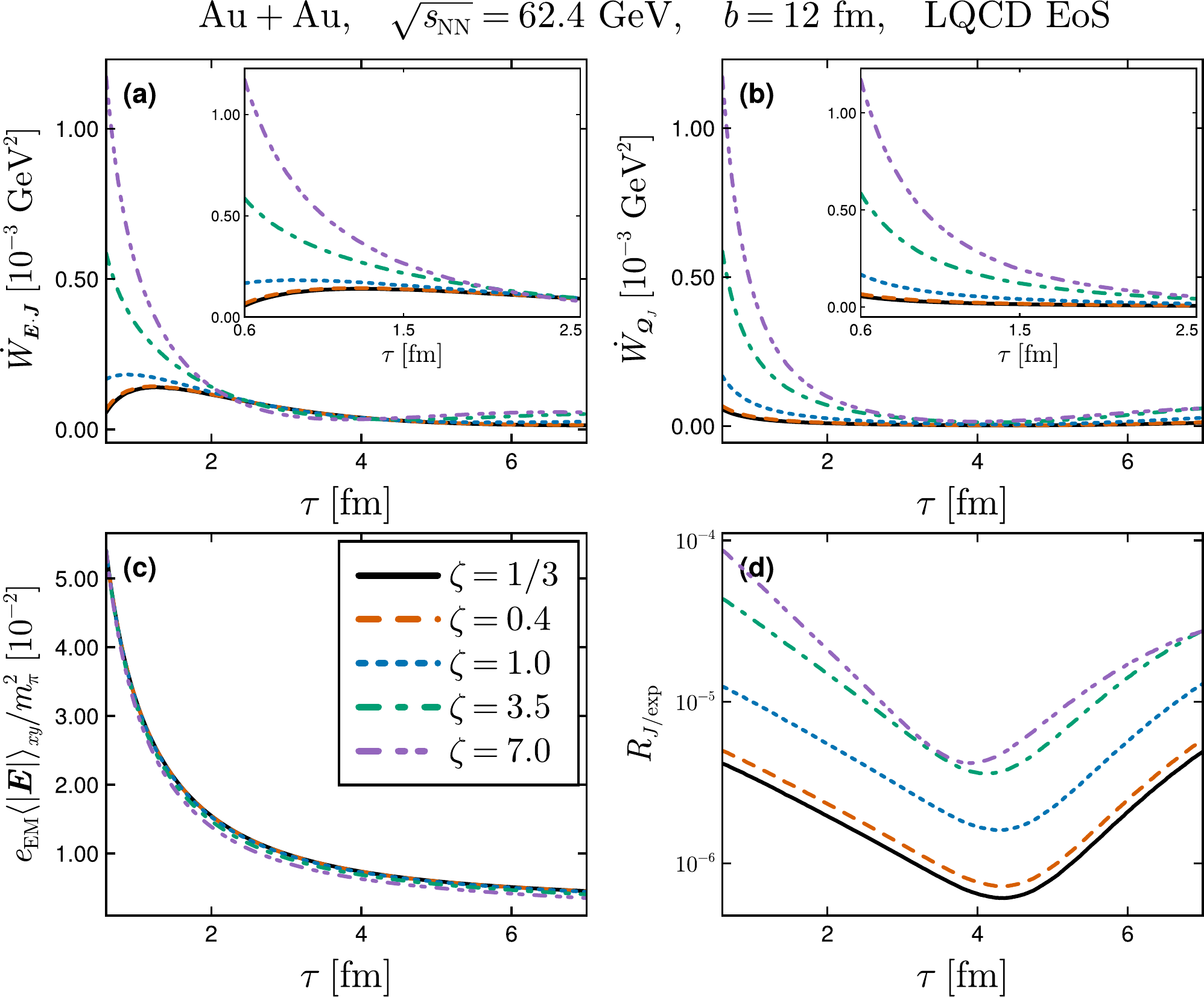}
    \end{minipage}

    \caption{
    The four figures are arranged according to impact parameter and equation of state (EoS). Each figure contains four panels showing: (a) $\dot{W}_{\mathbf{E}\cdot\mathbf{J}}$, (b) $\dot{W}_{\mathcal{Q}_{J}}$, (c) $e_{\mathrm{EM}}\langle |\mathbf{E}| \rangle_{xy}/m_\pi^2$, and (d) $R_{J/\mathrm{exp}}$. The upper row corresponds to $b=7~\mathrm{fm}$, with the ultra-relativistic EoS shown on the left and the lattice-QCD EoS on the right. The lower row shows the corresponding results for $b=12~\mathrm{fm}$, again with the ultra-relativistic EoS on the left and the lattice-QCD EoS on the right.
    }
    \label{fig:Au_Au_low_Ultra_LQCD}
\end{figure*}

We first discuss Au+Au collisions at $\sqrt{s_{\rm NN}}=62.4$~GeV, for which we
also examine the sensitivity to the equation of state by comparing the conformal
EoS, $P=e/3$, with the lattice-QCD EoS. The results for $b=7$ and $12$~fm are
shown in Fig.~\ref{fig:Au_Au_low_Ultra_LQCD}. Several features of this figure are
common to all systems considered here. At $\tau_0$ the fluid has no flow other than
the Bjorken flow, so that $\gamma=1$, $\mathbf{v\cdot f_{\rm EM}}=0$ and
$\dot{W}_{\mathbf{E}\cdot\mathbf{J}}=\dot{W}_{\mathcal{Q}_J/\gamma}$. Since the
initial fields do not depend on $\sigma$, both quantities are then proportional to
$\zeta$. A larger conductivity drains the electric field faster, and the curves for
different $\zeta$ approach each other within $\tau\approx2.5$~fm (insets of panels
(a) and (b)). For small $\zeta$, $\dot{W}_{\mathbf{E}\cdot\mathbf{J}}$ first grows
and reaches a maximum at $\tau\approx1.0$~fm, while $\dot{W}_{\mathcal{Q}_J/\gamma}$
decreases monotonically. The difference,
$\dot{W}_{\mathbf{v}\cdot\mathbf{f}_{\rm EM}}=\dot{W}_{\mathbf{E}\cdot\mathbf{J}}-\dot{W}_{\mathcal{Q}_J/\gamma}>0$ for most of the cases,
is the reversible work done by the Lorentz force on the fluid as the flow develops,
and it dominates the electromagnetic energy transfer after the first fm for all
$\zeta$. The transverse-averaged field $\langle|\mathbf E|\rangle_{xy}$ (panels (c))
depends only weakly on $\zeta$. %

The EoS has little effect on $\langle|\mathbf E|\rangle_{xy}$, which is set mainly
by the initial spectator field. The electromagnetic powers are, however, larger with
the lattice-QCD EoS, by a factor of about $1.2$ ($b=7$~fm) and $1.4$ ($b=12$~fm) at
$\tau_0$ for $\zeta=7$. For the same initial energy density, the lattice-QCD EoS
gives a higher temperature than the conformal EoS, since
$e/T^4$ lies below its Stefan-Boltzmann limit, and therefore a larger $\sigma(T)$.
The EoS also changes the late-time behavior of $R_{J/{\rm exp}}$ (panel (d)). With the conformal
EoS, $R_{J/{\rm exp}}$ rises strongly after its minimum at $\tau\approx4$~fm and is
largest at $\tau_f$, partly because $\dot{W}_{\mathcal{Q}_J/\gamma}$ increases again
at late times for $\zeta\geq3.5$. With the lattice-QCD EoS this late increase is much
weaker, and $R_{J/{\rm exp}}$ is largest at $\tau_0$. In both cases $R_{J/{\rm exp}}$
lies between about $10^{-7}$ and $10^{-5}$ for $b=7$~fm and between about
$4\times10^{-7}$ and $2\times10^{-4}$ for $b=12$~fm.

\begin{figure}[h]
		\hspace{- 0.5 cm}
		\centering
		\includegraphics[width=1.0\linewidth]{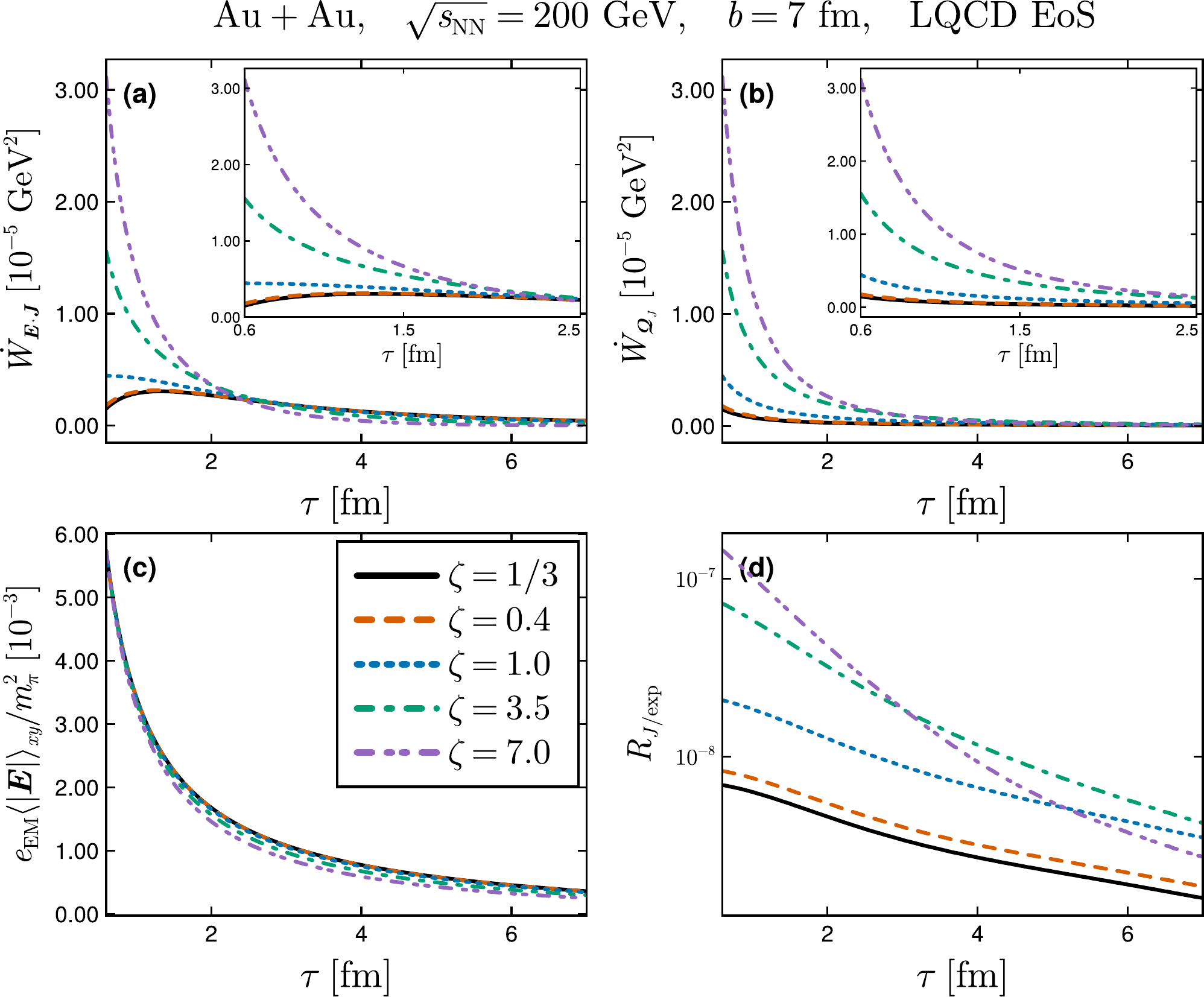}
		\captionsetup{font=small}
		\caption{Proper-time evolution of (a) $\dot{W}_{\mathbf{E}\cdot\mathbf{J}}$, (b) $\dot{W}_{\mathcal{Q}_{J}}$,
			(c) $e_{\mathrm{EM}}\langle|\mathbf{E}|\rangle_{xy}/m_\pi^2$, and
			(d) $R_{J/\mathrm{exp}}$ for Au+Au collisions at $b=7 \; \rm{fm}$.
			Different color curves denote different $\zeta$ parametrization, and panel (d) uses a logarithmic scale.}
		\label{fig:Au_Au_b7}
	\end{figure}

\begin{figure}[h]
		\hspace{- 0.0 cm}
		\centering
		\includegraphics[width=1.0\linewidth]{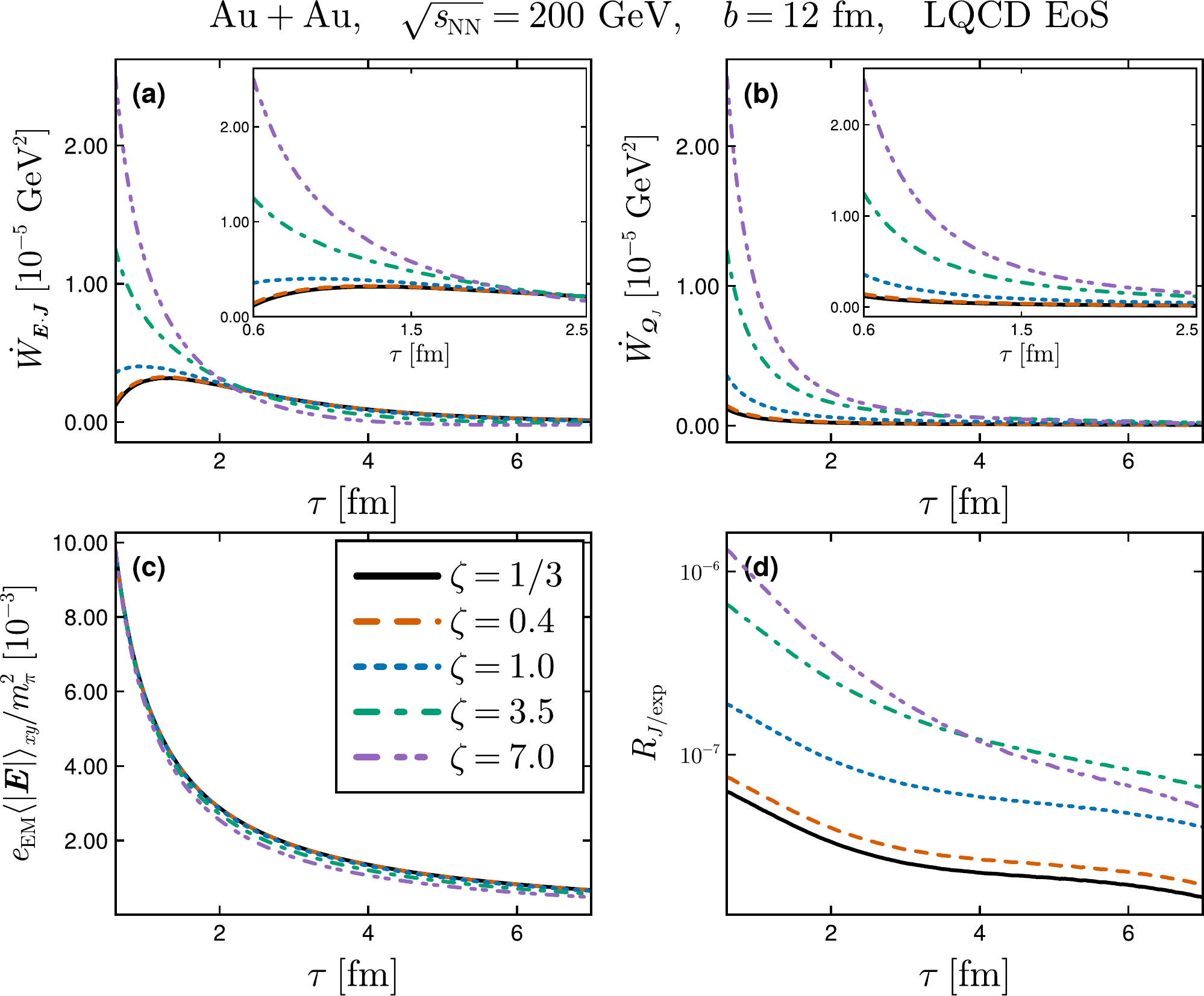}
		\captionsetup{font=small}
		\caption{Proper-time evolution of (a) $\dot{W}_{\mathbf{E}\cdot\mathbf{J}}$, (b) $\dot{W}_{\mathcal{Q}_{J}}$,
			(c) $e_{\mathrm{EM}}\langle|\mathbf{E}|\rangle_{xy}/m_\pi^2$, and
			(d) $R_{J/\mathrm{exp}}$ for Au+Au collisions at $b=12 \; \rm{fm}$.
			Different color curves denote different $\zeta$ parametrization, and panel (d) uses a logarithmic scale.}
		\label{fig:Au_Au_b12}
	\end{figure}

\begin{figure}[h]
		\hspace{-0.6 cm}
		\centering
		\includegraphics[width=1.0\linewidth]{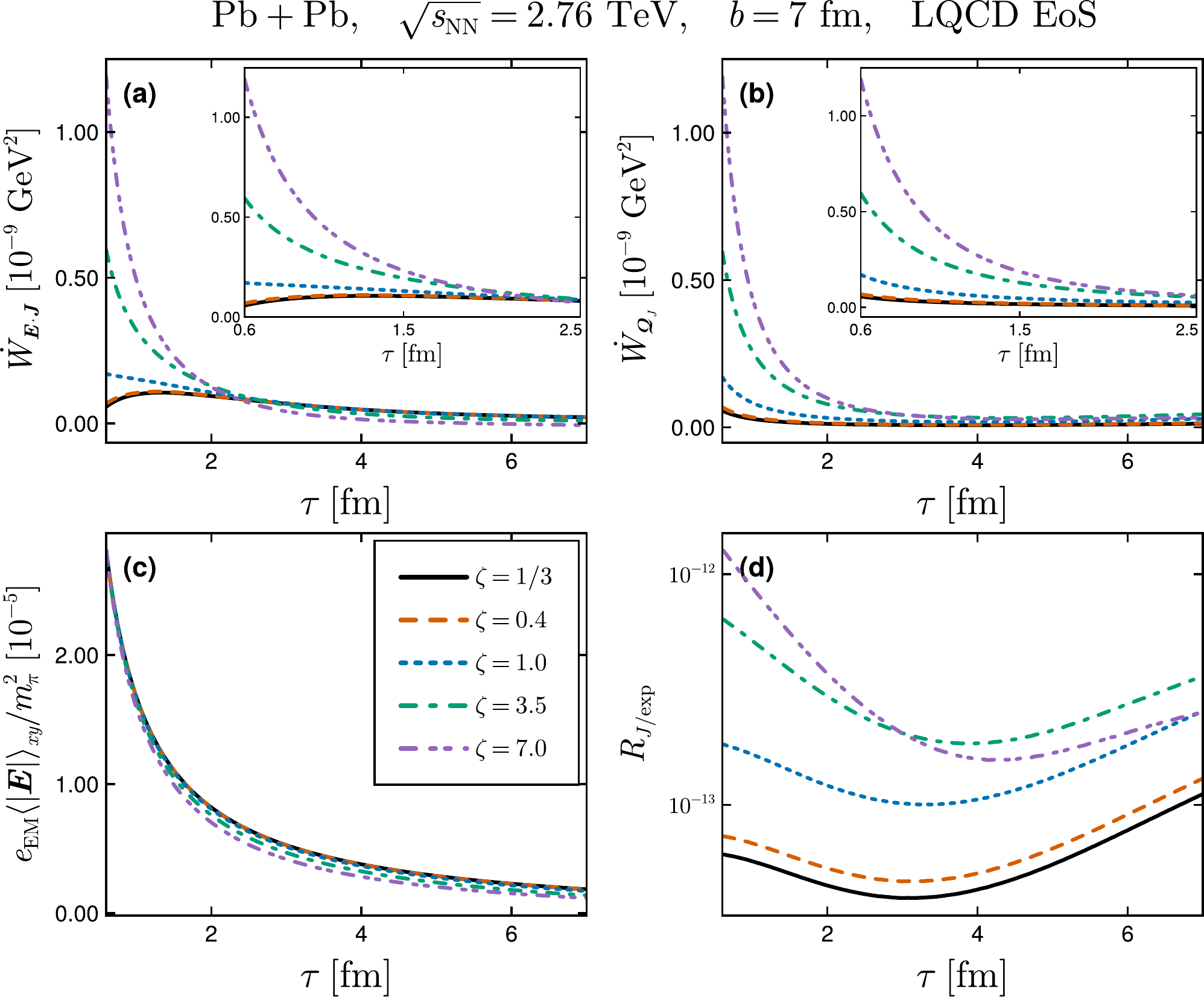}
		\captionsetup{font=small}
		\caption{Proper-time evolution of (a) $\dot{W}_{\mathbf{E}\cdot \mathbf{J}}$, (b) $\dot{W}_{\mathcal{Q}_{J}}$,
			(c) $e_{\mathrm{EM}}\langle|\mathbf{E}|\rangle_{xy}/m_\pi^2$, and
			(d) $R_{J/\mathrm{exp}}$ for Pb+Pb collisions at $b=7 \; \rm{fm}$.
			Different color curves denote different $\zeta$ parametrization, and panel (d) uses a logarithmic scale.}
		\label{fig:Pb_Pb_7}
	\end{figure}

\begin{figure}[h]
		\hspace{-0.0 cm}
		\centering
		\includegraphics[width=1.0\linewidth]{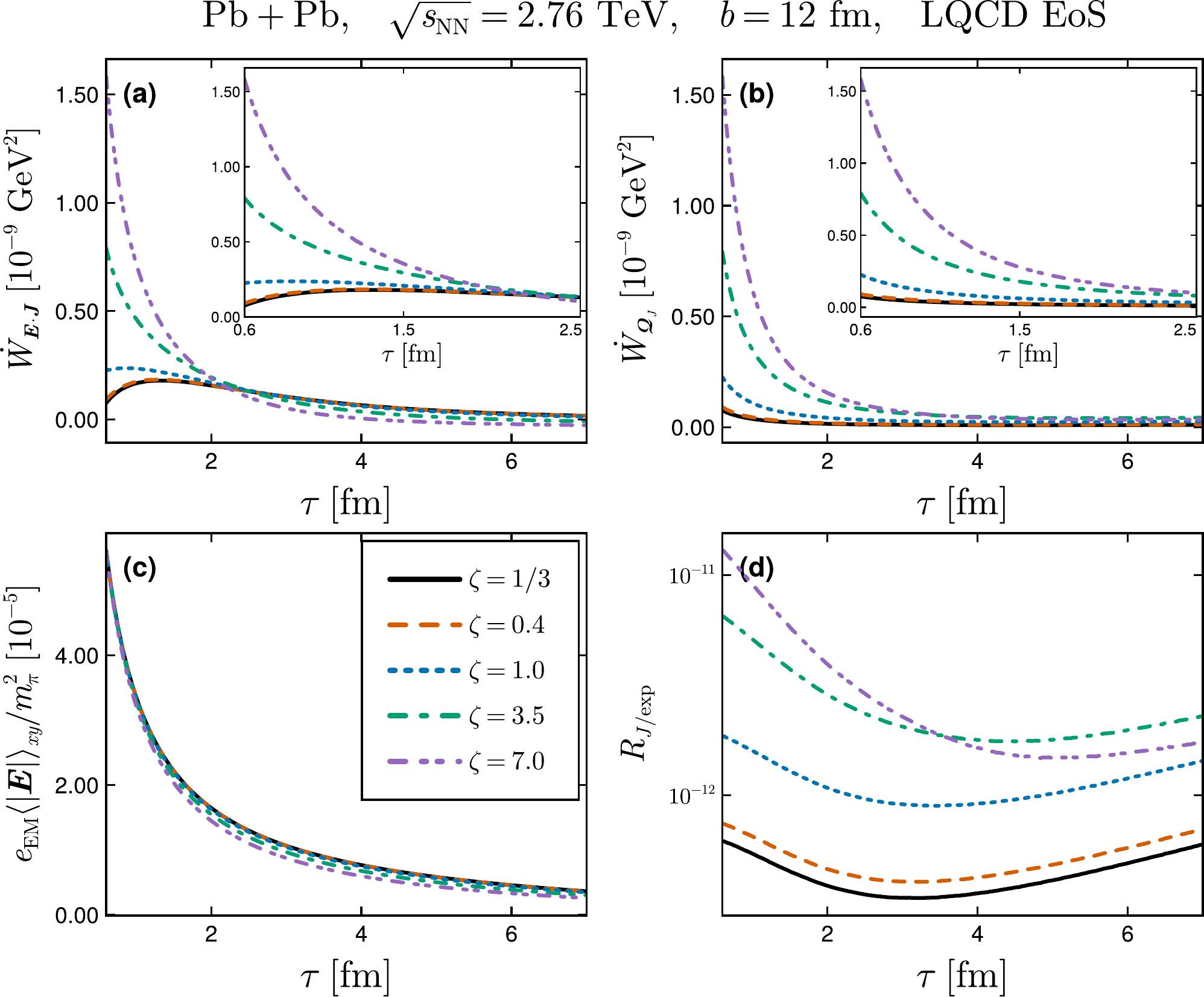}
		\captionsetup{font=small}
		\caption{Proper-time evolution of (a) $\dot{W}_{\mathbf{E}\cdot\mathbf{J}}$, (b) $\dot{W}_{\mathcal{Q}_{J}}$,
			(c) $e_{\mathrm{EM}}\langle|\mathbf{E}|\rangle_{xy}/m_\pi^2$, and
			(d) $R_{J/\mathrm{exp}}$ for Pb+Pb collisions at $b=12 \; \rm{fm}$.
			Different color curves denote different $\zeta$ parametrization, and panel (d) uses a logarithmic scale.}
		\label{fig:Pb_Pb_12}
\end{figure}

Figures~\ref{fig:Au_Au_b7}--\ref{fig:Pb_Pb_12} show the corresponding results with
the lattice-QCD EoS for Au+Au collisions at $\sqrt{s_{\rm NN}}=200$~GeV and Pb+Pb
collisions at $\sqrt{s_{\rm NN}}=2.76$~TeV. The qualitative features are the same as
at $62.4$~GeV, but the magnitudes are much smaller: the electromagnetic powers are of
order $10^{-5}$~GeV$^2$ at $200$~GeV and $10^{-9}$~GeV$^2$ at $2.76$~TeV, compared with
$10^{-3}$~GeV$^2$ at $62.4$~GeV. This strong energy dependence follows from the
initialization. At $\tau_0$ the spectators are already at a distance $v\tau_0$ from
the transverse plane $\eta=0$, and for $\gamma v\tau_0\gg R_A$ the field of each
spectator there behaves as $|\mathbf{E}|\propto 1/(\gamma^2v^3\tau_0^3)$. The field therefore
scales approximately as $\gamma^{-2}$ and the powers, which are quadratic in the
field, as $\gamma^{-4}$, up to the dependence on $\sigma(T)$ and on the size of the
overlap region. Between $200$~GeV and $2.76$~TeV, the ratio of
$\langle|\mathbf E|\rangle_{xy}$ at $\tau_0$ is about $190$, in agreement with
$(\gamma_{2760}/\gamma_{200})^2\approx190$. At $62.4$~GeV, the scaling holds only approximately. Accordingly,
$R_{J/{\rm exp}}$ lies between about $7\times10^{-9}$ and $1.5\times10^{-6}$ at
$200$~GeV and between about $5\times10^{-14}$ and $2\times10^{-11}$ at $2.76$~TeV.
At $62.4$~GeV it is about two orders of magnitude larger than at $200$~GeV and six
to seven orders of magnitude larger than at $2.76$~TeV.


Increasing the impact parameter from $b=7$ to $12$~fm increases
$\langle|\mathbf E|\rangle_{xy}$, because more protons are spectators. The integrated
powers nevertheless decrease at $62.4$ and $200$~GeV, since the temperature is much cooler, which reduces $\sigma(T)$. Only for
Pb+Pb collisions do the powers increase with $b$. %

At $62.4$~GeV and, more weakly, at $2.76$~TeV, $R_{J/{\rm exp}}$ increases at late
times. This arises mainly because the expansion cooling decreases faster than the
Joule power, with an additional contribution from the late-time increase of
$\dot{W}_{\mathcal{Q}_J/\gamma}$ for $\zeta\gtrsim3.5$. At $200$~GeV,
$R_{J/{\rm exp}}$ decreases monotonically up to $\tau_f$. The Joule-heating power is
non-negative at all times, as required by $\mathcal{Q}_J=\sigma e^\mu e_\mu\geq0$,
whereas $\dot{W}_{\mathbf{E}\cdot\mathbf{J}}$ approaches zero and can become slightly
negative at late times for $\zeta\geq3.5$, most visibly for $b=12$~fm. A negative
value means $\dot{W}_{\mathbf{v}\cdot\mathbf{f}_{\rm EM}}<-\dot{W}_{\mathcal{Q}_J/\gamma}$: the
expanding fluid does work against the Lorentz force and returns part of its energy to
the electromagnetic field.

However, the very small values of $R_{J/{\rm exp}}$ at high collision energies
show that irreversible Joule heating is dynamically negligible compared with
expansion cooling, for the spectator-field initialization and the range of
conductivities considered here. At $\sqrt{s_{\rm NN}} = 62.4$~GeV,
$R_{J/{\rm exp}}$ is larger by six to seven orders of magnitude than at
$\sqrt{s_{\rm NN}} = 2.76$~TeV, and by about two orders of magnitude than at
$200$~GeV (Fig.~\ref{fig:Au_Au_low_Ultra_LQCD} compared with
Figs.~\ref{fig:Au_Au_b7}--\ref{fig:Pb_Pb_12}). Even so, it remains at most of
order $10^{-4}$, so Joule heating stays a small correction to the energy balance
and produces no appreciable modification of the bulk thermal evolution. The strong
growth of $R_{J/{\rm exp}}$ with decreasing collision energy nevertheless suggests
that electromagnetic dissipation may become non-negligible at still lower energies,
such as those probed in the RHIC Beam Energy Scan and at FAIR and NICA. A
quantitative study in that regime would, however, require a finite baryon chemical
potential, a treatment beyond boost invariance, and an initialization that accounts
for the finite passage time of the nuclei, which is no longer small compared with
$\tau_0$.

These results show that, for the collision energies considered here, Joule
heating produces no appreciable modification of the bulk thermal evolution. The
present calculation therefore primarily establishes the conductivity dependence
of the electromagnetic energy transfer within the RRMHD framework. We have also
examined the conductivity dependence of the matter and total momentum
anisotropies, $\varepsilon_p^{M}$ and $\varepsilon_p^{\rm tot}$, for the Au+Au and
Pb+Pb configurations (not shown). Within the conductivity range considered,
neither quantity varies appreciably with $\sigma$. This is in contrast with the
controlled elliptic-fireball calculation of Sec.~\ref{sec:benchmarks_results},
where increasing the initial magnetic-field strength modifies
$\varepsilon_p^{\rm tot}$, mainly through the direct electromagnetic-stress
contribution, while changing $\varepsilon_p^{M}$ only slightly. The difference
arises because, for the spectator-field initial conditions employed here, the
electromagnetic stress is small compared with the fluid pressure. Changing the
conductivity then affects the relaxation of the electric field and the
electromagnetic energy transfer, but is not sufficient to produce a visible
modification of the momentum anisotropies.

\section{\label{sum_and_con}Summary and conclusions}

We have developed \mshiva~1.0, an IMEX finite-volume solver for relativistic
resistive magnetohydrodynamics in Cartesian and boost-invariant Milne
coordinates. The coupled fluid and electromagnetic fields are evolved in
conservative form, with generalized Lagrange-multiplier fields enforcing the
Maxwell constraints. The conductivity-induced relaxation of the electric field is
treated through a local implicit update, and primitive variables are recovered
using a Lorentz-boosted enthalpy inversion after the explicit step, together with
a complementary pressure-based inversion after the implicit update.

In Cartesian coordinates, the code reproduces the characteristic structures of
the Brio--Wu shock tube, the magnetized cylindrical blast wave and the
relativistic rotor. High-conductivity blast-wave calculations further demonstrate
the robustness of the post-implicit pressure recovery: the standard magnetized
blast remains stable up to $\sigma = 2000$, while a weaker-field configuration is
evolved at $\sigma = 10^{6}$ (in code units) without primitive-recovery failure.
In Milne coordinates, the implementation reproduces the expected transverse
dynamics of an expanding rotor.

We then considered initial conditions relevant for heavy-ion collisions. In an
elliptic Gaussian fireball with a conformal EoS and constant conductivity, the
magnetic field yields a visible electromagnetic contribution to the total momentum
anisotropy, while the corresponding change in the matter momentum anisotropy is
comparatively small. We decomposed the electromagnetic work into mechanical
Lorentz-force work and non-negative Joule heating,
$\dot{W}_{\mathbf{E}\cdot\mathbf{J}}=\dot{W}_{\mathbf{v}\cdot\mathbf{f}_{\rm EM}}+\dot{W}_{\mathcal{Q}_J/\gamma}$,
and evaluated these contributions for different initial magnetic-field strengths.
For Au+Au collisions at $\sqrt{s_{\rm NN}} = 62.4$ and $200$~GeV and Pb+Pb
collisions at $\sqrt{s_{\rm NN}} = 2.76$~TeV, initialized with optical-Glauber
matter profiles and spectator-proton electromagnetic fields, and evolved with the
lattice-QCD EoS and $\sigma(T)=\zeta C_{\rm EM}T$, increasing the conductivity
enhances the early Ohmic contribution but also accelerates the relaxation of the
electric field. In this setup, the ratio of Joule heating to expansion cooling
lies approximately in the range $10^{-14}$--$10^{-11}$ for Pb+Pb at $2.76$~TeV,
$10^{-8}$--$10^{-6}$ for Au+Au at $200$~GeV and $10^{-7}$--$10^{-4}$ for Au+Au at
$62.4$~GeV. Since the spectators have already receded from the transverse plane at
$\tau_0$, the initial field scales approximately as $\gamma^{-2}$ and the
electromagnetic powers roughly as $\gamma^{-4}$. The relative importance of Joule
heating therefore grows by about two orders of magnitude from $200$ to $62.4$~GeV,
and by six to seven orders of magnitude from $2.76$~TeV to $62.4$~GeV, but remains
far below unity in all cases. At $62.4$~GeV, replacing the lattice-QCD EoS by a
conformal EoS leaves the field evolution nearly unchanged, but reduces the
electromagnetic powers and strengthens the late-time increase of the ratio.
Consequently, irreversible Joule heating does not appreciably modify the bulk
thermal evolution for the initialization and conductivity range considered here,
and neither the matter nor the total momentum anisotropy shows an appreciable
dependence on $\sigma$. The late-time increase of the ratio, seen at $62.4$~GeV
and, more weakly, at $2.76$~TeV, arises primarily because the expansion-cooling
term decreases faster than the integrated Joule power, with a smaller contribution
from a late-time increase of the Joule power at the largest conductivities. The
present heavy-ion calculation therefore primarily establishes the conductivity
dependence of the electromagnetic energy transfer within the RRMHD framework,
together with its strong growth toward lower collision energies.

The present study is limited to (2+1)-dimensional boost-invariant evolution of an inviscid resistive fluid at vanishing baryon chemical potential, with smooth optical-Glauber matter profiles combined with event-by-event spectator fields, a temperature-dependent electrical conductivity, and spectator fields applied only at the initial time.
Natural extensions include rapidity-dependent (3+1)-dimensional evolution, shear and bulk viscosity, finite baryon density, continuously evolving spectator and participant currents, event-by-event fluctuating matter initial conditions, consistent with the fluctuating spectator fields used here, together with event-by-event evolution and ensemble averaging, freeze-out, and charge-dependent experimental observables. In particular, finite
baryon density, evolution beyond boost invariance and a time-dependent spectator source are required to address the lower collision energies of the RHIC Beam Energy
Scan, FAIR and NICA, where our results suggest that Joule heating becomes increasingly important.
	
	\begin{acknowledgments}
V.R. acknowledges support from the ANRF Core Research Grant (grant no.\ \texttt{CRG/2023/001309}). We would also like to thank Reghukrishnan Gangadharan for the valuable discussions.  K.S, and V.R also acknowledges Department of Atomic Energy (DAE), India for financial support.
    \end{acknowledgments}

	\newpage
	\appendix
	
	\clearpage
	\onecolumngrid
	
	\section{Ideal MHD limit}
	\label{app:Ideal_MHD_Limit}
	
	Here, we show how the usual ideal-MHD condition follows from the covariant form of Ohm's law. In Cartesian coordinates, and using the metric convention adopted in this work, the resistive Ohm law is written as
	\begin{equation}
		J^\mu
		=
		\sigma F^{\mu\nu}u_\nu
		+
		q_0 u^\mu ,
		\label{eq:covariant_ohm_law}
	\end{equation}
	where \(\sigma\) is the electrical conductivity, \(q_0\) is the charge density measured in the local rest frame of the fluid, and \(u^\mu=\gamma(1,\mathbf{v})\) is the fluid four-velocity.
	
	The laboratory-frame charge density is defined as
	\begin{equation}
		q := J^0 .
	\end{equation}
	Taking the \(\mu=0\) component of Eq.~\eqref{eq:covariant_ohm_law}, we obtain
	\begin{equation}
		q
		=
		\sigma F^{0i}u_i
		+
		q_0 \gamma .
	\end{equation}
	With the electromagnetic-field convention used in this work, this gives
	\begin{equation}
		q
		=
		\sigma \gamma \left(\mathbf{E}\cdot\mathbf{v}\right)
		+
		q_0\gamma .
	\end{equation}
	Therefore,
	\begin{equation}
		q_0
		=
		\frac{q}{\gamma}
		-
		\sigma\left(\mathbf{E}\cdot\mathbf{v}\right).
		\label{eq:q0_from_q}
	\end{equation}
	
	Next, the spatial part of Eq.~\eqref{eq:covariant_ohm_law} gives
	\begin{equation}
		J^i
		=
		\sigma F^{i\nu}u_\nu
		+
		q_0 u^i .
	\end{equation}
	In three-vector notation this becomes
	\begin{equation}
		\mathbf{J}
		=
		\sigma \gamma
		\left[
		\mathbf{E}
		+
		\mathbf{v}\times\mathbf{B}
		\right]
		+
		q_0\gamma\mathbf{v}.
	\end{equation}
	Substituting Eq.~\eqref{eq:q0_from_q} into the above expression gives
	\begin{equation}
		\mathbf{J}
		=
		\sigma\gamma
		\left[
		\mathbf{E}
		+
		\mathbf{v}\times\mathbf{B}
		-
		\left(\mathbf{E}\cdot\mathbf{v}\right)\mathbf{v}
		\right]
		+
		q\mathbf{v}.
		\label{eq:spatial_current_ohm}
	\end{equation}
	This is the form of Ohm's law used in the numerical implementation. The first term on the right-hand side is the conduction current, while the second term is the advective charge current.
	
	If we substitute,
	\begin{equation}
		\mathbf{v}=0,
		\implies 
		\gamma=1,
	\end{equation}
	Equation~\eqref{eq:spatial_current_ohm} then reduces to
	\begin{equation}
		\mathbf{J}
		=
		\sigma \mathbf{E},
	\end{equation}
	which is the usual Ohm law if the fluid is rest in the lab frame.
	
	We now consider the ideal-MHD limit i.e. $R_{m} >> 1$ that is achieved by tending $\sigma \rightarrow \infty$. Then
	the conduction current remains finite only if
	\begin{equation}
		\mathbf{E}
		+
		\mathbf{v}\times\mathbf{B}
		-
		\left(\mathbf{E}\cdot\mathbf{v}\right)\mathbf{v}
		=
		0 .
		\label{eq:ideal_limit_intermediate}
	\end{equation}
	
	To show that Eq.~\eqref{eq:ideal_limit_intermediate} gives the usual ideal-MHD condition, take the scalar product with \(\mathbf{v}\). Since
	\begin{equation}
		\mathbf{v}\cdot(\mathbf{v}\times\mathbf{B})=0,
	\end{equation}
	we obtain
	\begin{equation}
		\mathbf{E}\cdot\mathbf{v}
		-
		\left(\mathbf{E}\cdot\mathbf{v}\right)v^2
		=
		0 .
	\end{equation}
	Therefore,
	\begin{equation}
		\left(\mathbf{E}\cdot\mathbf{v}\right)
		\left(1-v^2\right)
		=
		0 .
	\end{equation}
	For a physical subluminal velocity, \(v^2<1\), this implies
	\begin{equation}
		\mathbf{E}\cdot\mathbf{v}=0 .
		\label{eq:E_dot_v_zero}
	\end{equation}
	Substituting Eq.~\eqref{eq:E_dot_v_zero} back into Eq.~\eqref{eq:ideal_limit_intermediate}, we finally obtain
	\begin{equation}
		\mathbf{E}
		+
		\mathbf{v}\times\mathbf{B}
		=
		0 .
	\end{equation}
	Thus, in the ideal-MHD limit,
	\begin{equation}
		\boxed{
			\mathbf{E}
			=
			-
			\mathbf{v}\times\mathbf{B}
		} .
		\label{eq:ideal_mhd_condition}
	\end{equation}
	
	Equivalently, decomposing the electric field into components parallel and perpendicular to the fluid velocity,
	\begin{equation}
		\mathbf{E}
		=
		\mathbf{E}_{\parallel}
		+
		\mathbf{E}_{\perp},
	\end{equation}
	one finds from Eq.~\eqref{eq:E_dot_v_zero} that
	\begin{equation}
		\mathbf{E}_{\parallel}=0 .
	\end{equation}
	The remaining perpendicular part satisfies
	\begin{equation}
		\mathbf{E}_{\perp}
		+
		\mathbf{v}\times\mathbf{B}
		=
		0 .
	\end{equation}
	Since the parallel component vanishes, this is identical to Eq.~\eqref{eq:ideal_mhd_condition}. Therefore, the high-conductivity limit of the resistive relativistic Ohm law consistently reproduces the ideal-MHD constraint.
	\section{Metric Choice}
    \label{app:metric choice}
We perform the $3+1$ decomposition of spacetime with respect to spacelike
hypersurfaces of constant time. The line element can be written as
\begin{equation}
    ds^{2}
    =
    (dx^{0})^{2}
    -
    \gamma_{ij}\,dx^{i}dx^{j},
    \label{eq:metric_3p1}
\end{equation}
where $\gamma_{ij}$ denotes the positive-definite spatial metric
and Latin indices $i,j=1,2,3$ label the spatial coordinates. For Cartesian coordinates,
\begin{equation}
    ds^{2} = dt^{2} - dx^{2} - dy^{2} - dz^{2}
\end{equation}
with
$\sqrt{-g}=1$.
For the Milne
coordinates,
\begin{equation}
    ds^{2} = d\tau^{2} - dx^{2} - dy^{2} - \tau^{2}d\eta^{2},
    \label{eq:milne_metric}
\end{equation}
with $\sqrt{-g}=\tau$.
These two spatial metric has no spatial-coordinate dependence,
$\partial_{j}\gamma_{ik}=0$.
Now we consider Eq~\ref{Modified_Maxwell_GLM_with_source} and Eq~\ref{Modified_Maxwell_GLM_without_source} we can write them as 
\begin{align}
    \partial_{\mu}\bigg[ \sqrt{-g} (F^{\mu \nu} + g^{\mu \nu} \psi) \bigg] & = \psi \partial_{\mu}(\sqrt{-g}g^{\mu \nu}) + \sqrt{-g}\bigg[ J^{\nu} - \kappa n^{\nu} \psi \bigg], \\
    \partial_{\mu}\bigg[ \sqrt{-g} (\tilde{F}^{\mu \nu} + g^{\mu \nu} \phi) \bigg] & = \phi \partial_{\mu}(\sqrt{-g}g^{\mu \nu})  - \sqrt{-g}\kappa n^{\nu} \phi , \\
\end{align}
respectively. Further we can write
\begin{equation}
    \partial_{\mu}\bigg[ \sqrt{-g} (F^{\mu \nu} + g^{\mu \nu} \psi) \bigg]  = \sqrt{-g} \bigg[ J^{\nu} - \kappa n^{\nu} \psi -\psi g^{\alpha \beta} \Gamma^{\nu}_{\alpha \beta}\bigg] 
    \label{eq:Final_modified_Maxwell_with_source}
\end{equation}   
\begin{equation}
    \partial_{\mu}\bigg[ \sqrt{-g} (\tilde{F}^{\mu \nu} + g^{\mu \nu} \phi) \bigg]  = \sqrt{-g}\bigg[ -\kappa n^{\nu} - \phi g^{\alpha \beta} \Gamma^{\nu}_{\alpha \beta}\bigg].
    \label{eq:Final_modified_Maxwell_without_source}
\end{equation}
For Cartesian coordinate $\partial_{\mu}(\sqrt{-g}g^{\mu \nu})$ term is zero and for Milne coordinate it will be 
\begin{equation}
    \partial_{\mu}(\sqrt{-g}g^{\mu 0}) = 1, \qquad \partial_{\mu}(\sqrt{-g} g^{\mu i}) = 0.
    \label{eq:Milne GLM extra terms}
\end{equation}
We have used the results of Eq~\ref{eq:Final_modified_Maxwell_with_source},\ref{eq:Final_modified_Maxwell_without_source} and ~\ref{eq:Milne GLM extra terms} and wrote the equation Eq~\ref{eq:Evolving_equations}.
	\section{Pressure-based primitive recovery}
	\label{app:pressure_based_recovery}
	
	An alternative way to recover the primitive variables is to use the thermodynamic pressure itself as the root-finding variable. This method is useful because, once the pressure is known, the auxiliary variable \(Z\), the velocity, the Lorentz factor, and the thermodynamic variables can be reconstructed directly.
	
	The conserved energy density is
	\begin{equation}
		\varepsilon = Z - P + \varepsilon_{\rm em},
	\end{equation}
	where
	\begin{equation}
		\varepsilon_{\rm em}
		=
		\frac{1}{2}
		\left(
		\mathbf{E}^2+\mathbf{B}^2
		\right).
	\end{equation}
	Therefore, for a given trial pressure \(P\), the auxiliary variable is
	\begin{equation}
		Z(P) = \varepsilon - \varepsilon_{\rm em} + P .
	\end{equation}
	The velocity magnitude is then
	\begin{equation}
		v^2(P) = \frac{Q^2}{Z(P)^2}.
	\end{equation}
	
	The rest-frame number density follows from
	\begin{equation}
		N_c = \gamma n,
	\end{equation}
	so that
	\begin{equation}
		n(P)
		=
		\frac{N_c}{\gamma(P)}
		=
		N_c\sqrt{1-v^2(P)} .
	\end{equation}
	The matter energy density is obtained from
	\begin{equation}
		Z = (e+P)\gamma^2 ,
	\end{equation}
	which gives
	\begin{equation}
		e(P)
		=
		\frac{Z(P)}{\gamma(P)^2}
		-
		P .
	\end{equation}
	Using \(1/\gamma^2=1-v^2\), this can be written as
	\begin{equation}
		e(P)
		=
		Z(P)\left[1-v^2(P)\right] - P .
	\end{equation}
	Since \(v^2(P)=Q^2/Z(P)^2\), we obtain
	\begin{equation}
		e(P)
		=
		Z(P)
		-
		\frac{Q^2}{Z(P)}
		-
		P .
	\end{equation}
	Using \(Z(P)=\varepsilon-\varepsilon_{\rm em}+P\), the above expression reduces to
	\begin{equation}
		e(P)
		=
		\varepsilon-\varepsilon_{\rm em}
		-
		\frac{Q^2}{Z(P)} .
	\end{equation}
	
	The pressure is then determined by solving the scalar equation
	\begin{equation}
		f(P)
		=
		P_{\rm EoS}\left(e(P),n(P)\right)
		-
		P
		=
		0 .
	\end{equation}
	For the ideal-gas equation of state,
	\begin{equation}
		P_{\rm EoS}(e,n)
		=
		(\Gamma-1)(e-n),
	\end{equation}
	the function becomes
	\begin{equation}
		f(P)
		=
		(\Gamma-1)\left[e(P)-n(P)\right]
		-
		P .
	\end{equation}
	To solve this equation using Newton's method, we need \(df/dP\):
	\begin{equation}
		\frac{df}{dP}
		=
		(\Gamma-1)
		\left(
		\frac{de}{dP}
		-
		\frac{dn}{dP}
		\right)
		-
		1 .
	\end{equation}
	
	The derivative of \(e(P)\) is obtained from
	\begin{equation}
		e(P)
		=
		\varepsilon-\varepsilon_{\rm em}
		-
		\frac{Q^2}{Z(P)} ,
		\implies
		\frac{de}{dP}
		=
		\frac{Q^2}{Z(P)^2}
		=
		v^2(P).
	\end{equation}
	
	Next, from
	\begin{equation}
		n(P)
		=
		N_c\sqrt{1-v^2(P)} , 
		\implies
		\frac{dn}{dP}
		=
		\frac{N_c\gamma(P)v^2(P)}{Z(P)} .
	\end{equation}
	Combining these results, the derivative required for the Newton iteration is
	\begin{equation}
		\frac{df}{dP}
		=
		(\Gamma-1)
		\left[
		v^2(P)
		-
		\frac{N_c\gamma(P)v^2(P)}{Z(P)}
		\right]
		-
		1 .
	\end{equation}
	The pressure is then updated according to
	\begin{equation}
		P_{k+1}
		=
		P_k
		-
		\frac{f(P_k)}
		{(df/dP)|_{P_k}} .
	\end{equation}
	
	Once the root \(P=P_0\) is obtained, all primitive variables are reconstructed as
	\begin{equation}
		Z_0
		=
		\varepsilon-\varepsilon_{\rm em}+P_0,
	\end{equation}
	\begin{equation}
		v^j
		=
		\frac{Q^j}{Z_0},
		\qquad
		v^2
		=
		\frac{Q^2}{Z_0^2},
		\qquad
		\gamma
		=
		\frac{1}{\sqrt{1-v^2}},
	\end{equation}
	and
	\begin{equation}
		n
		=
		\frac{N_c}{\gamma},
		\qquad
		e
		=
		\varepsilon-\varepsilon_{\rm em}
		-
		\frac{Q^2}{Z_0}.
	\end{equation}
	The recovered primitive state is therefore
	\begin{equation}
		\left(
		n,\,
		P,\,
		v^j,\,
		\mathrm{E}^j,\,
		\mathrm{B}^j,\,
		q,\,
		\psi,\,
		\phi
		\right).
	\end{equation}
	
	In the numerical implementation, the admissible solution must satisfy
	\begin{equation}
		Z(P) > |Q|,
		\qquad
		0 \leq v^2(P) < 1,
		\qquad
		P>0 .
	\end{equation}
	These conditions ensure a real Lorentz factor and a physical thermodynamic state.

	\section{IMEX Runge--Kutta coefficients}
	\label{app:imex_coefficients}
	The Butcher tableaux of the explicit and implicit Runge--Kutta
	schemes used in the IMEX update are listed below. 
    We follow the notation $\mathrm{SSP}k(s,\alpha,p)$, where
$k$ denotes the order of the SSP explicit scheme, $s$ and
$\alpha$ are the numbers of implicit and explicit stages,
respectively, and $p$ is the overall order of the IMEX scheme. In the notation of
	Eqs.~\eqref{eq:imex_stage}--\eqref{eq:imex_final}, the coefficients
	\(\tilde{a}_{ij}\), \(\tilde{\omega}_i\) define the explicit update of
	the non-stiff terms, while \(a_{ij}\), \(\omega_i\) define the
	diagonally implicit update of the stiff source terms.
	
	For the second-order SSP2(3,3,2) IMEX scheme, the coefficients are
	\begin{equation}
		\tilde{A}
		=
		\begin{bmatrix}
			0 & 0 & 0 \\
			1/2 & 0 & 0 \\
			1/2 & 1/2 & 0
		\end{bmatrix},
		\qquad
		A
		=
		\begin{bmatrix}
			1/4 & 0 & 0 \\
			0 & 1/4 & 0 \\
			1/3 & 1/3 & 1/3
		\end{bmatrix}.
	\end{equation}
	The corresponding weights are
	\begin{equation}
		\tilde{\omega}
		=
		\left[
		\frac{1}{3},\frac{1}{3},\frac{1}{3}
		\right],
		\qquad
		\omega
		=
		\left[
		\frac{1}{3},\frac{1}{3},\frac{1}{3}
		\right].
	\end{equation}
	
	Equivalently, the double Butcher tableau is
	\begin{equation}
		\begin{array}{c|ccc}
			0 & 0 & 0 & 0 \\
			1/2 & 1/2 & 0 & 0 \\
			1 & 1/2 & 1/2 & 0 \\
			\hline
			& 1/3 & 1/3 & 1/3
		\end{array}
		\qquad
		\begin{array}{c|ccc}
			1/4 & 1/4 & 0 & 0 \\
			1/4 & 0 & 1/4 & 0 \\
			1 & 1/3 & 1/3 & 1/3 \\
			\hline
			& 1/3 & 1/3 & 1/3
		\end{array}.
	\end{equation}
	
	For the SSP3(3,3,2) IMEX scheme, define
	\begin{equation}
		\chi := 1-\frac{1}{\sqrt{2}}.
	\end{equation}
	The explicit and implicit coefficient matrices are
	\begin{equation}
		\tilde{A}
		=
		\begin{bmatrix}
			0 & 0 & 0 \\
			1 & 0 & 0 \\
			1/4 & 1/4 & 0
		\end{bmatrix},
		\qquad
		A
		=
		\begin{bmatrix}
			\chi & 0 & 0 \\
			1-2\chi & \chi & 0 \\
			1/2-\chi & 0 & \chi
		\end{bmatrix}.
	\end{equation}
	The corresponding weights are
	\begin{equation}
		\tilde{\omega}
		=
		\left[
		\frac{1}{6},\frac{1}{6},\frac{2}{3}
		\right],
		\qquad
		\omega
		=
		\left[
		\frac{1}{6},\frac{1}{6},\frac{2}{3}
		\right].
	\end{equation}
	
	The associated double Butcher tableau is
	\begin{equation}
		\begin{array}{c|ccc}
			0 & 0 & 0 & 0 \\
			1 & 1 & 0 & 0 \\
			1/2 & 1/4 & 1/4 & 0 \\
			\hline
			& 1/6 & 1/6 & 2/3
		\end{array}
		\qquad
		\begin{array}{c|ccc}
			\chi & \chi & 0 & 0 \\
			1-\chi & 1-2\chi & \chi & 0 \\
			1/2 & 1/2-\chi & 0 & \chi \\
			\hline
			& 1/6 & 1/6 & 2/3
		\end{array}
	\end{equation}
	\section{Extra cylindrical blast and rotor tests}
	\label{app:Extra cylindrical blast and rotor tests}
	\begin{figure}[h]
		\centering
		
		\subfloat[Cylindrical blast test with initial background magnetic field
		$\mathbf{B}_{x}=0.05$ and conductivity $\sigma=10^{6}$.%
		\label{fig:cyc_blast_sigma_1e6}]
		{
			\includegraphics[width=0.48\textwidth]{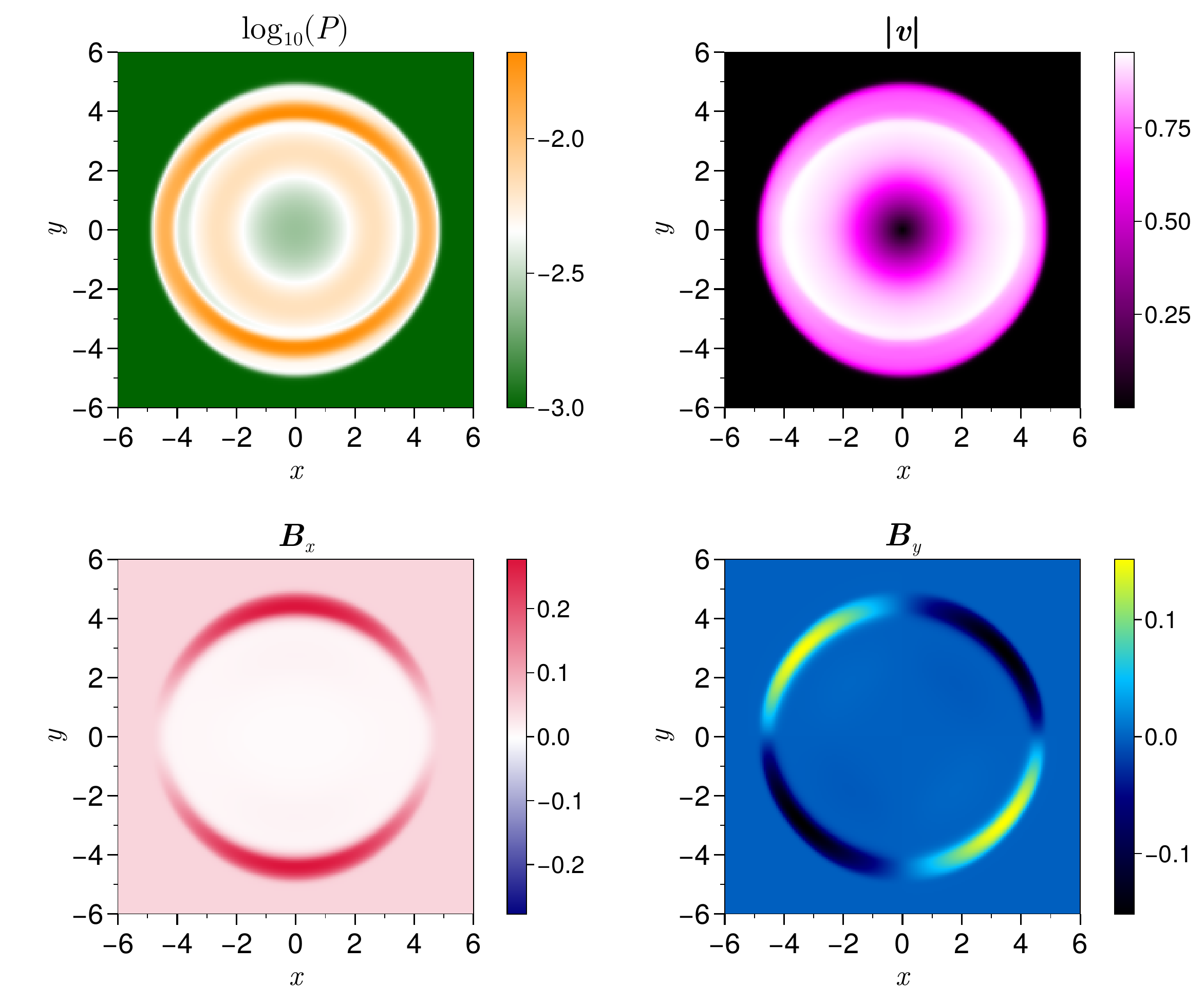}
		}
		\hfill
		\subfloat[Cylindrical blast test with initial background magnetic field
		$\mathbf{B}_{x}=0.1$ and conductivity $\sigma=2000$.%
		\label{fig:cyc_blast_sigma_2k}]
		{
			\includegraphics[width=0.48\textwidth]{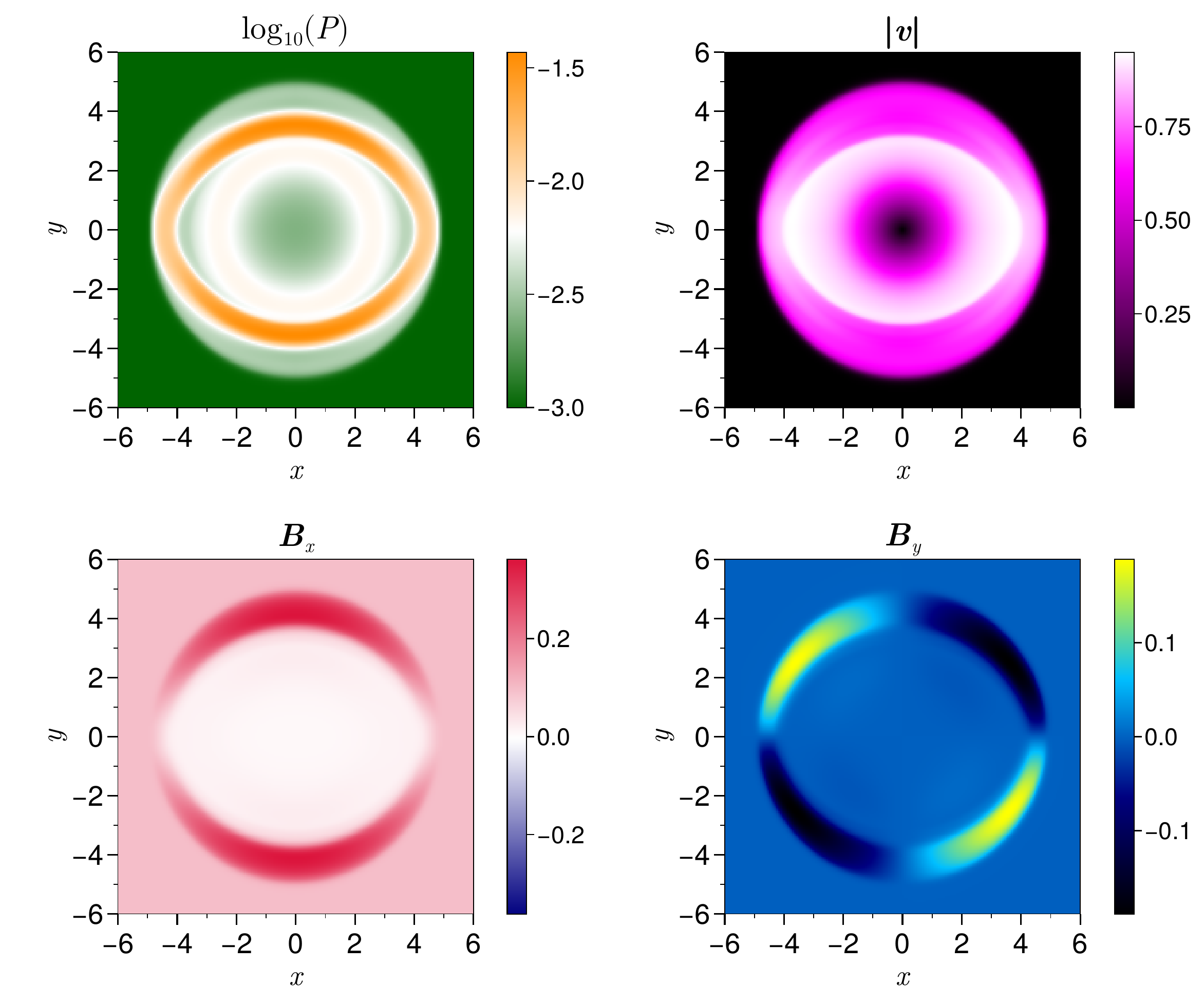}
		}
		
		\caption{Results of the extra cylindrical blast tests}
		\label{fig:Extra_cyc_blast_tests}
	\end{figure}

    	\begin{figure}[h]
		\centering
		
		\subfloat[ These four panel figures corresponds to $\sigma = 10$%
		\label{fig:Milne_rotor_sigma_10}]
		{
			\includegraphics[width=0.48\textwidth]{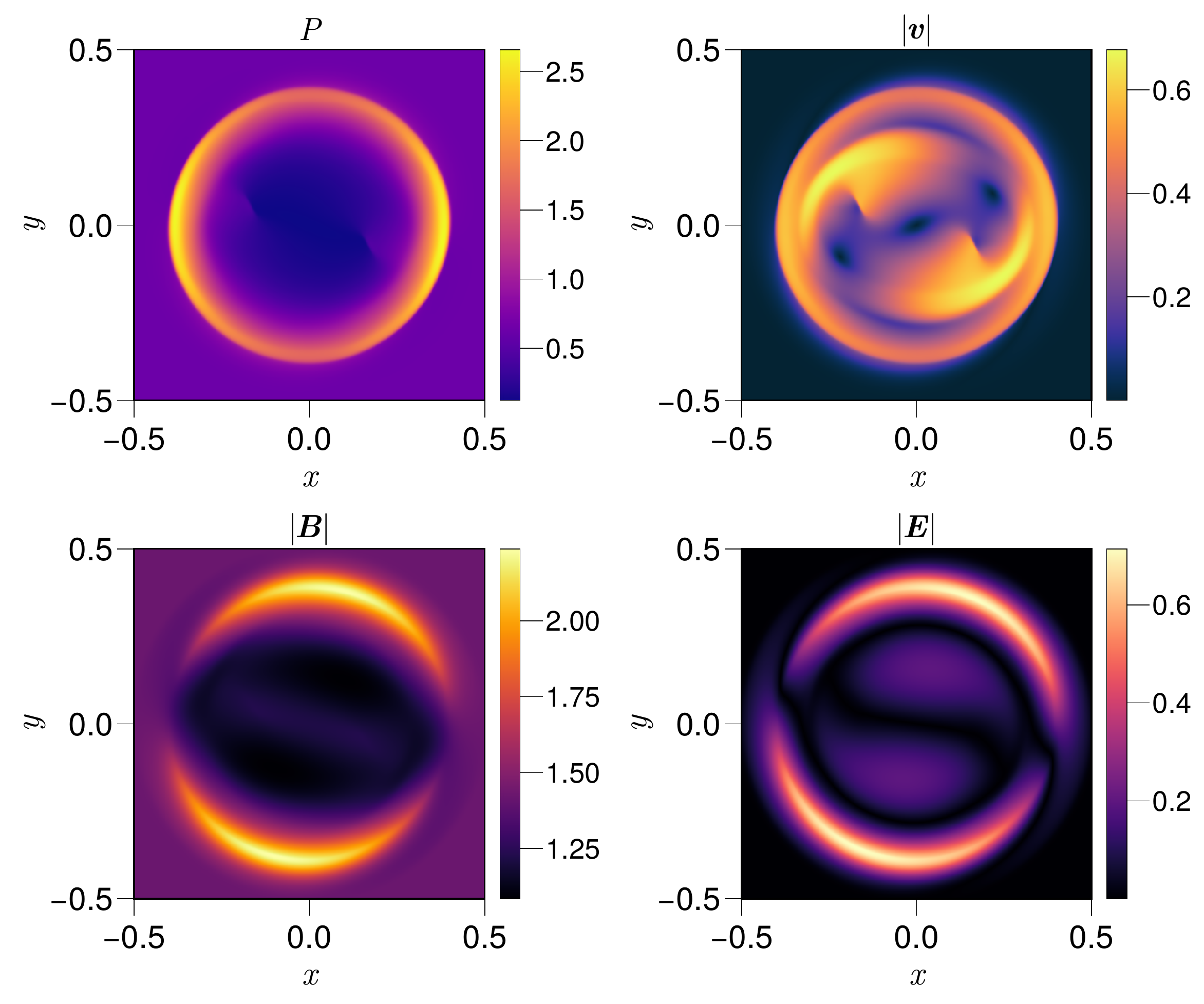}
		}
		\hfill
		\subfloat[These four panel figures corresponds to $\sigma = 1000$.%
		\label{fig:Milne_rotor_sigma_1k}]
		{
			\includegraphics[width=0.48\textwidth]{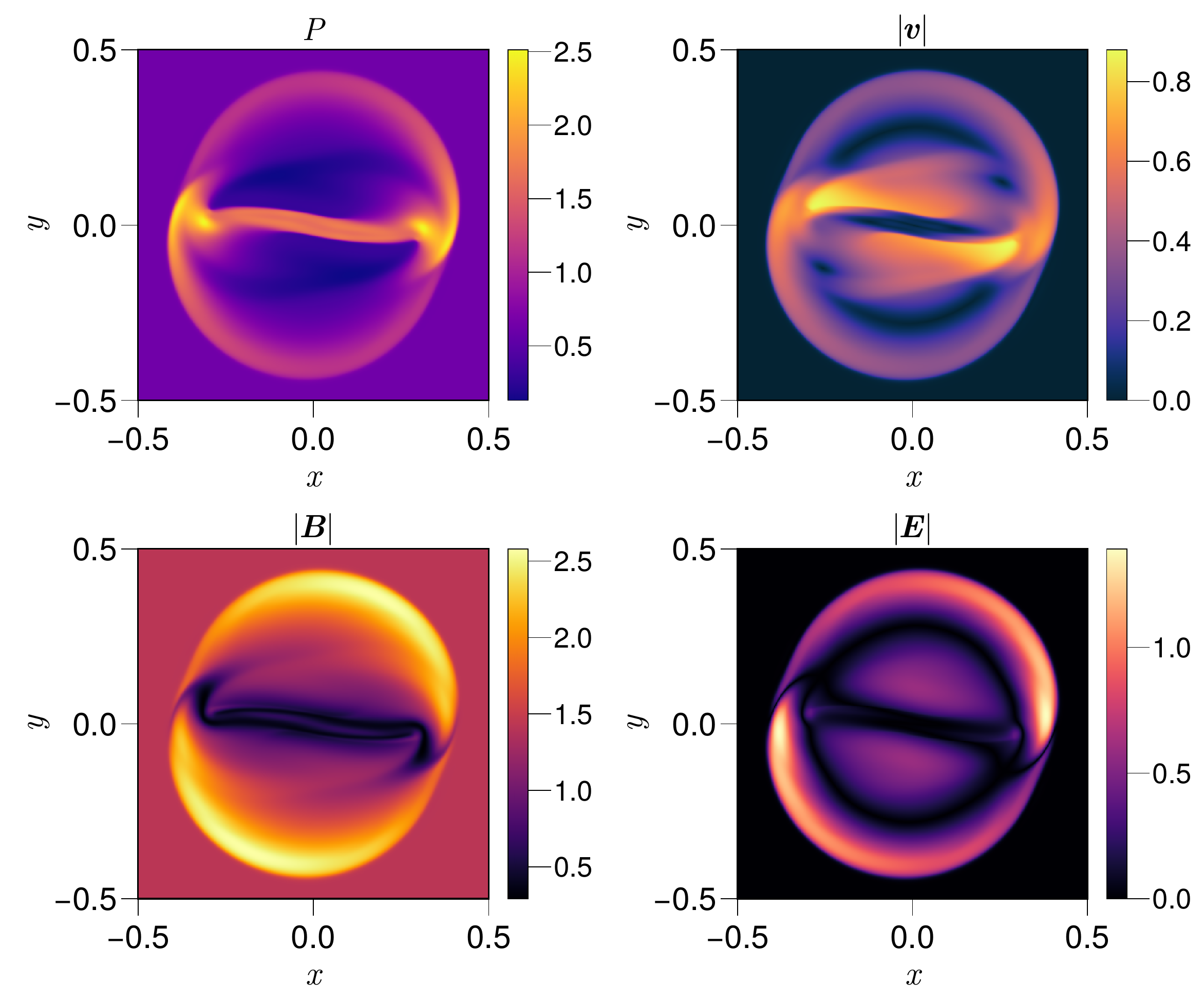}
		}
		
		\caption{In these Milne rotor tests we set $r_{1} = 0.0$ and $\Omega = 0.97$ all the other thermodynamic quantities, electric and magnetic field initialized with the same value as reported in Fig~\ref{fig:milne_rotor_sigma1e6}.}
		\label{fig:Extra_Milne_rotor_tests.}
	\end{figure}

	\newpage
	\twocolumngrid
	
	\bibliography{References}

\end{document}